\documentclass[11pt, a4paper]{article}
\usepackage{jheppub}

\usepackage{graphics}

\usepackage{epsf}
\usepackage{color}
\usepackage{amsmath}
\usepackage{amssymb}
\usepackage{latexsym}
\usepackage{slashed}
\usepackage{float}
\usepackage{cases}
\usepackage{multirow}

\newcommand{\fbinv}{\,\text{fb}^{-1}}
\newcommand{\GeV}{\,\text{GeV}}

\newcommand{\al}[1]{\begin{align}#1\end{align}}

\newcommand{\bp}{\begin{pmatrix}}
\newcommand{\ep}{\end{pmatrix}}
\newcommand{\bb}{\begin{bmatrix}}
\newcommand{\eb}{\end{bmatrix}}

\newcommand{\paren}[1]{\left(#1\right)}
\newcommand{\sqbr}[1]{\left[#1\right]}

\newcommand{\br}[1]{\left\{#1\right\}}

\newcommand{\MKK}{{M_{\text{KK}}}}

\newcommand{\beq}{\begin{equation}}
\newcommand{\eeq}{\end{equation}}
\newcommand{\bea}{\begin{eqnarray}}
\newcommand{\eea}{\end{eqnarray}}
\newcommand{\half}{\frac{1}{2}}

\newcommand{\pal}{\partial}

{

\title{Higgs production and decay processes via loop diagrams in various 6D Universal Extra Dimension Models at LHC}
\author[a]{Kenji Nishiwaki}
\affiliation[a]{Department of Physics, Kobe University,\\
1-1 Rokkodai-Cho, Nada-Ku, Kobe 657-8501, JAPAN}
\emailAdd{nishiwaki@stu.kobe-u.ac.jp}
\abstract{We calculate loop-induced Higgs production and decay processes which are relevant for
the LHC in various six-dimensional Universal Extra Dimension models. 
More concretely, we focus on {the} Higgs production through gluon fusion and {the} Higgs decay into two photons
induced by loop diagrams. 
They are one-loop leading processes and the contribution of Kaluza-Klein particles is considered to be significant.
These processes are divergent in six dimensions. 
Therefore, we employ {a} momentum cutoff, whose size is fixed from the
validity of perturbative calculation through naive dimensional analysis.
{In these six-dimensional Universal Extra Dimension models, the Higgs production cross section through gluon fusion {is highly enhanced} and the Higgs decay width into two photons {is suppressed}.
In particular in the case of the compactification on 
{Projective Sphere}, these effects are remarkable.}
The deviation of the $h^{(0)} \rightarrow 2 \gamma$ signal from
the prediction of the Standard model is much {greater} than that in the case of
the five-dimensional minimal UED model.
{We also consider threshold corrections in the two processes and these effect are noteworthy {even when we take a higher cutoff and/or a heavy KK scale}.
Comparing our calculation to the {recent} LHC results which were published at the Lepton-Photon 2011 {and at the December of 2011} is performed briefly.}}
\arxivnumber{1101.0649 [hep-ph]}
\dedicated{KOBE-TH-10-04}

\begin{document}

\maketitle

\section{Introduction}  

After a long shutdown, the LHC (Large Hadron Collider) {restarted} and new era of particle physics comes.
Stimulated by the advent of two renowned works~\cite{ArkaniHamed:1998rs,Randall:1999ee}, phenomenology in extra dimension has been well studied.
Universal Extra Dimension (UED) is one of the interesting possibility along this direction
and has been studied very well. 
In this model, all the fields describing particles of the Standard Model (SM) propagate in the bulk space.\footnote{
This possibility is first considered within string theory context~\cite{Antoniadis:1990ew}.
}
The minimal UED (mUED) model is constructed with one extra spacial dimension of 
orbifold $S^1/Z_2$~\cite{Appelquist:2000nn}.
This orbifold {imposes} the identification between the extra spacial coordinate $y$ and
$-y$ and there are two fixed points at $y = 0 , \pi R$, where $R$ is the radius of $S^1$.
Due to this identification four-dimensional (4D) chiral fermions describing the SM fermions appear.
One of the interesting points of UED model is that the constraints from the current experiments are very loose. The Kaluza-Klein (KK) mass scale ${M_{\text{KK}}}$, which is defined by
the inverse of the compactification radius $R${,} is constrained~\cite{Appelquist:2000nn,Agashe:2001ra,Agashe:2001xt,Appelquist:2001jz,
Appelquist:2002wb,Oliver:2002up,Chakraverty:2002qk,Buras:2002ej,Colangelo:2006vm,
Gogoladze:2006br} {in the mUED case}. In UED model, the zero mode profile takes constant value and 
the overlap integral between zero mode{s} and KK mode{s} does not generate large deviation 
from the SM result. Therefore we can take the lower KK mass scale
than in {the} other types of extra dimensional models.
{In addition,} the existence of dark matter candidate is naturally explained by the KK parity, which is the remnant
of the translational invariance along the extra spacial direction.
The particle cosmology {in} the five-dimensional (5D) UED models has been studied strenuously~{\cite{Cheng:2002ej,Servant:2002aq,Kakizaki:2005en,Matsumoto:2005uh,Burnell:2005hm,
Kakizaki:2006dz,Kong:2005hn,Matsumoto:2007dp,
Kakizaki:2005uy,Belanger:2010yx,Hisano:2010yh}}.
The collider signature of the 5D UED models is similar to the one of the supersymmetric
theory with neutralino dark matter~\cite{Cheng:2002ab}. The discrimination between these models is {also} well studied~\cite{Datta:2005zs,Matsumoto:2009tb}.

And another thing, UED models with two spacial dimensions have been studied energetically.
Six-dimensional (6D) UED models have remarkable theoretical properties, for example,
prediction of the number of matter generations imposed by the condition of (global) anomaly cancellation~\cite{Dobrescu:2001ae},
{ensuring} proton stability~\cite{Appelquist:2001mj},
generating electroweak symmetry breaking~\cite{ArkaniHamed:2000hv,Hashimoto:2003ve,Hashimoto:2004xz}.
These topics drive us into considering such a class of models.
In phenomenological point of view, there are also interesting aspects in 6D UED model.
In 6D case, the KK mass spectrum is not equally-spaced, up to 
radiative corrections~\cite{Cheng:2002iz,Ponton:2005kx}.
And a 4D new scalar particle named ``spinless adjoint" emerges in the model corresponding to a 6D gauge boson. These are un-eaten physical scalars associated to
the 4D vector components of the 6D gauge bosons.
These two points exert considerable influence on collider physics and particle 
cosmology~\cite{Burdman:2006gy,Dobrescu:2007xf,Dobrescu:2007ec,Freitas:2007rh,
Freitas:2008vh,Ghosh:2008dp,Bertone:2009cb,Blennow:2009ag}.
These studies are executed on the 6D UED model based on two-torus $T^2$~{\cite{
Dobrescu:2004zi,Burdman:2005sr,Cacciapaglia:2009pa}}. It is noted that recently the UED models based on two-sphere $S^2$ are proposed and these models have interesting properties
\cite{Maru:2009wu,Dohi:2010vc}.
In the $S^2$-based models, the KK mass spectrum is totally different from
that of the ordinary $T^2$-based models and we consider that this difference would 
have an impact
on collider and cosmological phenomenology.\footnote{In 5D case, there are also many approaches of considering the non-minimal UED
models~\cite{Flacke:2008ne,Park:2009cs,Csaki:2010az,Haba:2009uu,Haba:2009pb,
Haba:2009wa,Haba:2010xz,Nishiwaki:2010te}.}

In this paper, we focus on the Higgs boson production and decay sequences through
one-loop leading processes expected to occur at the LHC.
In one-loop leading processes, the contribution of KK particles is considered to be significant.
More concretely,
we consider the Higgs production by gluon fusion and the Higgs decay to two photons.
The former process is very important because it is the dominant Higgs production process
at the LHC.
The latter process becomes important in the case when Higgs boson mass is {about from}
$120$ {GeV} to $150$ GeV.
{Actually, the ATLAS and CMS experiments at the CERN LHC have presented their latest results for the $\simeq 2\fbinv$ of data at the center of mass energy 7\,TeV at the Lepton-Photon 2011, Mumbai, India, 22-27 August
2011 and the Higgs decay to two photons process play{s} a significant role at this range~\cite{ATLAS-CONF-2011-135,CMS_PAS_HIG-11-022}.}{\footnote{
{During revising this paper, both the ATLAS and CMS have published the new results, which claim
that there is a peak around 125 GeV~\cite{ATLAS:2012ae,Chatrchyan:2012tx}.}
} }
In the SM, the branching ratio of the decay into two photons is too small, but the signal of this
process is very clear at the LHC experiments.
Using the result of the above two processes, we can perform {a} crude estimation
of the difference of the number of the decay events to two photons from the SM 
expectation value. By naive power counting argument, the production and {the} decay
processes are known to be divergent logarithmically. We adopt the regularization scheme by use of KK momentum cutoff, which is determined by naive dimensional analysis. 
{We also consider threshold corrections in the two processes and these effect are noteworthy{, especially} when we choose low cutoff scale in 6D UED.}

As the end of the introduction, we show the organization of this paper briefly.
In Section~\ref{Universal Extra Dimension on T2Z4}, we give a brief review of 6D UED model on $T^2/Z_4$, which is one of
the $T^2$-based 6D UED model and has been studied well.
In Section~\ref{Calculation of one loop Higgs production and decay processes}, we calculate the rate of the Higgs production process through gluon fusion and
the Higgs decay process to two photons via loop diagrams in the 6D UED model on $T^2/Z_4$.
These results can be applied for the $S^2$-based 6D UED cases with some modifications.
In Section~\ref{Universal Extra Dimension Models based on $S^2$}, we get an overview of gauge theory on $S^2$ and give a brief review
of the two types of $S^2$-based 6D UED models.
{In Section~\ref{Naive Dimensional Analysis}, we estimate the maximal cutoff scale, where the validity of perturbation
will break down.}  
In Section~\ref{The deviation of the rates of Higgs production and its decay from the standard model predictions}, we estimate the deviation of the rate of the Higgs production and decay processes
and evaluate the difference of the event number from the SM results
{with/without threshold corrections}.
Section~\ref{Summary} is devoted to summary and discussions.

\section{Universal Extra Dimension on $T^2/Z_4$
\label{Universal Extra Dimension on T2Z4}}

We give a brief review of UED model on $T^2/Z_4$.
A detailed construction of the minimal 5D UED based on $S^1/Z_2$ is studied in \cite{Appelquist:2000nn}.
We consider a gauge theory on six-dimensional spacetime $M^4 \times T^2/Z_4$, which is a direct product of the four-dimensional Minkowski spacetime $M^4$ and two-torus $T^2$ {with $Z_4$ orbifolding}.
We use the coordinate of six-dimensional spacetime defined by $x^M = (x^{\mu},y,z)$ and the mostly-minus metric convention 
 $\eta_{MN} = \text{diag}(1,-1,-1,-1,-1,-1)$.\footnote{Latin indices ($M,N$) run for $0,1,2,3,y,z$ and Greek indices ($\mu,\nu$) run
for $0,1,2,3$.}
{The representation of} Clifford algebra which we adopt is
\beq
\Gamma^{\mu} = \gamma^{\mu} \otimes I_2 =  \begin{bmatrix} \gamma^{\mu} & 0 \\ 0 & \gamma^{\mu} \end{bmatrix} \ , \ 
\Gamma^y = \gamma^5 \otimes i \sigma_1 = \begin{bmatrix} 0 & i \gamma^5 \\ i \gamma^5 & 0 \end{bmatrix}\ ,\ 
\Gamma^z = \gamma^5 \otimes i \sigma_2 = \begin{bmatrix} 0 & \gamma^5 \\ - \gamma^5 & 0 \end{bmatrix},
\label{eq:gammas}
\eeq
where $\gamma^5$ is 4D chirality operator and
$\sigma_i\ (i=1,2,3)$ are Pauli matrices.
To obtain 4D Weyl fermion from 6D Weyl fermion, we choose the type of orbifold as $Z_4$, not as $Z_2$ in 5D case~\cite{Dobrescu:2004zi,Burdman:2005sr}. 
$Z_4$ symmetry is realized as the rotation on the $y-z$ plane by an angle $\frac{\pi}{2}$ on $T^2$.
This means a bulk scalar field $\Phi(x;y,z)$ obeys the following relation:
\beq
\Phi_t(x,-z,y) = t \Phi_t(x,y,z).
\eeq
$t$ is $Z_4$ parity which takes the possible values $t = \pm 1 , \pm i$ and all the fields are classified according to their parity. Following the general prescription~\cite{Georgi:2000ks}, mode functions of $T^2/Z_4$ $f^{(m,n)}_t(y,z)$ are obtained as follows:\footnote{For simplicity, we drop the overall $-i$ factor for $t = \pm i$ cases.}
\beq
\displaystyle
f^{(m,n)}_t(y,z) =
\left\{
\begin{array}{ll}
\displaystyle  \frac{1}{2 \pi R} \frac{1}{\sqrt{1 + 3 \delta_{m,0} \delta_{n,0}}}
\Big[ \cos\left( \frac{my+nz}{R} \right) + \cos\left( \frac{ny-mz}{R} \right)  \Big] & \text{for} \ t=1, \\
\displaystyle \frac{1}{2 \pi R} \Big[ \cos\left( \frac{my+nz}{R} \right) - \cos\left( \frac{ny-mz}{R} \right)  \Big] & \text{for} \ t=-1, \\
\displaystyle  \frac{1}{2 \pi R} \Big[ \sin\left( \frac{my+nz}{R} \right) -i \sin\left( \frac{ny-mz}{R} \right)  \Big] & \text{for} \ t=i, \\
\displaystyle\frac{1}{2 \pi R} \Big[ \sin\left( \frac{my+nz}{R} \right) +i \sin\left( \frac{ny-mz}{R} \right)  \Big] & \text{for} \ t=-i,
\end{array}
\right.
\label{eq:mode_function}
\eeq
where $m$ and $n$ are $y$ and $z$ directional KK numbers, respectively and take the values
$m \geq 1 , n \geq 0$ or $m=n=0$ (only for $t=1$).\footnote{{The complex factor $i$ in $f_{t=\pm i}^{(m,n)}$ generates
CP violating interactions after KK expansion in KK sector~\cite{Lim:2009pj}}.}

And realizing cancellation of 6D gravitational and SU(2)$_{L}$ global anomalies requires the choice 
of 6D chiralities, for example, as follows~\cite{Dobrescu:2001ae}:
\beq
(\mathcal{Q}_{-} , \mathcal{U}_{+} , \mathcal{D}_{+} , \mathcal{L}_{-} , \mathcal{E}_{+} , \mathcal{N}_{+}),
\eeq
whose zero modes form single generation of the standard model; $\mathcal{Q}_{-}^{(0)} =
(u,d)_L  ,\  \mathcal{U}_{+}^{(0)} = u_R  ,\  \mathcal{D}_{+}^{(0)} = d_R  ,\  \mathcal{L}_{-}^{(0)} = (\nu , l)_L  ,\  \mathcal{E}_{+}^{(0)} =   l_R  ,\  \mathcal{N}_{+}^{(0)} = \nu_{R}$.
The $\pm$ suffixes represent 6D chirality of each field and 6D chirality operator is defined as
\beq
\Gamma_7 = \gamma^5 \otimes \sigma_3.
\eeq
Using 6D chiral projective operator {$\Gamma_{\pm} \equiv \frac{1}{2} (1 \pm \Gamma_7)$}, 6D Weyl fermions $\Psi_{\pm}$ are described as follows;
\beq
\Psi_{+} = \begin{pmatrix} \psi_{+R} \\ \psi_{+L} \end{pmatrix}, \quad \Psi_{-} = \begin{pmatrix} \psi_{-L} \\ \psi_{-R} \end{pmatrix},
\eeq
where $\psi_{L(R)}$ is a left(right)- handed 4D Weyl fermion.
We can take the boundary condition of 6D fermion $\Psi_6 = (\psi,\Psi)^{\mathrm{T}}$ 
($\mathrm{T}:$ transpose) as {in}~\cite{Scrucca:2003ut}{:}
\al{
\Psi_6 (x, -z,y) &= (i)^{\half + r} \left( \frac{1+\Gamma^y \Gamma^z}{\sqrt{2}} \right) P \Psi_6 (x, y,z) 
\notag \\
{\Longleftrightarrow}
\begin{pmatrix} \psi \\ \Psi \end{pmatrix}(x , -z,y) &= 
\begin{pmatrix} i^r & 0 \\ 0 & i^{r+1} \end{pmatrix} P
\begin{pmatrix} \psi \\ \Psi \end{pmatrix}(x , y,z).
}
$r$ is $Z_4$ twist factor which can takes the values ($r=0,1,2,3$) and $P$ is group twist matrix for
fundamental representation with {the}
$Z_4$ identification $(P^4 =1)$, which we discuss soon later.
When we choose {the values of $r$} as $0$ or $3$, zero mode sectors of $\Psi_{\pm}$ become 4D chiral.

Next, we go on to the gauge sector.
The boundary condition of this part is {as in}~\cite{Scrucca:2003ut}{:}
\beq
A_{\mu}(x, i \omega) = P A_{\mu}(x, \omega) P^{-1}, \quad A_{\omega}(x, i \omega) = (-i) P A_{\omega}(x,\omega) P^{-1}.
\eeq
Here we define a complexified coordinate and a vector field component for clarity as
\beq
\omega \equiv \frac{y+iz}{\sqrt{2}}, \quad A_{\omega} \equiv \frac{A_y -i A_z}{\sqrt{2}}.
\eeq 
In UED model, we do not break the gauge symmetry by boundary condition.
Then the matrix $P$ is selected as $P = \mathbf{1}$.
This means that none of the fields belonging to $A_{\omega}$ (or $A_{\bar{\omega}}$) takes zero mode, which is {an} would-be exotic SM particle. 
Finally we discuss the 6D scalar $\Phi$.
The boundary condition for this field is very simple:
\beq
\Phi(x, i \omega) = P \Phi(x, \omega).
\eeq 
Choosing $P = \mathbf{1}$, $\Phi$'s zero mode remains and can be identified as the SM Higgs field. 
From above discussion, we can form the zero mode sector just as the SM one.

We write down the part of the 6D UED {Lagrangian} which is requisite for {our} calculation.
{The 6D action takes the form as follows:}
\begin{align}
S &= \int_0^{2 \pi R} dy {\int_0^{2 \pi R}} dz \int d^4 x \Bigg\{ - \frac{1}{2} \sum_{i=1}^{3} \mathrm{Tr} \big[ F_{MN}^{(i)} F^{(i)MN}  \big] \notag \\
& \qquad {+ (D^M H)^{\dagger} (D_M H) + \bigg[ \mu^2 |H|^2 - \frac{\lambda_6^{(H)}}{4} |H|^4 \bigg]} \notag \\
& \qquad +i \bar{\mathcal{Q}}_{3-} \Gamma^M D_M \mathcal{Q}_{3-} + i  \bar{\mathcal{U}}_{3+} \Gamma^M D_M \mathcal{U}_{3+} - 
\bigg[ \lambda_{6}^{(t)} \bar{\mathcal{Q}}_{3-} (i \sigma_2 H^{\ast}) \mathcal{U}_{3+} + \mathrm{h.c.} \bigg]
\Bigg\}.
\label{eq:6Daction}
\end{align}
$F^{(i)}_{MN}$ are the field strength{s} of gauge fields, where
$F^{(i)}_{MN} = \pal_M A_N^{(i)} - \pal_N A_M^{(i)} -i g_6^{(i)} [A_M^{(i)} , A_N^{(i)}]$, 
and the gauge groups
are those for $U(1)_{Y} \ (i=1), SU(2)_{L} \ (i=2)$ and $SU(3)_{C} \ (i=3)$ in the SM. 
The covariant derivatives $D_M$ are expressed in our convention as
\beq
D_M = \pal_M -i \sum_{i=1}^{3} g_6^{(i)} T^{(i)a} A^{(i)a}_M,
\eeq
where {} $g_6^{(i)}$ are the six-dimensional gauge couplings and {} $T^{(i)a}$ are
the group generators of each corresponding gauge group.
$H$ is the Higgs doublet, and $\mu$, $\lambda_6^{(H)}$ and $\lambda_6^{(t)}$ are
the usual Higgs mass, Higgs self coupling and Yukawa coupling of {the} top quark in 6D theory, respectively.\footnote{All the six-dimensional couplings are dimensionful.
After the KK expansion, corresponding 4D couplings become dimensionless as
they should be so.}
$\mathcal{Q}_{3-}$ is the quark doublet in third generation and $\mathcal{U}_{3+}$ is the top quark singlet.

We are ready to derive the four-dimensional effective action by expanding all the 6D fields by use of {Eq.}~(\ref{eq:mode_function}).
The concrete forms of KK expansion are as follows:
\begin{align}
A^{(i)}_{\mu} (x; y,z) &= \frac{1}{2 \pi R} \bigg\{ A^{(i)(0)}_{\mu}(x) \notag \\
& \qquad {+ {\sum_{m \geq 1, n \geq 0}} A_{\mu}^{(i)(m,n)}(x)  \Big[ \cos\left( \frac{my+nz}{R} \right) + \cos\left( \frac{ny-mz}{R} \right)  \Big]   \bigg\} },
 \\
A^{(i)}_{\omega} (x; y,z) &= \frac{1}{2 \pi R} \bigg\{  {\sum_{m \geq 1, n \geq 0}} A_{\omega}^{(i)(m,n)}(x)  \Big[ \sin\left( \frac{my+nz}{R} \right) +i \sin\left( \frac{ny-mz}{R} \right)  \Big]   \bigg\},
\\
H (x; y,z) &= \frac{1}{2 \pi R} \bigg\{ H^{(0)}(x) + {\sum_{m \geq 1, n \geq 0}} H^{(m,n)}(x)  \Big[ \cos\left( \frac{my+nz}{R} \right) + \cos\left( \frac{ny-mz}{R} \right)  \Big]   \bigg\},
 \\
\mathcal{Q}_{3-} (x; y,z) &=  \frac{1}{2 \pi R}
\begin{pmatrix}
\displaystyle   Q_{3L}^{(0)}(x) + {\sum_{m \geq 1, n \geq 0}} Q_{3L}^{(m,n)} \Big[ \cos\left( \frac{my+nz}{R} \right) + \cos\left( \frac{ny-mz}{R} \right)  \Big] \\
 \qquad \displaystyle  {\sum_{m \geq 1, n \geq 0}} Q_{3R}^{(m,n)}
 \Big[ \sin\left( \frac{my+nz}{R} \right) -i \sin\left( \frac{ny-mz}{R} \right)  \Big]   
 \end{pmatrix},
\\
\mathcal{U}_{3+} (x; y,z) &= \frac{1}{2 \pi R}
\begin{pmatrix}
\displaystyle   t_{R}^{(0)}(x) + {\sum_{m \geq 1, n \geq 0}} t_{R}^{(m,n)} \Big[ \cos\left( \frac{my+nz}{R} \right) + \cos\left( \frac{ny-mz}{R} \right)  \Big] \\
 \qquad \displaystyle  {\sum_{m \geq 1, n \geq 0}} t_{L}^{(m,n)}
 \Big[ \sin\left( \frac{my+nz}{R} \right) -i \sin\left( \frac{ny-mz}{R} \right)  \Big]   
 \end{pmatrix}.
\end{align}
{In} the fermionic part, we choose all the twist factors as $r=0$.
Now we can find the SM fields $A^{(0)(i)}_{\mu}, H^{(0)}, Q^{(0)}_{3L} \big(= \big( t^{(0)}_L, b^{(0)}_L \big)^{\mathrm{T}} \big)$ and $t^{(0)}_R$ in the zero mode sectors. 
{Here} we focus on the {5D} Higgs doublet in terms of {4D} component fields:
\beq
H^{(0)} = \begin{pmatrix} \phi^{+(0)} \\
\frac{1}{\sqrt{2}} \big( v + h^{(0)} + i \chi^{(0)} \big)
\end{pmatrix}, \quad 
H^{(m,n)} = \begin{pmatrix} \phi^{+(m,n)} \\
\frac{1}{\sqrt{2}} \big( h^{(m,n)} + i \chi^{(m,n)} \big)
\end{pmatrix}.
\eeq
At the zero mode part, $v$ and $h^{(0)}$ are the ordinary four-dimensional Higgs 
Vacuum Expectation Value
(VEV) and the
usual SM physical Higgs field.
$\phi^{+(0)}$ is the would-be Nambu-Goldstone boson of $W_{\mu}^{+(0)}$ 
and generate the longitudinal d.o.f.
for $W^{+(0)}_{\mu}$ and $\chi^{(0)}$ is for $Z^{(0)}_{\mu}$.
Subsequently, we take notice of the (4D) scalar KK excitation modes.
In addition to the Higgs KK excitation modes ${\{}h^{(m,n)},\phi^{+(m,n)},\chi^{(m,n)}{\}}$,
there are other excitation modes {closely related to the {(zero mode)} massive gauge bosons}, which are $y$ and $z$ components of 6D gauge fields.

Throughout this paper, we use information about W boson zero mode and its KK particles and their associative particles, which are zero and KK modes of $\phi^{{+}}$, $W^{{+y}}$ and $W^{{+z}}$. 
In what follows, we discuss only the free {Lagrangian} with respect to the non-zero
KK modes
of W boson and their associative particles 
since the zero mode part is the same with the SM one.

\noindent
From {Eq.} (\ref{eq:6Daction}), we can read off the free {Lagrangian} part $S^W|_{{\text{free}}}$ {as}
\begin{align}
&S^W |_{{\text{free}}}= \int d^4 x {\sum_{m \geq 1, n \geq 0}} \Bigg\{
- \frac{1}{2} \Big[ F_{\mu \nu}^{W(m,n)} F^{W (m,n) \mu \nu} \Big]_{{\text{quad}}}
\notag \\
& \ + \frac{1}{2} \Big[ (\pal_{\mu} \phi^{+(m,n)}) (\pal^{\mu} \phi^{-(m,n)})
+ (\pal_{\mu} W^{+(m,n)y}) (\pal^{\mu} W^{-(m,n)y}) 
+ (\pal_{\mu} W^{+(m,n)z}) (\pal^{\mu} W^{-(m,n)z}) \Big] \notag \\
& \ + \big(m_W^2 + m_{(m,n)}^2 \big) W_{\mu}^{+(m,n)} W^{\mu -(m,n)} 
- m_{(n)}^2 W^{+(m,n)y} W^{-(m,n)y} \notag \\
& \ - m_{(m)}^2 W^{+(m,n)z} W^{-(m,n)z} + m_{(m)} m_{(n)} 
\Big[ W^{+(m,n)y} W^{-(m,n)z} + W^{-(m,n)y} W^{+(m,n)z}\Big] \notag \\
& \ - m_{(m,n)}^2 \phi^{+(m,n)} \phi^{-(m,n)}
- i m_W \phi^{-(m,n)} \Big[ m_{(m)} W^{+(m,n)y} +  m_{(n)} W^{+(m,n)z}  \Big] \notag \\
& \ + i m_W \phi^{+(m,n)} \Big[ m_{(m)} W^{-(m,n)y} + m_{(n)} W^{-(m,n)z}  \Big]
\notag \\ 
& \ - m_W^2 \Big[ W^{+(m,n)y} W^{-(m,n)y} + W^{+(m,n)z} W^{-(m,n)z}  \Big] \notag \\
& \ -i m_W \Big[  (\pal^{\mu} \phi^{-(m,n)}) W_{\mu}^{+(m,n)} -  (\pal^{\mu} \phi^{+(m,n)}) W_{\mu}^{-(m,n)} \Big] \notag \\
& \ - \Big[ m_{(m)} (\pal^{\mu} W^{+(m,n)y}) + m_{(n)} (\pal^{\mu} W^{+(m,n)z}) \Big] W^{-(m,n)}_{\mu} \notag \\
& \
- \Big[   m_{(m)} (\pal^{\mu} W^{-(m,n)y}) + m_{(n)} (\pal^{\mu} W^{-(m,n)z}) \Big] W^{+(m,n)}_{\mu}\Bigg\},
\label{eq:W-free}
\end{align}
where  $\Big[ F_{\mu \nu}^{W(m,n)} F^{W (m,n) \mu \nu} \Big]_{{\text{quad}}} = \big(\pal^{\mu} W^{+(m,n)\nu} - \pal^{\nu} W^{+(m,n)\mu} \big) \big(\pal_{\mu} W^{-(m,n)}_{\nu} - \pal_{\nu} W^{-(m,n)}_{\mu} \big)$ is the KK W-boson{'}s kinetic term,
$m_W$ is {the} W-boson mass; $m_{(m)} = \frac{m}{R}$ and $m_{(m,n)}^2 = m_{(m)}^2 + m_{(n)}^2$ are describing the KK masses.

{Here} we adopt the following type of gauge-fixing term about W boson to eliminate cross terms in {Eq.} (\ref{eq:W-free}) {as}
\al{
S^{W}_{{\text{gf}}} &= - \frac{1}{\xi} \int_{0}^{2 \pi R} dy {\int_0^{2\pi R}} dz  \int d^4 x \Big[ \pal_{\mu} W^{+ \mu} + \xi \big( 
\pal_y W^{+y} + \pal_z W^{+z} -i m_W \phi^{+} \big) \Big] \notag \\ 
&\qquad \times \Big[ \pal_{\mu} W^{- \mu} + \xi \big( 
\pal_y W^{-y} + \pal_z W^{-z} +i m_W \phi^{-} \big) \Big].
}
From {Eq.} (\ref{eq:W-free}), the mass of {the field} $W^{+(m,n)}_{\mu}$ is determined as $m_{W,(m,n)}^2 = m_W^2 + m_{(m,n)}^2$.
Meanwhile, we have to diagonalize the scalar mass terms about $\phi^{+(m,n)}$, $W^{+(m,n)y}$ and $W^{+(m,n)z}$ to execute  perturbative calculation{s}.
When we focus on this part $S^{W}_{\text{scalar-mass}}$ out of $S^W + S^W_{{\text{gf}}}$,
\beq
S^{W}_{\text{scalar-mass}}  = - \int d^4 x {\sum_{m \geq 1, n \geq 0}} 
\Big( W^{+(m,n)y} , W^{+(m,n)z} , \phi^{+(m,n)} \Big)
\mathcal{M}_{(m,n)}
\begin{pmatrix}
W^{-(m,n)y} \\ W^{-(m,n)z} \\ \phi^{-(m,n)}
\end{pmatrix},
\eeq
\beq
\mathcal{M}_{(m,n)} =
\begin{bmatrix}
m_W^2 + \xi m_{(m)}^2 + m_{(n)}^2 && - (1 - \xi) m_{(m)} m_{(n)} && -i (1 + \xi) m_W m_{(m)} \\
- (1 - \xi) m_{(m)} m_{(n)} && m_W^2 + m_{(m)}^2 + \xi m_{(n)}^2 && -i (1 + \xi) m_W m_{(n)} \\
+i (1 + \xi) m_W m_{(m)} && +i (1 + \xi) m_W m_{(n)} && \xi m_W^2 + m_{(m)}^2 + m_{(n)}^2
\end{bmatrix}.
\eeq
{By} using those mass eigenstates ${\{} G^{+(m,n)} , a^{+(m,n)} , H^{+(m,n)} {\}}$, we can diagonalize the matrix $\mathcal{M}_{(m,n)}$ to the following form{:}
\beq
\begin{pmatrix}
G^{\pm(m,n)} \\ a^{\pm(m,n)} \\ H^{\pm(m,n)} 
\end{pmatrix}
= N_{(m,n)}^{\pm}
\begin{pmatrix}
W^{\pm (m,n)y} \\ W^{\pm (m,n)z} \\ \phi^{\pm (m,n)}
\end{pmatrix},
\eeq
\beq
N_{(m,n)}^{\pm} = \frac{1}{m_{W,(m,n)} m_{(m,n)}}
\begin{bmatrix}
m_{(m)} m_{(m,n)} && m_{(n)} m_{(m,n)} && \mp i m_W m_{(m,n)} \\
\mp i m_W m_{(m)} && \mp i m_W m_{(n)} && m_{(m,n)}^2 \\
- m_{(n)} m_{W,(m,n)} && + m_{(m)} m_{W,(m,n)} && 0
\end{bmatrix},
\eeq
\beq
 N^{-}_{(m,n)} \mathcal{M}_{(m,n)} \big( N^{-}_{(m,n)}  \big)^{\dagger} = \text{diag}
\big( \xi m_{W,(m,n)}^2 \ ,\  m_{W,(m,n)}^2 \ ,\  m_{W,(m,n)}^2 \big).
\eeq
This result means that $G^{+(m,n)}$ is the would-be Nambu-Goldstone boson of $W^{+(m,n)}_{\mu}$
and the others $a^{+(m,n)},H^{+(m,n)}$ are physical 4D scalars.
It is noted that ${H}^{+(m,n)}$ is called ``spinless adjoint" because ${H}^{+(m,n)}$
is constructed only {by} extra spacial {components} of the 6D gauge boson $W^{+(m,n)}_{\mu}$.
They contribute to $h^{(0)} \rightarrow 2 \gamma$ Higgs decay process via loop diagrams.

Next, we derive the mass eigenstates of fermions.
Just like the case mentioned above, we again consider {the} KK part only.
The kinetic terms are diagonal, {and} therefore there is no need to discuss the part.
The mass term of $(m,n)$-th KK mode fermions arising from {Eq.} (\ref{eq:6Daction}) is
\beq
\Big( \bar{t}^{(m,n)}_R \ ,\  \bar{Q}^{(m,n)}_{tR} \Big)
\begin{pmatrix}
- m_{(m)} + im_{(n)} & m_t \\
m_t & m_{(m)} +i m_{(n)}
\end{pmatrix}
\begin{pmatrix}
t_L^{(m,n)} \\ Q_{tL}^{(m,n)}
\end{pmatrix}
+ \text{h.c.},
\eeq
where $m_t$ is the zero mode top quark mass and $Q_{t}^{(m,n)}$ is the upper component of the SU(2) doublet $Q_{3}^{(m,n)}$.
By using the following unitary transformation including chiral rotation, we can derive the ordinary diagonalized Dirac mass term as follows:
\beq
\begin{pmatrix}
t^{(m,n)} \\ Q_{t}^{(m,n)} 
\end{pmatrix}
=
\begin{pmatrix}
e^{\frac{i}{2} \gamma^5 \varphi_{(m,n)}} & 0 \\
0 & e^{-\frac{i}{2} \gamma^5 \varphi_{(m,n)}} 
\end{pmatrix}
\begin{pmatrix}
- \cos{\alpha_{(m,n)}} \gamma^5 & \sin{\alpha_{(m,n)}} \\
\sin{\alpha_{(m,n)}} \gamma^5 & \cos{\alpha_{(m,n)}}
\end{pmatrix}
\begin{pmatrix}
t^{'(m,n)} \\ Q_{t}^{'(m,n)} 
\end{pmatrix},
\eeq
where $t^{'(m,n)}$ and $Q_{t}^{'(m,n)} $ are mass eigenstates of their corresponding fields 
with degenerate
$(m,n)$-th level masses{;} $m_{t,(m,n)}^2 = m_t^2 + m_{(m,n)}^2$.
The mixing angles $\varphi_{(m,n)}$ and $\alpha_{(m,n)}$ are determined as 
\beq
\tan{\varphi_{(m,n)}} = - \frac{m_{(n)}}{m_{(m)}}, \quad \cos{{2} \alpha_{(m,n)}} = \frac{m_{(m,n)}}{m_{t,(m,n)}},
\eeq
from the condition to obtain the ordinary diagonalized Dirac mass matrix.
Now we are ready to calculate the rates of Higgs processes at the LHC.
{Some} requisite interactions {in this paper} are discussed at the next section.


\section{Calculation of one loop Higgs production and decay processes
\label{Calculation of one loop Higgs production and decay processes}}


We calculate {some} virtual effect{s} of KK particle via loop diagrams in the Higgs production process through gluon ($g$) fusion $2 g \rightarrow h^{(0)}$
and the Higgs decay process to two photon ($\gamma$) $h^{(0)} \rightarrow 2 \gamma$.
Those processes are 1-loop leading and it is expected that 
the effects of
massive KK particles are significant.
In addition, there is another 1-loop leading Higgs decay process to photon and Z-boson ($Z$) $h^{(0)} \rightarrow \gamma Z$, which we do not discuss in this paper.
Before the concrete discussion about interactions, we have to understand the general structure of interactions which is needed for our study.
In the scope of this paper, all external particles are SM particles, which are described by zero modes. 
This means the effective couplings which we use are obtained by the following type of integrals concerning mode
functions $f_t^{(m,n)}$,
\begin{align}
\int_0^{2\pi R} dy {\int_0^{2\pi R}} dz \left\{ f_{t_i}^{(0,0)} f_{t_j}^{(m,n)} f_{t_k}^{(m',n')} \right\}
\ [\text{3-point}], \\
\int_0^{2\pi R} dy {\int_0^{2\pi R}} dz \left\{f_{t_l}^{(0,0)} f_{t_i}^{(0,0)} f_{t_j}^{(m,n)} f_{t_k}^{(m',n')} \right\} \ [\text{4-point}],
\end{align}
where the Latin indices $i,j,k,l$ indicate  type{s} of the particles.
The $Z_4$-parities $t_l,t_i$ are determined as $t_l =t_i=1$
and the condition $(t_j)^{\ast} = t_k$ is required from $Z_4$ invariance of the action {in Eq.}~(\ref{eq:6Daction}).
Because of orthonormality of mode functions, we know that
the integrals are non-vanishing only when
$(m,n) = (m',n')$ and the 
integrals can be reduced to the ones for the zero modes alone. 
In other words, the value of the vertex containing KK modes,
which are described by the above integrals, 
is exactly the same with the value of the corresponding vertex for the zero mode alone in the basis of gauge eigenstates.

We give a comment on the Higgs mass $m_h$ and the lowest KK mass ${M_{\text{KK}}}${, which is defined as $1/R$ on the geometry of $T^2$}.
In UED model, those two parameters are free, which means they are not determined by the theory,
but there are some constraints on these parameters.
From the result of LEP{2} experiment, $m_h$ is bounded from below as $m_h > 114\  \text{GeV}$.
And recently another bound is announced from {the LHC experiments~{\cite{ATLAS-CONF-2011-135,CMS_PAS_HIG-11-022,ATLAS:2012ae,Chatrchyan:2012tx}}. We discuss this point {in Section~\ref{The deviation of the rates of Higgs production and its decay from the standard model predictions}}.}

We ignore the graviton contributions.
In any 6D UED model, 6D Planck scale $M_{\ast}$ is related to 4D Planck scale $M_{\text{pl}}$
through {a} KK mass scale ${M_{\text{KK}}}$ as follows:
\beq
M_{\ast}^2 \sim {M_{\text{KK}}} M_{\text{pl}}.
\eeq
$M_{\text{pl}}$ is approximately $10^{18}$ GeV and we are interested in the case ${M_{\text{KK}}} \sim
\mathcal{O}(1)$ TeV.
Then the magnitude of $M_{\ast}$ is estimated easily as $\sim 10^{10}$ GeV and
{gravitons are} still weekly coupled to other fields.

\subsection{$2g \rightarrow h^{(0)}$ process}

This gluon fusion process gets contribution only from the fermion triangle loops at 1-loop level.
The SM contribution is calculated in~{Eq.}~\cite{Georgi:1977gs,Rizzo:1979mf}.
In UED model, the intermediate fermions are not only SM ones (zero modes) but also
their KK excitations.
Studies of the production process for the case of 5D minimal UED~\cite{Petriello:2002uu} and 6D $S^2/Z_2$ UED~\cite
{Maru:2009cu} are
made.
We consider only contributions from the top quark and its KK states.
The reason why we ignore other {types of} quarks and its KK modes is that the coupling of fermions to the Higgs
is proportional to each zero mode quark mass{,} and {thereby} those effects are negligible in our analysis.
In terms of {the} fermion mass eigenstates, the interactions of KK quarks {are} 
\begin{align}
S^{t}_{{\text{int}}} &= \int d^4 x {\sum_{m \geq 1, n \geq 0}} \notag \\
&  \times \Bigg\{ \Big( \bar{t}^{'(m,n)} \ ,\  \bar{Q}^{'(m,n)}_{t} \Big)
\Bigg[
\begin{pmatrix}
1 & 0 \\
0 & 1
\end{pmatrix}
g^{(3)} \gamma^{\mu} g_{\mu} \notag \\
& \qquad -
\frac{m_t}{v}h^{(0)}
\begin{pmatrix}
\sin{2\alpha_{(m,n)}} & \cos{2\alpha_{(m,n)}} \gamma^5 \\
- \cos{2\alpha_{(m,n)}} \gamma^5 & \sin{2\alpha_{(m,n)}}
\end{pmatrix}
\Bigg]
\begin{pmatrix}
t^{'(m,n)} \\ Q_{t}^{'(m,n)}
\end{pmatrix}
\Bigg\}.
\label{eq:top-couplings}
\end{align}
The production cross section of $2g \rightarrow h^{(0)}$ process is given as follows:
\beq
{\sigma_{2g \rightarrow h^{(0)}} = \frac{\sqrt{2}\pi G_F}{64} \paren{\frac{\alpha_s}{\pi}}^2 |F_{\text{{gluonfusion}}}|^2,}
\eeq
where $G_F$ is the Fermi constant and $\alpha_s $ is the QCD coupling.
$F_{\text{{gluonfusion}}}$ is the loop function, {which consists of}
the SM top quark effect $F_t^{{\text{SM}}}$, the KK top quark effect $F_t^{{\text{KK}}}$ {and
the threshold correction $F_{\text{gluonfusion}}^{\text{TC}}$.}
Then we can write $F_{\text{{gluonfusion}}} = F_t^{{\text{SM}}} + F_{t}^{{\text{KK}}} {+ F_{\text{gluonfusion}}^{\text{TC}}}$ and $F_t^{{\text{SM}}}$ is given in {Ref.}~\cite{Rizzo:1979mf}
in our notation as
\beq
F_t^{\text{{SM}}}
= {-2} \lambda(m_t^2) {+} \lambda(m_t^2) (1-4\lambda(m_t^2)) J\paren{\lambda(m_t^2)},
\label{eq:FtSM}
\eeq
{where $\lambda(m^2)$ and the loop function $J(\lambda)$ are defined as}
\al{
{\lambda(m^2)} &{= m^2/m_h^2,} \label{defoflambda} \\
{J(\lambda)} &{= \int_0^1 {dx\over x} \ln \sqbr{{x(x-1)\over\lambda}+1-i\epsilon}} \notag \\
&{=	\begin{cases}
		\displaystyle -2\sqbr{\arcsin{1\over\sqrt{4\lambda}}}^2
			&	\text{(for $\lambda\geq{1\over4}$)},\\
		\displaystyle {1\over2}\sqbr{
			\ln{1+\sqrt{1-4\lambda}\over1-\sqrt{1-4\lambda}}
			-i\pi
			}^2
			&	\text{(for $\lambda<{1\over4}$)},
		\end{cases} 
		} \label{eq:scalarPVfunction}
}
{respectively.}\footnote{{Our loop function $J$ is related to the three-point scalar Passarino-Veltman function $C_0$ in Refs.~\cite{Passarino:1978jh,Denner:1991kt} as $J = m_h^2 C_0$.}}
%
The KK top quark coupling to the gluon and the zero mode Higgs is shown in {Eq.} (\ref{eq:top-couplings}).
After some calculation, we can get the form of $F_t^{{\text{KK}}}$ in 6D UED model, where {the concrete form is}
\al{
F_t^{{\text{KK}}} &= 2 \sum_{m \geq 1, n \geq 0} \paren{\frac{m_t}{m_{t,(m,n)}}}^2 \notag \\
& \quad \times \br{{-2} \lambda(m_{t,(m,n)}^2) {+} \lambda(m_{t,(m,n)}^2) (1-4\lambda(m_{t,(m,n)}^2)) J\paren{\lambda(m_{t,(m,n)}^2)}} .
\label{eq:FtKK}
}
{It is noted that $F_t^{\text{SM}}$ and $F_t^{\text{KK}}$ contain the $(-1)$ factor due to fermionic loop.}
Our result is directly related to the minimal 5D case in {Ref.}~\cite{Petriello:2002uu}.
The reason is that the only difference between 5D and 6D case is the KK top quark mass
spectrum and the structure of the Feynman diagrams itself describing this process is
completely the same.
The $m^2$ in ${\lambda (= {m^2}/{m_h^2})}$ indicates the intermediate mass scale propagating
in the loops and we consider the situation that KK scale $m_{(m,n)}$ is much greater than the Higgs scale $m_h$.
{It is noted that we only focus on the light Higgs possibility; $120\, \text{GeV}  \lesssim m_h \lesssim 150\, \text{GeV}$
and thereby we have to only consider the ${\lambda} \geq 1/4$ case.}
{
Finally the contribution from threshold correction is obtained as
\beq
F_{\text{gluonfusion}}^{\text{TC}} = \sqbr{\paren{\frac{\alpha_s}{\pi}} \frac{1}{v}}^{-1} C'_{hgg},
\label{FtTC}
\eeq
where $C'_{hgg}$ is a dimensionful coefficient describing the threshold correction and is related to the dimensionless constant in {a part of the} Lagrangian $C_{hgg}$ with the UED cutoff scale ${\Lambda_{\text{UED}}}$ as
\beq
C'_{hgg} = \frac{C_{hgg}\paren{v \over \sqrt{2}}}{{\Lambda_{\text{UED}}}^2}.
\label{TCcoefficient_gluonfusion}
\eeq
We see the details in Appendix~{\ref{Detail on threshold correction}}.
}

From naive power counting, this result is logarithmically divergent.
The reason is the following.
Higgs decay through gluon fusion is described with dimension-six operator {in four-dimensional picture after KK reduction}.
{In} UED model, there is no shift symmetry alleviating divergence, 
then this process obeys the above simple estimation.\footnote{In 5D UED, we can calculate this process without cutoff dependence.}
Therefore, we introduce a cutoff scale ${\Lambda_{\text{UED}}}$ to regularize {the $F_t^{\text{KK}}$ in Eq.~(\ref{eq:FtKK})}.
We estimate an upper bound of ${\Lambda_{\text{UED}}}$ by use of Naive Dimensional Analysis (NDA) technique~\cite{Appelquist:2000nn} {in Section~\ref{Naive Dimensional Analysis}}.



\subsection{$h^{(0)} \rightarrow 2 \gamma$ process}


Now we turn to the Higgs decay process $h^{(0)} \rightarrow 2 \gamma$, which is the experimentally
favorable at the LHC with Higgs mass region {$120\, \text{GeV} \lesssim m_h \lesssim 150\, \text{GeV}$}.
The Feynman diagrams describing $h^{(0)} \rightarrow 2 \gamma$ process
due to the contribution of W boson and its associated particles
are shown in Fig.~{1, 2, 3, and 4}.
$\omega_W^{(m,n)}$ and $\bar{\omega}_W^{(m,n)}$ indicate $(m,n)$-th ghost and anti-ghost modes
originated from $W^{(m,n)}_{\mu}$ boson{, respectively}.
We also need to consider a flipped $(\mu \leftrightarrow \nu)$ one for each diagram if it exists. 
It is noted that there are another triangle loop diagrams contributing to this process, whose intermediate
particles are the top quark and its KK states.
But we can take these effects into account by use of {the} previous result in {Eq.}~(\ref{eq:FtKK})
with some modifications.
The decay width can be written as
\beq
{\Gamma_{h^{(0)} \rightarrow 2 \gamma} = \frac{\sqrt{2}G_F}{16\pi} \paren{\frac{\alpha_{\text{EM}}}{\pi}}^2 m_h^3 |F_{\text{decay}}|^2,}
\eeq
where $\alpha_{\text{EM}}$ is the electromagnetic coupling strength.
In this process, the function describing loop effects $F_{\text{decay}}$ is written by
\beq
F_{\text{decay}} = F_W + 3 Q_t^2 \paren{ {F_t^{\text{SM}} + F_t^{\text{KK}}} } {+ F^{\text{TC}}_{\text{decay}}},
\label{Fdecay}
\eeq
where the {first} term represents the effect of W boson and its associated particles, and
the {second} term represents that of the top quark and its KK states.\footnote{3 is color factor and $Q_t$ is the electromagnetic charge of the top quark $(=\frac{2}{3})$.}
{The third term describes the threshold correction.}
The SM result for $F_t^{{\text{SM}}}$ is previously obtained in Eq.~(\ref{eq:FtSM})
and the concrete form of $F_W^{{\text{SM}}}$ is derived in~\cite{Ellis:1975ap} as
\beq
{F_W^{{\text{SM}}} = \frac{1}{2} + 3 \lambda{(m_W^2)} - 3 \lambda{(m_W^2)} (1 - 2 \lambda{(m_W^2)})
J\paren{\lambda(m_W^2)}},
\eeq 
where $J$ is given in {Eq.}~(\ref{eq:scalarPVfunction}).
We set  $F_W = F_W^{{\text{SM}}} + F^{{\text{KK}}}_{W}$, where 
$F_t^{{\text{KK}}}$ has been already discussed and $F_W^{{\text{KK}}}$ 
represents the contribution of KK W boson and its associated KK particles.
And we decompose $F_W^{{\text{KK}}}$ into four pieces as
\beq
F_W^{{\text{KK}}} = F^{{\text{KK}}}_{\text{gauge}} + F^{{\text{KK}}}_{\text{NG}} + F^{{\text{KK}}}_{\text{scalar1}} + F^{{\text{KK}}}_{\text{scalar2}},
\eeq
where each term $F_W^{{\text{KK}}}$ indicates the loop effects coming from gauge, would-be NG boson, scalar particles, respectively
and corresponding Feynman diagrams are found in {Fig.~1, 2, 3, and 4, respectively}.
The four sets of diagrams are $U(1)_{EM}$ gauge invariant and we can check this fact by use of Ward identity.

\begin{figure}[H]
\centering
\includegraphics[width=140mm, clip]{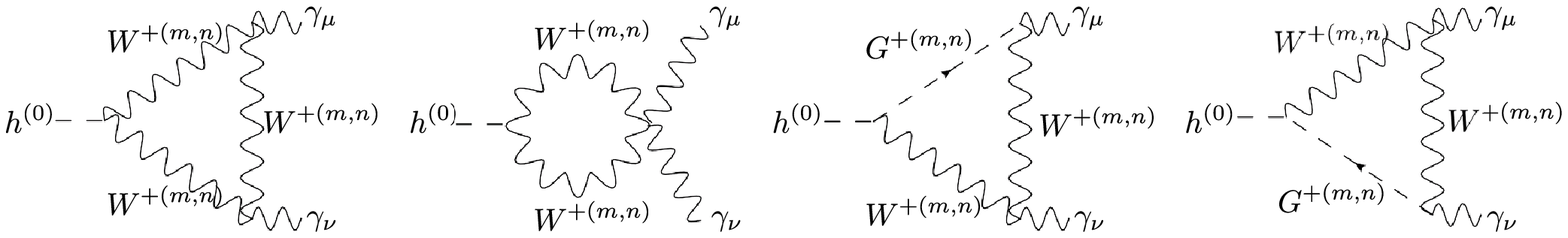}\\
\includegraphics[width=140mm, clip]{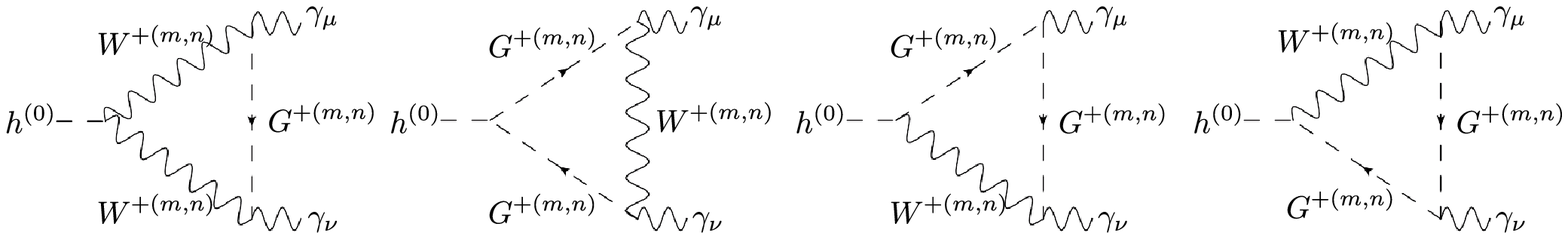}\\
\includegraphics[width=130mm, clip]{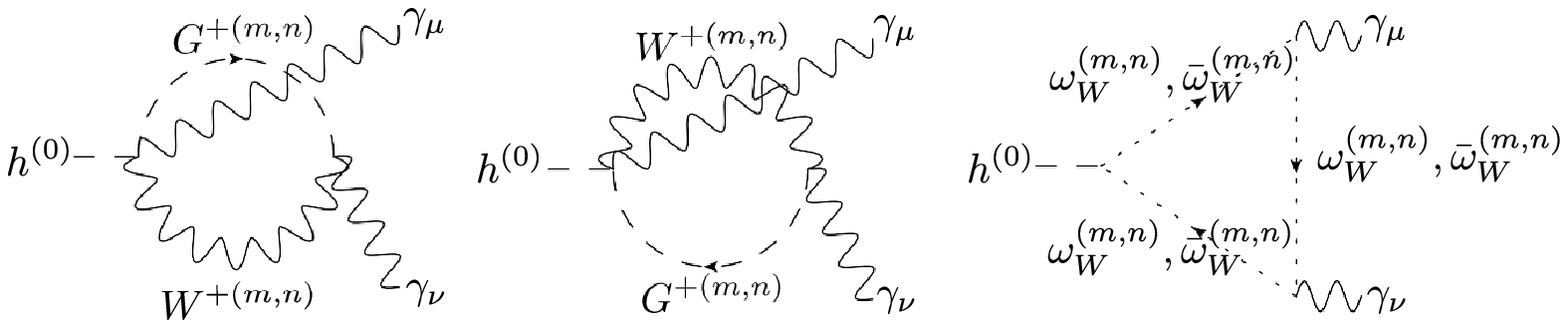}
\caption{
Feynman diagrams of 4D gauge sector describing $h^{(0)} \rightarrow 2 \gamma$ process.
}
\label{decayfig1}
\end{figure}
\begin{figure}[H]
\centering
\includegraphics[width=70mm , clip]{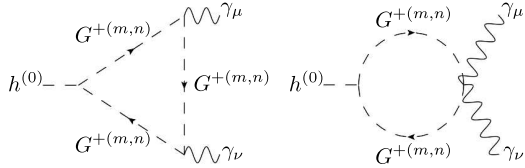}
\caption{
Feynman diagrams of 4D would-be NG boson sector describing $h^{(0)} \rightarrow 2 \gamma$ process.
}
\label{decayfig2}
\end{figure}
\begin{figure}[H]
\centering
\includegraphics[width=70mm, clip]{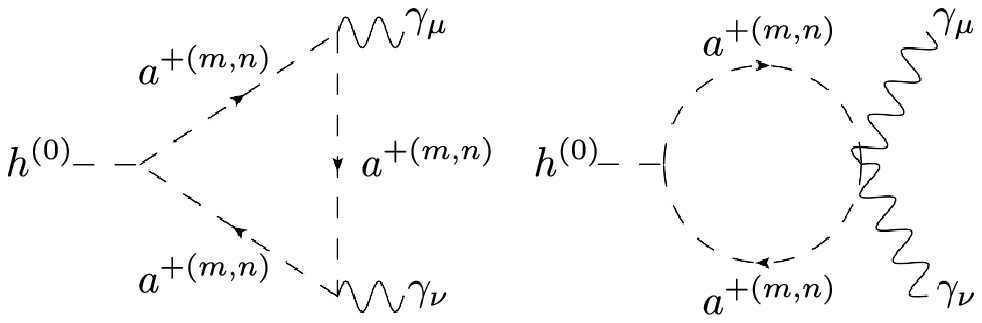}
\caption{
Feynman diagrams of 4D scalar sector describing $h^{(0)} \rightarrow 2 \gamma$ process.
}
\label{decayfig3}
\end{figure}
\begin{figure}[H]
\centering
\includegraphics[width=70mm, clip]{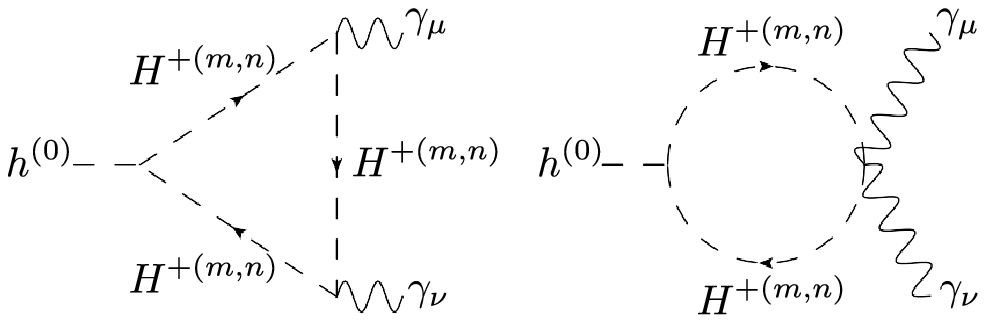}
\caption{
Feynman diagrams of 4D scalar (``spinless adjoint") sector describing $h^{(0)} \rightarrow 2 \gamma$ process.
}
\label{decayfig4}
\end{figure}

\noindent
After some tedious but straightforward calculation, we can get the result as follows:\footnote{In Appendix~\ref{Feynman Rules containing scalar particle}, we write down some Feynman rules to calculate this process.}
\begin{align}
F^{{\text{KK}}}_{\text{gauge}} &= \sum_{m \geq 1, n \geq 0}
\left\{ 3 \lambda(m_W^2) + 2 \lambda(m_W^2) \big(3 \lambda(m_{W,(m,n)}^2) - 2 \big) J\paren{\lambda(m_{W,(m,n)}^2)} \right\},
\label{eq:FWKK1} \\
F^{{\text{KK}}}_{\text{NG}} &= \sum_{m \geq 1, n \geq 0}
\left( \frac{1}{2} \frac{m_h^2}{m_{W,(m,n)}^2} \right)
\lambda(m_W^2) \br{ 1 + 2 \lambda(m_{W,(m,n)}^2) J\paren{\lambda(m_{W,(m,n)}^2)} }, 
\label{eq:FWKK2}\\
F^{{\text{KK}}}_{\text{scalar1}} &= \sum_{m \geq 1, n \geq 0}
\left( \frac{1}{2} \frac{1}{m_{W,(m,n)}^2} \right)
\left[ \frac{m_h^2}{m_W^2} m_{(m,n)}^2 + {2} m_{W,(m,n)}^2  \right] \notag \\
& \quad \times \lambda(m_W^2) \br{ 1 + 2 \lambda(m_{W,(m,n)}^2) J\paren{\lambda(m_{W,(m,n)}^2)} }
\label{eq:FWKK3}, \\
F^{{\text{KK}}}_{\text{scalar2}} &= \sum_{m \geq 1, n \geq 0}
\lambda(m_W^2) \br{ 1 + 2 \lambda(m_{W,(m,n)}^2) J\paren{\lambda(m_{W,(m,n)}^2)} }
\label{eq:FWKK4},
\end{align}
By adding up {Eqs.}~(\ref{eq:FWKK1})-(\ref{eq:FWKK4}),
the concrete form of $F_W^{{\text{KK}}}$ is given as
\al{
F_{W}^{{\text{KK}}} &= \sum_{m \geq 1, n \geq 0}\Big\{
\frac{1}{2}  + 5 \lambda(m_W^2)
- \Big[
\lambda(m_W^2) (4 - 10 \lambda(m_{W,(m,n)}^2)) \notag \\
& \qquad \qquad - \lambda(m_{W,(m,n)}^2)
\Big] J \paren{\lambda(m_{W,(m,n)}^2)}
\Big\},
}
where we use the relation $m_{W,{(m,n)}}^2 = m_W^2 + m_{(m,n)}^2$.
{This loop-induced process is also described by dimension-six operator in 4D point of view and we have to introduce the cutoff scale ${\Lambda_{\text{UED}}}$ to regularize the summations.
The concrete form of the third term in Eq.~(\ref{Fdecay}), which originates form threshold correction, is as follows:
\beq
F^{\text{TC}}_{\text{decay}} = \sqbr{\paren{\frac{\alpha_\text{EM}}{\pi}} \frac{2}{v}}^{-1} C'_{h\gamma\gamma},
\label{FdecayTC}
\eeq 
where $C'_{h\gamma\gamma}$ is a dimensionful coefficient describing the threshold correction and is related to the dimensionless constant in {a part of the} Lagrangian $C_{h\gamma\gamma}$ with the UED cutoff scale ${\Lambda_{\text{UED}}}$ as
\beq
C'_{h\gamma\gamma} = \frac{C_{h\gamma\gamma}\paren{v \over \sqrt{2}}}{{\Lambda_{\text{UED}}}^2}.
\label{TCcoefficient_twophoton}
\eeq
We also see the details in Appendix~{\ref{Detail on threshold correction}}.
}

\section{Universal Extra Dimension Models based on $S^2$
\label{Universal Extra Dimension Models based on $S^2$}}

Recently, Universal Extra Dimension Models based on $S^2$ are proposed in {Refs.}~\cite{Maru:2009wu,Dohi:2010vc}.
After an overview of gauge theory on $S^2$, we give a brief review of these models. 
 
\subsection{Gauge Theory on $S^2$}

We consider a gauge theory on six-dimensional spacetime $M^4 \times S^2$, which is a direct product of the four-dimensional Minkowski spacetime $M^4$ and two-sphere $S^2$.
We use the coordinate of six-dimensional spacetime defined by $x^M = (x^{\mu},\theta,\phi)$.
$\theta \ (\phi)$ is zenith (azimuthal) angle of $S^2$, respectively and we use the same coordinate conventions
as in {Section~\ref{Universal Extra Dimension on T2Z4}}.
The metric ansatz of $M^4 \times S^2$ is 
\beq
g_{MN} = \text{diag}(1,-1,-1,-1,-R^2,-R^2 \sin^2{\theta})
\eeq
and
we also need to introduce the vielbein $e_M^{\ \  \underline{N}}  = \text{diag}(1,1,1,1,R,R \sin{\theta})$ 
to describe tangent space which fermions live in.
{In this tangent space, 
the coordinate is expressed with barred letters and
we choose the same representation of Clifford algebra as in Eq.~(\ref{eq:gammas})}.
$S^2$ has a positive curvature and then a radius of $S^2$ described by $R$ only can take 
an infinite value by the consistency with {the 6D} Einstein equation.
To stabilize the system, we introduce a $U(1)_X$ gauge field which has a monopole-like configuration 
in classical level $X^c_M$~\cite{RandjbarDaemi:1982hi}.
This configuration is defined as follows:
\beq
[X^c_{\phi}(x^{\mu},\theta,\phi)]^{{N} \atop {S}} =  { n \over 2g^{(X)}_6 }(\cos{\theta} \mp 1),
\quad
(\text{other components}) = 0,
\label{eq:monopoleconfig}
\eeq
where $g^{(X)}_6$ is {a} $U(1)_X$ gauge coupling and $n$ is {a} monopole index.
The superscript $N \atop S$ indicates that the field is given in north (involving the $\theta = 0$ point) and
south (involving the $\theta = \pi$ point) patches, respectively and we use this notation throughout 
the rest of this paper.
The gauge transformation from {the} north to {the} south patch is given by
\beq
[X_M(x^{\mu},\theta,\phi)]^{S} = [X_M(x^{\mu},\theta,\phi)]^{N} + \frac{1}{g_6^{(X)}} \pal_M \alpha(x^{\mu},
\theta,\phi){,}
\label{eq:gaugepatchtransf}
\eeq
where the function $\alpha(x^{\mu},\theta,\phi) = n \phi$.
Because of the monopole-like configuration, the radius of $S^2$ is stabilized spontaneously as
\beq
R^2 = \left( {n \over 2 g_6^{(X)} M_{\ast}^2} \right)^2.
\label{eq:radius}
\eeq

{Every} 6D field $\Phi$ on $S^2$ is KK expanded by use of the spin-weighted spherical harmonics 
${}_s Y_{jm}(\theta,\phi)$ as follows:\footnote{Newman-Penrose edth formalism~\cite{Newman:1966ub} is useful for description of  spin weighted spherical harmonics.}
\beq
\Phi(x,\theta,\phi)^{N \atop S} = \sum_{j = |s|}^{\infty} \sum_{m=-j}^{j} \varphi^{(j,m)}(x) f_{\Phi}^{(j,m)}(\theta,\phi)
^{N \atop S}, \quad
f_{\Phi}^{(j,m)}(\theta,\phi)
^{N \atop S} := {{}_s Y_{jm}(\theta,\phi) e^{\pm is \phi} \over R},
\eeq
where $s$ is the spin weight of the field $\Phi$.
The spin-weighted spherical harmonics ${}_s Y_{jm}(\theta,\phi)$ matches the orthonormal condition as 
\beq
{
\int_{0}^{2\pi} d\phi \int_{-1}^{1} d\cos{\theta} \overline{{}_s Y_{jm}(\theta,\phi)}
{}_s Y_{j'm'}(\theta,\phi) = \delta_{jj'} \delta_{mm'}.
}
\eeq
{A spin} weight of {a} fermion is closely related to its $U(1)_X$ charge.
When we assign $U(1)_X$ charges of 6D Weyl fermions $\Psi_{\pm}$ as $q_{\Psi_{\pm}}$,
the corresponding spin weights of 4D Weyl fermions $\{ \psi_{+ {R \atop L}},\psi_{- {R \atop L}} \}$ {are} given as follows in our convention:
\beq
s_{+ {R \atop L}} = - \left( {nq_{\Psi_+} \mp 1 \over 2 } \right), \quad
s_{- {R \atop L}} = - \left( {nq_{\Psi_-} \pm 1 \over 2 } \right).
\eeq
We can find the fact that if a 6D Weyl fermion takes the $s=0$ spin weight,
{one zero mode $(j=0)$ appears as a 4D Weyl fermion with no KK mass}.
This means we can get the SM fermions without orbifolding in the case of $S^2$.
When we take the values {as} $(s_{+R},s_{+L},s_{-R},s_{-L}) = (0,-1,-1,0)$, we can create the same situation
as in $T^2/Z_4$ which we discussed before.
{A spin} weight of {a} 4D vector component of {a} 6D gauge boson is $s=0$ and then there is a zero mode which we can assign as {a} SM gauge boson. However, extra dimensional components of 6D gauge boson are expanded
by the $|s|=1$ spin-weighted spherical harmonics.
This is because these parts are closely related to $S^2$ structure.
Concretely speaking, the combinations of components
$A_{\pm} = \frac{1}{\sqrt{2}} (A_{\underline{\theta}} \pm i A_{\underline{\phi}})$
are KK expanded with $s=\pm 1$ spin weighted spherical harmonics, respectively,
where
$A_{\underline{M}}$ is a gauge field on tangent space defined as $A_{\underline{M}} = e_{\underline{M}}
^{\ \ N} A_{N}$.
Then there is no zero mode in these parts.
After the introduction of gauge fixing term concerning a gauge field $A_{M}$, whose concrete form is
\beq
- \frac{1}{\xi} \text{tr} \left( \eta^{\mu \nu} \pal_{\mu} A_{\nu} - \frac{\xi}{R^2 \sin{\theta}} \pal_{\theta} \sin{\theta} A_{\theta}
- \frac{\xi}{R^2 \sin^2{\theta}} A_{\phi} \right)^2 \quad (\xi:\text{gauge fixing parameter}),
\eeq  
the mass eigenstates are obtained as follows:
\beq
\begin{pmatrix} A_{\underline{\theta}} \\ A_{\underline{\phi}} \end{pmatrix}
=
\begin{pmatrix}
\pal_{\theta} & - \csc{\theta} \pal_{\phi} \\
\pal_{\theta} & + \csc{\theta} \pal_{\phi}
\end{pmatrix}
\begin{pmatrix} \phi_{1}^{(A)} \\ \phi_{2}^{(A)} \end{pmatrix}.
\label{eq:vectoscalarmassbases}
\eeq
$\phi_{1}^{(A)}$ and $\phi_{2}^{(A)}$ are 4D physical scalar field and unphysical would-be Nambu-Goldstone 
mode, respectively.
{A} 6D scalar field can take {a} nonzero spin weight through the interaction with the $U(1)_X$ gauge boson.
But we would like to regard the zero mode of {a} 6D scalar field as the SM Higgs, then the value of the spin weight must
be $s=0$.

In our configuration, any $(j,m)$-th KK mode has the KK mass,
\beq
m_{(j,m)}^2 = \frac{j(j+1)}{R^2}.
\label{eq:KKmass_S2}
\eeq
An important point is that the form of the above KK mass is independent of the index of $m$.
{This means there are $2j+1$ degenerated modes for each $j$.
It is noted that each KK mode summation {over} $j$ begins {from} one.}
{In contrast to the $T^2$ case, the value of the first KK mass is represented as $M_{\text{KK}} = \sqrt{2}/R$.}

We can construct an Universal Extra Dimension model on $S^2$ along the direction
which we have discussed.
But  there are two problems in this model.
One is   absence of KK parity.
In usual UED models based on orbifold, there are fixed points of orbifold discrete symmetry and
KK parity is realized as a remnant  of extra spatial symmetry, which is {an} invariance of system in exchange
of fixed points.
It ensures the existence of dark matter candidate in these models.
But the geometry of $S^2$ do not have fixed point and {thereby} the UED on $S^2$ cannot possess KK parity.
The other is more serious.
As we discussed before, {a} 4D vector component of {a} 6D gauge boson has zero mode {in $S^2$}.
In case of the $U(1)_X$ gauge boson, which has the monopole-like configuration, this is true. 
We should notice that the gauge coupling of {an} extra massless gauge boson is severely
constrained to be $g^{(X)} \lesssim 10^{-23}$
by a torsion balance experiment~\cite{Smith:1999cr}.
$g^{(X)}$ is the 4D effective coupling of the 6D $U(1)_X$ gauge coupling $g^{(X)}_6$ and 
is described as $g^{(X)} = {{g^{(X)}_6}/{\sqrt{4 \pi R^2}}}$.
By use of (\ref{eq:radius}), we can estimate the value of $g^{(X)}$ in the UED model on $S^2$ as
\beq
g^{(X)} {\simeq} \frac{n {M_{\text{KK}}}}{M_{\text{pl}}}{{.}}
\eeq
In the viewpoint of our phenomenological motivation, ${M_{\text{KK}}}$ must be $\sim \mathcal{O}(1)$ TeV.
In such a situation, $g^{(X)}$ becomes $\sim 10^{-15} \cdot n$ and its value is far from the experimental bound. Since monopole charge $n$ only can take integer value, we cannot 
resolve this pathology by tuning of {the} parameter $n$.

Fortunately, we can solve these problems by some modifications {in the} $S^2$ geometry.
In the rest of this section, we follow {some} essential points of these ideas.


\subsection{UED on $S^2/Z_2$}

Following {Ref.}~{\cite{Maru:2009wu}}, we take {a} $Z_2$ orbifold {on the geometry of $S^2$}.
On {this} orbifold, the point ${(\theta,\phi)}$ is identified with $(\pi - \theta , - \phi)$.
The 6D action is as follows:
\begin{align}
S &  =  \int_{0}^{\pi} d \theta \int_0^{2 \pi} d \phi \int d^4 x \sqrt{-g} \Bigg\{ - \frac{1}{2} \sum_{i=1}^{3} \mathrm
g^{MN} g^{KL}
\text{Tr} \big[ F_{MK}^{(i)} F^{(i)}_{NL}  \big]
- \frac{1}{4} g^{MN} g^{KL}
 \big[ F_{MK}^{(X)} F^{(X)}_{NL}  \big] \notag \\
& \qquad + g^{MN} (D_M H)^{\dagger} (D_N H) + \bigg[ \mu^2 |H|^2 - \frac{\lambda_6^{(H)}}{4} |H|^4 \bigg] \notag \\
& \qquad +i \bar{\mathcal{Q}}_{3-} \Gamma^M D_M \mathcal{Q}_{3-} + i  \bar{\mathcal{U}}_{3+} \Gamma^M D_M \mathcal{U}_{3+} - 
\bigg[ \lambda_{6}^{(t)} \bar{\mathcal{Q}}_{3-} (i \sigma_2 H^{\ast}) \mathcal{U}_{3+} + \mathrm{h.c.} \bigg]
\Bigg\}
\label{eq:6DactionofS2Z2}
\end{align}
where $\sqrt{-g} = R^2 \sin{\theta}$.
In this model, {the form of 6D action and matter content are almost the same with {these of} the $T^2/Z_4$ except
{the existence of} the $U(1)_X$ gauge field}
and $F_{MN}^{(X)}$ has the classical part arising from the monopole-like configuration as
\beq
F_{\theta \phi} = - \frac{n}{2 g_6^{(X)}} \sin{\theta},\quad (\text{other components}) =0.
\eeq
The covariant derivative of {the} Higgs is given in an ordinary form {as}
\beq
D_M = \pal_M -i \sum_{i=1}^{3} g_6^{(i)} T^{(i)a} A^{(i)a}_M,
\eeq
and the covariant derivatives of fermions are obtained as follows:
\beq
D_M = \pal_M -i \sum_{i=1}^{3} g_6^{(i)} T^{(i)a} A^{(i)a}_M 
-i g_6^{(X)} q_{\Psi} (X^c_M + X_M) + \Omega_M.
\eeq
$q_{\Psi}$ is {a} $U(1)_X$ charge of {a} fermion and $X^c_M$ is the monopole-like classical configuration
in {Eq.}~(\ref{eq:monopoleconfig}).
The other additional term $\Omega_M$ is the spin connection in $S^2$, {whose concrete form is}
\beq
(\Omega_{\phi})^{N \atop S} = \frac{i}{2} (\cos{\theta} \mp 1)
\begin{pmatrix}
1_4 & 0 \\ 0 & -1_4
\end{pmatrix},
\quad
(\text{other components}) = 0,
\eeq
where $1_4$ is {a} four-by-four unit matrix.
We can easily construct mode functions of $S^2/Z_2$ $f_{s,t}^{(j,m)}(\theta,\phi)$ with spin weight $s$
in both north and south patches
following the general prescription {in Ref.}~\cite{Georgi:2000ks} as follows:
\beq
f_{s,t}^{(j,m)}(\theta,\phi)^{N \atop S} =
\left\{
\begin{array}{ll}
\displaystyle \frac{1}{2R}
\left[ {}_s Y_{jm}(\theta,\phi) + (-1)^{j-s} {}_s Y_{j-m}(\theta,\phi) \right] e^{\pm is \phi} & 
\text{for} \ t=+1 \\
\displaystyle \frac{1}{2R}
\left[ {}_s Y_{jm}(\theta,\phi) - (-1)^{j-s} {}_s Y_{j-m}(\theta,\phi) \right] e^{\pm is \phi} & 
\text{for} \ t=-1
\end{array}
\right. ,
\eeq
where $t=\pm 1$ is {the} $Z_2$ parity.
These mode functions have the property that
$f_{s,t=\pm 1}^{(j,m)}(\pi - \theta, - \phi)^{N \atop S} = \pm f_{s,t=\pm 1}^{(j,m)}(\theta,\phi)^{S \atop N}$.
To realize {the} $Z_2$ symmetry, we identify a field at $(\theta,\phi)$ in {the} north patch with
the same field at $(\pi - \theta , - \phi)$ in {the} south patch.
The conditions are as follows:
\begin{align}
H(x,\pi - \theta,- \phi)^{N \atop S} &= + H(x,\theta,\phi)^{S \atop N}, \label{eq:Higgsid}\\
\{A^{(i)}_{\mu},X_{\mu}\}(x,\pi - \theta,- \phi)^{N \atop S} &= + \{A^{(i)}_{\mu},X_{\mu}\}(x,\theta,\phi)
^{S \atop N}, \label{eq:gaugeid1} \\
\{A^{(i)}_{\theta,\phi},X_{\theta,\phi}\}(x,\pi - \theta,- \phi)^{N \atop S} &= - \{A^{(i)}_{\theta,\phi},X_{\theta,\phi}\}
(x,\theta,\phi)^{S \atop N}, \label{eq:gaugeid2} \\
\{ \mathcal{Q}_{3-}, \mathcal{U}_{3+} \}(x,\pi - \theta,- \phi)^{N \atop S} &= +i \Gamma^{\underline{y}} \Gamma^{\underline{z}} \{ \mathcal{Q}_{3-}, \mathcal{U}_{3+} \}(x,\theta,\phi)^{S \atop N} \label{eq:fermionid},
\end{align}
where we take the choice that all gauge twist matrices are trivial ($P={\bf 1}$).
And we define the transformation of 6D Weyl fermion $\Psi_{\pm}$ from {the} north to {the} south patch as
\beq
\Psi^{S}_{\pm}(x,\theta,\phi) = \exp(i q_{\Psi_{\pm}} \alpha + 2 \phi \Sigma^{\underline{y} \underline{z}})
\Psi^{N}_{\pm}(x,\theta,\phi){,} 
\eeq
where $\alpha$ is the $U(1)_X$ gauge transformation function in {Eq.}~(\ref{eq:gaugepatchtransf}) and
$ \Sigma^{\underline{y} \underline{z}}$ is the $(\underline{y} , \underline{z})$ component of
the local Lorentz generator of {a} 6D Weyl fermion.\footnote{In our notation, $ \Sigma^{\underline{y} \underline{z}} = \frac{-i}{2}
\begin{pmatrix} 1 & 0 \\ 0 & -1 \end{pmatrix}$.}
{The} Higgs does not transform along {the} patches because {the} Higgs does not have spin and interaction
with the $U(1)_X$ gauge field.
By use of the above facts and some specific information of this model,\footnote{We can find the details in {Ref.}~\cite{Dohi:2010vc}.} we can show that
the action {in Eq.}~(\ref{eq:6DactionofS2Z2}) is equal at both {the} north and {the} south patches.
Combining this result with {Eqs.} (\ref{eq:Higgsid})-(\ref{eq:fermionid}),
it is clear that the $Z_2$ symmetry is entailed on the action in {Eq.}~(\ref{eq:6DactionofS2Z2}). 

The specific forms of each KK expansion are as follows:
\begin{align}
 \{A^{(i)}_{\mu},X_{\mu}\}(x,\theta,\phi)^{N \atop S} &= \frac{1}{\sqrt{4\pi} R} \{A^{(i)(0)}_{\mu},X^{(0)}_{\mu}\}(x) \notag \\
& \quad + \sum_{j=1}^{\infty} \sum_{m=0}^{j} \{A^{(i)(j,m)}_{\mu},X^{(j,m)}_{\mu}\}(x) \cdot 
(\sqrt{2} (i)^{j+m})f_{s=0,t=+1}^{(j,m)}(\theta,\phi)^{N \atop S}, \label{eq:Z2Amuexpansion}\\
\{A^{(i)}_{\pm},X_{\pm}\}(x,\theta,\phi)^{N \atop S} &=  \sum_{j=1}^{\infty} \sum_{m=0}^{j} \{A^{(i)(j,m)}_{\pm},X^{(j,m)}_{\pm}\}(x) \cdot 
(\sqrt{2} (i)^{j+m+1}) f_{s=\pm 1,t=-1}^{(j,m)}(\theta,\phi)^{N \atop S}, \label{eq:Z2Apmexpansion} \\
H(x,\theta,\phi)^{N \atop S} &= \frac{1}{\sqrt{4\pi} R} H^{(0)}(x)
+ \sum_{j=1}^{\infty} \sum_{m=0}^{j} H^{(j,m)} (x) \cdot \sqrt{2} f_{s=0,t=+1}^{(j,m)}(\theta,\phi)^{N \atop S}, \\
\mathcal{Q}_{3-}(x,\theta,\phi)^{N \atop S} &=
\begin{pmatrix}
\displaystyle  \frac{1}{\sqrt{4\pi} R} Q_{3L}^{(0)}(x)
+ \sum_{j=1}^{\infty} \sum_{m=0}^{j} Q_{3L}^{(j,m)} (x) \cdot \sqrt{2} f_{s=0,t=+1}^{(j,m)}(\theta,\phi)^{N \atop S} \\
\displaystyle  \sum_{j=1}^{\infty} \sum_{m=0}^{j} Q_{3R}^{(j,m)} (x) \cdot \sqrt{2} f_{s=-1,t=-1}^{(j,m)}(\theta,\phi)^{N \atop S}
\end{pmatrix}, \\
\mathcal{U}_{3+}(x,\theta,\phi)^{N \atop S} &=
\begin{pmatrix}
\displaystyle  \frac{1}{\sqrt{4\pi} R} t_{R}^{(0)}(x)
+ \sum_{j=1}^{\infty} \sum_{m=0}^{j} t_{R}^{(j,m)} (x) \cdot \sqrt{2} f_{s=0,t=+1}^{(j,m)}(\theta,\phi)^{N \atop S} \\
\displaystyle  \sum_{j=1}^{\infty} \sum_{m=0}^{j} t_{L}^{(j,m)} (x) \cdot \sqrt{2} f_{s=-1,t=-1}^{(j,m)}(\theta,\phi)
^{N \atop S}
\end{pmatrix}.
\end{align}
{Here} we introduce suitable normalization factor $(\sqrt{2})$ in each KK modes and 
some phase factors ($(i)^{j+m} , (i)^{j+m+1}$) in {Eqs.}~(\ref{eq:Z2Amuexpansion},\ref{eq:Z2Apmexpansion})
to ensure the reality of these fields.
The range of the summation over $m$ shrinks from $[-j,j]$ to $[0,j]$ after {the} $Z_2$ identification.
This system has two fixed points of {the} $Z_2$ symmetry at $(\theta , \phi) = (\frac{\pi}{2},0) , (\frac{\pi}{2},\pi)$ and under the transformation of $(\theta,\phi) \rightarrow (\theta,\phi + \pi)$,
mode functions behave as 
\al{
f_{s=0,t=+1}^{(j,m)}(\theta,\phi + \pi)^{N \atop S} &= (-1)^{m} f_{s=0,t=+1}^{(j,m)}(\theta,\phi)^{N \atop S},
\notag \\
f_{s=\pm 1,t=-1}^{(j,m)}(\theta,\phi + \pi)^{N \atop S} &= -(-1)^m f_{s=\pm 1,t=-1}^{(j,m)}(\theta,\phi)^{N \atop S}.
}
{Thereby} after the fields redefinition {as}
\al{
 \{A^{(i)(j,m)}_{\pm},X^{(j,m)}_{\pm},Q^{(j,m)}_{3R},t^{(j,m)}_{L}\} \rightarrow
(-1) \{A^{(i)(j,m)}_{\pm},X^{(j,m)}_{\pm},Q^{(j,m)}_{3R},t^{(j,m)}_{L}\},
}
we can find that each KK field has KK parity $(-1)^m$, whose origin is considered to be
a remnant of KK angular momentum conservation. 

We focus on the $m=0$ modes of each $j$ level.
When we see the concrete forms of mode functions in $m=0$, which are
\begin{align}
f_{s=0,t=+1}^{(j,m=0)}(\theta,\phi)^{N \atop S} &= \frac{1}{2R} (1 + (-1)^j) \cdot
{}_0 Y_{j0}(\theta,\phi), \\
f_{s=+1,t=-1}^{(j,m=0)}(\theta,\phi)^{N \atop S} &= \frac{1}{2R} (1 + (-1)^j) \cdot
{}_1 Y_{j0}(\theta,\phi) e^{\pm i \phi}, \\
f_{s=-1,t=-1}^{(j,m=0)}(\theta,\phi)^{N \atop S} &= \frac{1}{2R} (1 + (-1)^j) \cdot
{}_{-1} Y_{j0}(\theta,\phi) e^{\mp i \phi},
\end{align}
we find that $m=0$ modes appear only in the case of even $j$. 
Then {the} degeneracy of KK masses is
\beq
\begin{array}{cl}
j+1 & \text{for} \quad j=\text{even}, \\
j & \text{for}  \quad j=\text{odd},
\end{array}
\eeq
since $m$ runs from $0$ to $j$. These results play an essential role at the Higgs production and decay
processes via loop diagrams.

After the $Z_2$ identification, the massless zero mode of $U(1)_X$ gauge boson survives.
In this model, it is assumed that the $U(1)_X$ symmetry is anomalous and it is broken at the quantum level~\cite{Scrucca:2003ra}.
Therefore gauge bosons should be heavy and decoupled from the low energy physics.


\subsection{UED on Projective {Sphere}}

We can also construct a UED model based on  a non-orbifolding idea {in Ref.}~\cite{Dohi:2010vc}.{\footnote{
{In~\cite{Dohi:2010vc} the terminology ``real projective plane'' is employed for the compactified space, the sphere with its antipodal points being identified.}}}
The {projective {sphere} $(PS)$} is a sphere $S^2$ with its antipodal points identified by
$(\theta,\phi) \sim (\pi - \theta , \phi + \pi)$.
In the UED model based on ${PS}$, the 6D action takes a different form from that of ordinary 6D UED model.
It is written as follows:
\begin{align}
S &  =  \int_{0}^{\pi} d \theta \int_0^{2 \pi} d \phi \int d^4 x \sqrt{-g} \Bigg\{ - \frac{1}{2} \sum_{i=1}^{3} \mathrm
g^{MN} g^{KL}
\text{Tr} \big[ F_{MK}^{(i)} F^{(i)}_{NL}  \big]
- \frac{1}{4} g^{MN} g^{KL}
 \big[ F_{MK}^{(X)} F^{(X)}_{NL}  \big] \notag \\
& \qquad + g^{MN} (D_M H)^{\dagger} (D_N H) + \bigg[ \mu^2 |H|^2 - \frac{\lambda_6^{(H)}}{4} |H|^4 \bigg] \notag \\
& \qquad + \frac{1}{2} \bigg[ i \bar{\mathcal{Q}}_{3-} \Gamma^M D_M \mathcal{Q}_{3-} 
+i \bar{\mathcal{Q}}_{3+} \Gamma^M D_M \mathcal{Q}_{3+} \bigg] 
+ \frac{1}{2} \bigg[
 i  \bar{\mathcal{U}}_{3+} \Gamma^M D_M \mathcal{U}_{3+}
+  i  \bar{\mathcal{U}}_{3-} \Gamma^M D_M \mathcal{U}_{3-} 
\bigg] \notag \\
& \qquad  - \frac{1}{2} \bigg[ \lambda_{6}^{(t)} \Big( \bar{\mathcal{Q}}_{3-} (i \sigma_2 H^{\ast}) \mathcal{U}_{3+} 
+ \bar{\mathcal{Q}}_{3+} (i \sigma_2 H^{\ast})^{\text{T}} \mathcal{U}_{3-} \Big)
+ \mathrm{h.c.} \bigg]
\Bigg\}.
\label{eq:6DactionofRP2}
\end{align}
{Here the $``1/2"$ factors are introduced for a later convenience.}
Like the $S^2/Z_2$ case, $F_{MN}^{(X)}$ has the classical part.
A new feature of this model is that we introduce ``mirror" 6D Weyl fermions 
$\{ \mathcal{Q}_{3+},\mathcal{U}_{3-} \}$, which have opposite 6D chirality and opposite
SM and $U(1)_X$ charges when compared with the fields $\{ \mathcal{Q}_{3-},\mathcal{U}_{3+} \}${, respectively.}
And {the} covariant derivatives in this model are given as
\begin{align}
D_M &= \pal_M -i \sum_{i=1}^{3} g_6^{(i)} T^{(i)a} A^{(i)a}_M, && \text{for} \quad H, \\
D_M &= \pal_M -i \sum_{i=1}^{3} g_6^{(i)} T^{(i)a} A^{(i)a}_M 
-i g_6^{(X)} q_{\Psi} (X^c_M + X_M) + \Omega_M, && \text{for} \quad \mathcal{Q}_{3-},\mathcal{U}_{3+},
\label{eq:ordinarycov} \\
D_M &= \pal_M -i \sum_{i=1}^{3} g_6^{(i)} \big[ -T^{(i)a} \big]^{\text{T}} A^{(i)a}_M 
-i g_6^{(X)} q_{\Psi} (X^c_M + X_M) + \Omega_M, && \text{for} \quad \mathcal{Q}_{3+},\mathcal{U}_{3-}.
\label{eq:mirrorcov}
\end{align}
The covariant derivative of {the} Higgs is the same with that in the $S^2/Z_2$ case, 
but there is a difference between fermions and these ``mirror" fermions.
We discuss these points shortly below.

${PS}$ is a non-orientable manifold and has no fixed point.
Therefore, we cannot perform identification like {the} $S^2/Z_2$ case.
We focus on the 6D $P$ and $CP$ transformations, which are defined as
\beq
[\text{6D}\ P] =
\left\{
\begin{array}{lcr}
A_{\mu}(x,\theta,\phi) & \rightarrow & A_{\mu}(x,\pi - \theta,\phi + \pi), \\
A_{\theta}(x,\theta,\phi) & \rightarrow & -A_{\theta}(x,\pi - \theta,\phi + \pi), \\
A_{\phi}(x,\theta,\phi) & \rightarrow & A_{\phi}(x,\pi - \theta,\phi + \pi), \\
\Psi(x,\theta,\phi) & \rightarrow & P \Psi (x,\pi - \theta,\phi + \pi), \\
H(x,\theta,\phi) & \rightarrow &H(x,\pi - \theta,\phi + \pi),
\end{array}
\right.
\eeq
\beq
[\text{6D}\ CP] =
\left\{
\begin{array}{lcr}
A_{\mu}(x,\theta,\phi) & \rightarrow & A_{\mu}^C(x,\pi - \theta,\phi + \pi), \\
A_{\theta}(x,\theta,\phi) & \rightarrow & -A_{\theta}^C(x,\pi - \theta,\phi + \pi), \\
A_{\phi}(x,\theta,\phi) & \rightarrow & A_{\phi}^C(x,\pi - \theta,\phi + \pi), \\
\Psi(x,\theta,\phi) & \rightarrow & P \Psi^C (x,\pi - \theta,\phi + \pi), \\
H(x,\theta,\phi) & \rightarrow &H^{\ast}(x,\pi - \theta,\phi + \pi).
\end{array}
\right.
\eeq
{Like before, we consider} $\Psi$ is a 6D fermion and the concrete shapes of 6D $C$ and $P$ transformations are
\beq
A_M^C = - A_M^{\text{T}} = - A_M^{\ast}, \quad
\Psi^C = \Gamma^{\underline{2}} \Gamma^{\underline{y}} \Psi^{\ast},\quad
P = \Gamma^{\underline{y}}.
\label{eq:6DPCP}
\eeq
It must be noted that the monopole-like configuration of {the} $U(1)_X$ gauge boson {in Eq.}~(\ref{eq:monopoleconfig})
behaves under the antipodal identification as
\beq
\{ X^c_{\phi} \}^{N \atop S}(x,\pi - \theta,\phi+\pi)
=- \{ X^c_{\phi} \}(x,\theta,\phi)^{S \atop N} = \{(X^c_{\phi})^C\}(x,\theta,\phi)^{S \atop N}.
\eeq
We use a property of $U(1)$ gauge field $(X_M^{\text{T}} = X_M)$.
We can notice that 
the monopole-like configuration is invariant under the 6D $CP$ transformation and transition between patches.
Then we consider the identification of the $U(1)_X$ gauge field as
\footnote{We pay attention the fact that identification conditions of classical field $(X_\phi^c)$
and quantum field $(X_{\phi})$ must be the same.}
\beq
\left\{
\begin{array}{lcc}
X_{\mu}(x,\pi - \theta,\phi + \pi)^{N \atop S} & = & X_{\mu}^C(x,\theta,\phi)^{S \atop N} , \\
X_{\theta}(x,\pi - \theta,\phi + \pi)^{N \atop S}  &= & -X_{\theta}^C(x,\theta,\phi)^{S \atop N}, \\
\{ X_{\phi}^c , X_{\phi} \}(x,\pi - \theta,\phi + \pi)^{N \atop S}  & = 
&\{ (X_{\phi}^c)^C , X_{\phi}^C \}(x,\theta,\phi)^{S \atop N}.
\end{array}
\right.
\label{eq:idRP2X}
\eeq
{These} conditions ensure the monopole-like configuration after {the} antipodal identification
and project{ed} out the non-desirable $U(1)_X$ zero mode. 
It is clearly understood by the additional minus factor coming from
the 6D CP transformation of gauge field in {Eq.}~(\ref{eq:6DPCP}).
In contrast, since we want the zero modes which describe the SM gauge bosons in UED model construction,
identification of $A^{(i)}_M$ should be done by another condition.
We adopt the 6D $P$ transformation and those identifications are written as
\beq
\left\{
\begin{array}{lcc}
A^{(i)}_{\mu}(x,\pi - \theta,\phi + \pi)^{N \atop S} & = & A^{(i)}_{\mu}(x,\theta,\phi)^{S \atop N} , \\
A^{(i)}_{\theta}(x,\pi - \theta,\phi + \pi)^{N \atop S}  &= & -A^{(i)}_{\theta}(x,\theta,\phi)^{S \atop N}, \\
A^{(i)}_{\phi}(x,\pi - \theta,\phi + \pi)^{N \atop S}  & = 
& A^{(i)}_{\phi}(x,\theta,\phi)^{S \atop N},
\end{array}
\right.
\label{eq:idRP2A}
\eeq
where it is evident that $A_{\mu}^{(i)}$'s zero mode survives.
We also identify {the} Higgs with the 6D $P$ transformation to obtain its zero mode as
\beq
H(x,\pi - \theta,\phi + \pi)^{N \atop S} = H(x,\theta,\phi)^{S \atop N}.
\label{eq:idRP2Higgs}
\eeq

Finally, we discuss the identification of 6D Weyl fermions.
Since 6D Weyl fermions have $U(1)_X$ charge and interact with {the} $U(1)_X$ gauge boson,
they should be identified by the 6D $CP$ transformation.
But if we do not consider {the} ``mirror" fermions, a fundamental problem arises.
The 6D $P$ transformation of fermion change{s} the 6D chirality like {the} ordinary 4D 
transformation. However,  the 6D $C$ transformation of fermion does not change {the 6D chirality} unlike {the} ordinary 4D case.
This means 6D chirality flips under {the} 6D $CP$ transformation and we should introduce {the}
 ``mirror" fermions with opposite 6D chirality and opposite
SM and $U(1)_X$ charges to perform identification. 
The specific forms are as follows:
\beq
\{ \mathcal{Q}_{3+} , \mathcal{U}_{3-} \}(x,\pi - \theta,\phi + \pi)^{N \atop S} =
P \{ \mathcal{Q}_{3-}^C , \mathcal{U}_{3+}^C \}(x,\theta,\phi)^{S \atop N} =
\Gamma^{\underline{2}}
\{ \mathcal{Q}_{3-}^{\ast} , \mathcal{U}_{3+}^{\ast} \}(x,\theta,\phi)^{S \atop N}.
\label{eq:idRP2fermions}
\eeq
And we {determine} the forms of {the} covariant derivatives in {Eqs.}~(\ref{eq:ordinarycov},\ref{eq:mirrorcov})
on the criterion of invariance of the action under the 6D $CP$ transformation {in advance}.
Using the identification conditions {in Eqs.}~(\ref{eq:idRP2X})-(\ref{eq:idRP2fermions}), we can see that {the} ``mirror" fermions vanish from the action {in Eq.}~(\ref{eq:6DactionofRP2})
after the identifications as
\begin{align}
S &  \longrightarrow  \int_{0}^{\pi} d \theta \int_0^{2 \pi} d \phi \int d^4 x \sqrt{-g} \Bigg\{ - \frac{1}{2} \sum_{i=1}^{3} \mathrm
g^{MN} g^{KL}
\text{Tr} \big[ F_{MK}^{(i)} F^{(i)}_{NL}  \big]
- \frac{1}{4} g^{MN} g^{KL}
 \big[ F_{MK}^{(X)} F^{(X)}_{NL}  \big] \notag \\
& \qquad + g^{MN} (D_M H)^{\dagger} (D_N H) + \bigg[ \mu^2 |H|^2 - \frac{\lambda_6^{(H)}}{4} |H|^4 \bigg] \notag \\
& \qquad + \bigg[ i \bar{\mathcal{Q}}_{3-} \Gamma^M D_M \mathcal{Q}_{3-} 
 \bigg] 
+  \bigg[
 i  \bar{\mathcal{U}}_{3+} \Gamma^M D_M \mathcal{U}_{3+}
\bigg] - \bigg[ \lambda_{6}^{(t)} \Big( \bar{\mathcal{Q}}_{3-} (i \sigma_2 H^{\ast}) \mathcal{U}_{3+} 
 \Big)
+ \mathrm{h.c.} \bigg]
\Bigg\},
\label{eq:projected6DactionofRP2}
\end{align}
and we obtain a usual type of UED model action.

Next we discuss the mass spectrum of the UED model on ${PS}$.
Roughly speaking, about a half of modes are projected out.
First, we focus on the $U(1)_X$ gauge boson.
By use of properties of spin weighted spherical harmonics, we can conclude that its
identification conditions in terms of 4D KK fields are as follows:
\begin{align}
X_{\mu}^{(j,m)}(x) &= (-1)^{j} (X^{(j,m)}_{\mu})^{\text{c}}(x) =  (-1)^{j+1} (X^{(j,m)}_{\mu})(x), 
\label{eq:Xmustateid}\\
X_{\pm}^{(j,m)}(x) &= (-1)^{j+1} (X^{(j,m)}_{\mp})^{\text{c}}(x),
\label{eq:Xpmstateid} \\
\phi^{(X)(j,m)}_{1}(x) &= (-1)^{j+1} (\phi^{(X)(j,m)}_{1})^{\text{c}}(x) = (-1)^{j} (\phi^{(X)(j,m)}_{1})(x), 
\label{eq:X1stateid}\\ 
\phi^{(X)(j,m)}_{2}(x) &= (-1)^{j} (\phi^{(X)(j,m)}_{2})^{\text{c}}(x) = (-1)^{j+1} (\phi^{(X)(j,m)}_{2})(x),
\label{eq:X2stateid} 
\end{align}
where the superscript ${}^{\text{c}}$ means 4D charge conjugation and has the property that
$(X_M^{(j,m)})^{\text{c}} (x) = - (X_M^{(j,m)})^{\text{T}}$.
$\phi_{1,2}^{(X)}$ are 
{a} 4D physical scalar field and {an} unphysical would-be Nambu-Goldstone 
mode of $U(1)_X$ gauge field, respectively in {Eq.}~(\ref{eq:vectoscalarmassbases}).
In {Eq.}~(\ref{eq:Xmustateid}), it is clear that its unwanted zero mode is projected out correctly.
In ${PS}$ case,  
the range of the summation over $m$ does not shrink under {the} identification
and is still $[-j,j]$.
This means that degeneracy of KK masses is $2j+1$ in this model.
But from {Eqs.}~(\ref{eq:Xmustateid})-(\ref{eq:X2stateid}),
we can find that
the even $j$ modes of both $X_{\mu}^{(j,m)}$ and $\phi^{(X)(j,m)}_{2}$
and
the odd $j$ modes of $\phi^{(X)(j,m)}_{1}$ are projected out.
The structure of these mass spectrums is one of the most characteristic feature in the UED model
on ${PS}$ and influences the rates of  the Higgs production and decay processes via loop diagrams.

Next, we go on to the gauge bosons $A_M^{(i)}$ and {the} Higgs {$H$}.
These field are identified by the 6D $P$ transformation and  
its identification conditions in terms of 4D KK fields are as follows:
\begin{align}
A_{\mu}^{(i)(j,m)}(x) &= (-1)^{j} (A^{(i)(j,m)}_{\mu})(x), 
\label{eq:Amustateid}\\
A_{\pm}^{(i)(j,m)}(x) &= (-1)^{j+1} ({A}^{(i)(j,m)}_{\mp})(x),
\label{eq:Apmstateid} \\
\phi^{(i)(j,m)}_{1}(x) &= (-1)^{j+1} (\phi^{(i)(j,m)}_{1})(x), 
\label{eq:A1stateid}\\ 
\phi^{(i)(j,m)}_{2}(x) &= (-1)^{j} (\phi^{(i)(j,m)}_{2})(x).
\label{eq:A2stateid} 
\end{align}
\beq
H^{(j,m)}(x) = (-1)^{j} H^{(j,m)}(x).
\label{eq:Higgsstateid}
\eeq
From {Eqs.}~(\ref{eq:Amustateid})-(\ref{eq:Higgsstateid}),
it is obvious that
the even $j$ modes of $\phi^{(i)(j,m)}_{1}$
and
the odd $j$ modes of $A_{\mu}^{(i)(j,m)},\phi^{(i)(j,m)}_{2}$ and $H^{(j,m)}$ are projected out.
As a previous argument, the zero modes of $A_{\mu}^{(i)(j,m)}$ do not vanish.

Finally, 6D Weyl fermion must be discussed.
It is important that {the} ``mirror" fermions are completely projected out from the action in {Eq.}~(\ref{eq:projected6DactionofRP2}) after the antipodal identification.
This is interpreted that all modes of {the} ``mirror" fermions $\{ \mathcal{Q}_{3+} , \mathcal{U}_{3-} \}$ are erased and no mode of $\{ \mathcal{Q}_{3-} , \mathcal{U}_{3+} \}$ is projected out.

We comment on the dark matter candidate briefly.
In this model, there is no KK-parity because of lack of fixed points.
But alternatively, the conservation of KK angular momentum exists and
it implies that the lightest KK particle is stable.

{
\section{{Naive Dimensional Analysis}
\label{Naive Dimensional Analysis}}

%
%
%
%
In 6D UED models, since the gluon fusion Higgs production and Higgs decay to two photons {processes} are logarithmically divergent, 
we must consider upper limit of {the summations} of KK number in such {models}.
We review Naive Dimensional Analysis (NDA) in these 6D models briefly.
Following the {concept of} NDA, a loop expansion parameter $\epsilon$ in D-dimensional
SU(N) gauge theory at a scale $\mu$ is obtained as
\al{
\epsilon{(\mu)}
	&=	\frac{1}{2} \frac{2 \pi^{D/2}}{(2\pi)^D \Gamma(D/2)}
		N_g\,g_{Di}^{ 2}(\mu)\,{\Lambda_{\text{UED}}}^{D-4},
}
where $N_g$ is a group index, $g_{Di}$ is a gauge coupling in $D$-dimensions and ${\Lambda_{\text{UED}}}$ is a
cutoff scale.
The index $i$ is introduced to express the type of gauge interaction
and the remaining part is originated from D-dimensional momentum loop integral.
We should mention that $g_{Di}({\mu})$, which is the effective running coupling, has {energy} dependency and
obeys power-of-two law scaling.
When we consider a 6D theory $(D=6)$ with two {compact} spacial directions, an effective 4D gauge
coupling $g_{i}$ emerges after KK decomposition and it is described with 
the volume of two extra dimensions $V_2$ as
$g_{i} = {g_{6i}}/{\sqrt{V_2}}$.
{
The cutoff scale $\mu$ is the scale where the perturbation breaks down $\epsilon({\Lambda_{\text{UED}}})\sim 1$.
}
It is obvious that the upper bound of ${\Lambda_{\text{UED}}}$ depends on the value of $V_2$,
whose value is $(2\pi R)^2$ in $T^2$ and $4\pi R^2$ in $S^2$, where $R$ is the radius of $T^2$ or $S^2$.

Next we would like to focus on the behavior of running of the 4D effective gauge coupling strength $\alpha_{i}({\Lambda_{\text{UED}}})$ along the energy.
We consider the following renormalization group equation:
	\al{{
	\alpha_{4i}^{{-1}}(\mu) 
		=	\alpha_{4i}^{{-1}}(m_Z) 
			-\frac{\textsf	{b}_i^{\text{SM}}}{2\pi}
				\ln{\mu \over m_Z}
			+2C\,{\textsf{b}_i^\text{6D}\over2\pi}
				\ln{\mu \over M_{\text{KK}}}
			-C\,\frac{\textsf{b}_i^{\text{6D}}}{2\pi}  
				\sqbr{ \left(  \frac{\mu}{M_\text{KK}} \right)^2 -1 },
			\label{6DcouplingRGequation}
	}}
where $C$ represents $\pi/2\,(1)$ in the case of $T^2$\,($S^2$) geometry. 
We note that the coefficient of the quadratic term for $T^2$ coincide{s} with that in Refs.~\cite{Dienes:1998vh,Dienes:1998vg}
obtained from a different regularization scheme.
The value of $C$ differs due to the structure of the background geometry.\footnote{{Readers who are interested in the details see Appendix in Ref.~\cite{Nishiwaki:2011gm}.}}
In Eq.~(\ref{6DcouplingRGequation}), we take a scheme of approximation;
masses of particles are almost degenerated in each KK level
regardless of type of the fields,
the effect of KK particles appears after the reference energy $\mu$ 
exceeds the value of $M_{\text{KK}}$.\footnote{
{When we consider PS model with non-orientable manifold, there are differences in KK spectrum of
gauge and Higgs fields compared to that of the other ``ordinary" UED models as we discussed before.
We ignore the effect coming from this in our analysis. }
}
The coefficients are summarized in Table.~\ref{RGequation_coefficients}.\footnote{
{Note that we do not employ the GUT normalization for the $U(1)_Y$ coupling and the beta function.}
}
\begin{table}
{
\begin{center}
\begin{tabular}{|c|c|c|}
\hline 
gauge group & SM contribution ($\textsf{b}^{\text{SM}}_i$) & KK contribution ($\textsf{b}^{\text{6D}}_i$) \\
\hline 
$SU(3)_C$ &  $\displaystyle -7$ & $-2$ \\
$SU(2)_W$ & $\displaystyle -{19}/{6}$ & $\displaystyle {3}/{2}$ \\
$U(1)_Y$ & $\displaystyle {41/6}$ & $\displaystyle {27}/{2}$ \\
\hline
\end{tabular}
\caption{Coefficients of renormalization group equation in Eq.~(\ref{6DcouplingRGequation}).}
\label{RGequation_coefficients}
\end{center}
}
\end{table}%

Considering only the quadratic term, Eq.~\eqref{6DcouplingRGequation} reads
\al{
\alpha_{4i}^{-1} (\Lambda_{\text{UED}}) \sim \alpha_{4i}^{-1} ({m_Z})
- \frac{C \textsf{b}^{\text{6D}}_i}{{2}\pi}
\frac{\Lambda_{\text{UED}}^2}{M_\text{KK}^2}.
\label{approximatedalpha4i}
}
From Eq.~\eqref{approximatedalpha4i} and $\epsilon({\Lambda_{\text{UED}}})\sim1$, we get
\al{
\Lambda_{\text{UED}}^2
	\sim	{
				{4\pi M_\text{KK}^2
				\over
				C\paren{N_g+2\textsf{b}_i^\text{6D}}\alpha_{4i}(m_Z)},
				}
\label{positionofcutoffscale}
}

\noindent
In the above analysis, we take values of $N_g$ as $3$, $2$ and $1$
in each case of $SU(3)_C$, $SU(2)_W$ and $U(1)_Y$, respectively, and
adopt some latest data announced by Particle Data Group\, (PDG) as
\begin{equation}
\left\{
\begin{array}{rcl}
\alpha_{U(1)_Y}(m_Z)^{-1}\big|_{\text{MS}} &=& 97.99, \\
\alpha_{SU(2)_W}(m_Z)^{-1}\big|_{\text{MS}} &=& 29.46, \\
\alpha_{SU(3)_C}(m_Z)^{-1}\big|_{\text{MS}} &=& 8.445, \\
m_Z &=& 91.18\, \text{[GeV]}.
\end{array}
\right.
\end{equation}
We do not consider ``TeV-scale gauge coupling unification" in this paper.

In the both $T^2$ and $S^2$ cases, the most stringent bounds come from the $U(1)_Y$ cutoff scales, which restrict the effective range of the perturbation the most severely.
Therefore we can conclude that the ``cutoff" scales are as follows:
\al{
{\Lambda_{\text{UED}}} &\lesssim {{5.3} \, M_{\text{KK}}},  &
&\text{for $T^2$-case} \ (V_2 = (2\pi R)^2, \, M_{\text{KK}} = 1/R),
\label{T2casecutoff} \\
{\Lambda_{\text{UED}}} &\lesssim {{6.6} \, M_{\text{KK}}}, &
&\text{for $S^2$-case} \ (V_2 = 4\pi R^2, \, M_{\text{KK}} = \sqrt{2}/R).
\label{S2casecutoff}
}
We truncate the KK mode summations up to these upper bounds in each case
to regularize the process. 
Before going on to the concrete calculation, we have to declare our choice of the UED cutoff scales.
We choose three patterns in $T^2$ and $S^2$ cases separately and the concrete forms are summarized in
Table~\ref{UEDcutoffvalues}.
We also list up the value of the QCD and electromagnetic coupling strengths ${\{}\alpha_s, \alpha_{\text{EM}}{\}}$ at the cutoff scales by use of Eq.~(\ref{approximatedalpha4i})  in Table~\ref{strengths_at_cutoff}.
It is noted that the values derived form Eq.~(\ref{approximatedalpha4i}) do not depend on the value of the
KK mass scale $M_{\text{KK}}$ up to our approximation in Eq.~(\ref{approximatedalpha4i}).
The electromagnetic coupling strength is defined by using $\alpha_{SU(2)_W}$ and $\alpha_{U(1)_Y}$ as
\al{
\alpha_{\text{EM}}(\mu)^{-1} = \alpha_{SU(2)_W}(\mu)^{-1} + \alpha_{U(1)_Y}(\mu)^{-1}.
}

\begin{table}
{
\begin{center}
\begin{tabular}{|c||cc|cc|}
\hline
& \multicolumn{2}{|c|}{$T^2$-based}
& \multicolumn{2}{|c|}{$S^2$-based}\\
 & high & low  & high & low \\
\hline
KK index &  $m^2+n^2 \leq 30$ & $m^2+n^2 \leq 10$ &
 $j(j+1) \leq 100$ & $j(j+1) \leq 30$\\
UV cutoff & 
	$\Lambda_{\text{UED}} \sim 5\MKK$ & 
	$\Lambda_{\text{UED}} \sim 3\MKK$ & 
	$\Lambda_{\text{UED}} \sim 7\MKK$ & 
	$\Lambda_{\text{UED}} \sim 4\MKK$ \\
\hline
\end{tabular}
\caption{{Two choices of high and low upper bounds for KK indices and for the corresponding UV cutoff scale.}}
\label{UEDcutoffvalues}
\end{center}
}
\end{table}

\begin{table}
{
\begin{center}
\begin{tabular}{|c||cc|cc|}
\hline
& \multicolumn{2}{|c|}{$T^2$-based}
& \multicolumn{2}{|c|}{$S^2$-based}\\
 & high & low &  high & low \\
\hline
$\alpha_s(\Lambda_{\text{UED}})^{{-1}}$  & $20.9$ & $12.9$ &
 $24.0$ & $13.5$\\
$\alpha_{\text{EM}}(\Lambda_{\text{UED}})^{{-1}}$ & 
	$33.7$ & 
	$93.7$ & 
	$10.5$ & 
	$89.3$ \\
\hline
\end{tabular}
\caption{{The value of the QCD and electromagnetic coupling strengths at the cutoff scales.}}
\label{strengths_at_cutoff}
\end{center}
}
\end{table}

}

\section{The deviation of the rates of Higgs production and its decay from the standard model predictions
\label{The deviation of the rates of Higgs production and its decay from the standard model predictions}}

\subsection{Formulation of calculation}

From the discussions which we have done,
we evaluate the ratio (fractional deviation) of the Higgs production cross section through gluon fusion 
and
the Higgs decay width into two photons to the SM ones in the three types of 6D UED models,
which are denoted by $\mathcal{R}_{2g \rightarrow h^{(0)}}$ and
$\mathcal{R}_{h^{(0)} \rightarrow 2\gamma}$, respectively.
These ratio are represented as
\beq
\mathcal{R}_{2g \rightarrow h^{(0)}} \equiv
\frac{\sigma(2g \rightarrow h^{(0)};\ \text{UED})}{\sigma(2g \rightarrow h^{(0)};\ \text{SM})}
=
\left(
1 + {F_t^{{\text{KK}}} {+ F_{\text{gluonfusion}}^{\text{TC}}} \over F_t^{{\text{SM}}}  }
\right)^2,
\eeq
\beq
\mathcal{R}_{h^{(0)} \rightarrow 2\gamma} \equiv
\frac{\Gamma(h^{(0)} \rightarrow 2\gamma;\ \text{UED})}{\Gamma(h^{(0)} \rightarrow 2\gamma;\ \text{SM})}
=
\left(
1 + {F_W^{{\text{KK}}} + 3Q_t^2 F_t^{{\text{KK}}} {+ F_\text{decay}^{\text{TC}}} \over F_W^{{\text{SM}}} + 3Q_t^2 F_t^{{\text{SM}}} }
\right)^2.
\eeq
We have obtained $F_W^{{\text{KK}}}, F_t^{{\text{KK}}}, {F_{\text{gluonfusion}}^{\text{TC}}}$ and ${F_\text{decay}^{\text{TC}}}$ in {Section}~\ref{Calculation of one loop Higgs production and decay processes} in the case of $T^2/Z_4$
by 1-loop calculation
and we can apply these results for {the} $S^2/Z_2$ and {the} ${PS}$ cases with some modifications.
It is important that {the} $U(1)_X$ gauge boson does not contribute to either the production process and 
the decay process at {the} 1-loop level.
Therefore no new type of diagram appears and
only difference {appears in} the KK mass spectrum and the multiplicity of each KK mode.
Once the $Z_2$ orbifolding or the antipodal identification is understood,
the {structure of KK} mass spectrum itself is the same as the case of $S^2$ {up to degeneracy}.
We summarize the information which is needed for the estimation in {Table~\ref{multiplicity_of_KKmass}}. 
\begin{table}[htbp]
\begin{center}
\begin{tabular}{c||c|c}
type of field & $S^2/Z_2$ case & ${PS}$ case \\
\hline
fermion & 
$\begin{array}{cl} j+1 & \text{for}\ j=\text{even} \\  j & \text{for}\ j=\text{odd} \end{array}$ 
& $\begin{array}{cl} 2j+1 & \text{for}\ j=\text{even} \\  2j+1 & \text{for}\ j=\text{odd} \end{array}$ \\
\hline
``mirror" fermion &
N/A 
& $\begin{array}{cl} 0& \text{for}\ j=\text{even} \\  0 &  \text{for}\ j=\text{odd} \end{array}$ \\
\hline
\begin{tabular}{c} gauge boson  \& would-be NG boson
\\ \& scalar(Higgs) \end{tabular}  & 
$\begin{array}{cl} j+1 & \text{for}\ j=\text{even} \\  j & \text{for}\ j=\text{odd} \end{array}$ 
& $\begin{array}{cl} 2j+1 & \text{for}\ j=\text{even} \\  0 & \text{for}\ j=\text{odd} \end{array}$ \\
\hline
scalar(``spinless adjoint") & 
$\begin{array}{cl} j+1 & \text{for}\ j=\text{even} \\  j & \text{for}\ j=\text{odd} \end{array}$ 
& $\begin{array}{cl} 0 & \text{for}\ j=\text{even} \\  2j+1 & \text{for}\ j=\text{odd} \end{array}$ \\
\hline
\end{tabular}
\caption{Multiplicities of fields at {j level} in $S^2$-based UED models.}
\label{multiplicity_of_KKmass}
\end{center}
\end{table}
In {the} $S^2/Z_2$ case, {the} KK state multiplicity is the same irrespective of 
the type of field. With the modification
\beq
{\sum_{m \geq 1, n \geq 0} \rightarrow
\sum_{j \geq 1}} \ n_{{S^2/Z_2}}(j)
,\quad
m_{(m,n)} \rightarrow m_{(j,m)},
\eeq
where $n_{S^2/Z_2}(j)$ shows the multiplicity of each level of KK modes {($j+1$ for $j=\text{even}$ or $j$ for $j=\text{odd}$)} and $m_{(j,m)}$ is the KK mass on $S^2$
{in Eq.}~(\ref{eq:KKmass_S2}), we can obtain the results as follows:
\begin{align}
F_t^{{\text{KK}}} &= 2 \sum_{j\geq 1} n_{S^2/Z_2}(j) \paren{\frac{m_t}{m_{t,(j,m)}}}^2 \notag \\
& \quad \times \br{{-2} \lambda(m_{t,(j,m)}^2) {+} \lambda(m_{t,(j,m)}^2) (1-4\lambda(m_{t,(j,m)}^2)) J\paren{\lambda(m_{t,(j,m)}^2)}} ,
\label{FtKKS2}\\
F_{W}^{{\text{KK}}} &= \sum_{j \geq 1} n_{S^2/Z_2}(j) \Big\{
\frac{1}{2}  + 5 \lambda(m_W^2)
- \Big[
\lambda(m_W^2) (4  - 10 \lambda(m_{W,(j,m)}^2)) \notag \\
& \qquad \qquad - \lambda(m_{W,(j,m)}^2)
\Big] J \paren{\lambda(m_{W,(j,m)}^2)}
\Big\} {,}
\label{KWKKsum_S2}
\end{align}
{where} we  use {$J(m^2)$ in Eq.~(\ref{eq:scalarPVfunction})}.
In ${PS}$ case, we should pay attention to the KK state multiplicity of each type of field.
There is no contribution from {the} ``mirror" fermions.
{The {concrete} forms}  are as follows:
\begin{align}
{F_t^{\text{KK}}} &= 2 \sum_{j\geq 1} (2j+1) \paren{\frac{m_t}{m_{t,(j,m)}}}^2 \notag \\
& \quad \times \br{{-2} \lambda(m_{t,(j,m)}^2) {+} \lambda(m_{t,(j,m)}^2) (1-4\lambda(m_{t,(j,m)}^2)) J\paren{\lambda(m_{t,(j,m)}^2)}} , \\
F^{{\text{KK}}}_{\text{gauge}} &= {\sum_{j \geq 1}
n_{PS\text{even}}(j)
\left\{ 3 \lambda(m_W^2) + 2 \lambda(m_W^2) \big(3 \lambda(m_{W,(j,m)}^2) - 2 \big) J\paren{\lambda(m_{W,(j,m)}^2)} \right\} },
 \\
F^{{\text{KK}}}_{\text{NG}} &= { \sum_{j \geq 1}
n_{PS\text{even}}(j)
\left( \frac{1}{2} \frac{m_h^2}{m_{W,(j,m)}^2} \right)
\lambda(m_W^2) \br{ 1 + 2 \lambda(m_{W,(j,m)}^2) J\paren{\lambda(m_{W,(j,m)}^2)} } }, 
\\
F^{{\text{KK}}}_{\text{scalar1}} &=  \sum_{j \geq 1}
n_{PS\text{even}}(j)
\Big( \frac{1}{2} \frac{1}{m_{W,(j,m)}^2} \Big)
\Big[ \frac{m_h^2}{m_W^2} m_{(j,m)}^2 + {2} m_{W,(j,m)}^2  \Big] \notag \\
& \quad \times \lambda(m_W^2) \br{ 1 + 2 \lambda(m_{W,(j,m)}^2) J\paren{\lambda(m_{W,(j,m)}^2)} } 
, \\
F^{{\text{KK}}}_{\text{scalar2}} &= { \sum_{j \geq 1}
n_{PS\text{odd}}(j)
\lambda(m_W^2) \br{ 1 + 2 \lambda(m_{W,(j,m)}^2) J\paren{\lambda(m_{W,(j,m)}^2)} } }
,
\end{align}
{where $n_{PS\text{even}}(j)$: $2j+1$ for $j=$even, $0$ for $j=$odd and $n_{PS\text{odd}}(j)$: $0$ for $j=$even, $2j+1$ for $j=$odd.
We have already discussed the cutoff scale in both {the} $T^2$ and $S^2$ case{s} concretely in Section~\ref{Naive Dimensional Analysis} and we are ready to estimate the ratio in the various 6D UED models.}

\subsection{{Results without threshold corrections}}

The numerical results of the ratios of the production cross section via gluon fusion
to the standard model prediction
$\mathcal{R}_{2g \rightarrow h^{(0)}}$ are given as functions of the first KK
mass scale {$(M_{\text{KK}})$} in a unit of GeV in Fig.~\ref{gluonfusiontotal}.
{In this paper, we consider two possibilities of Higgs mass; $m_h = 120\,\text{GeV}$ and $m_h = 145\,\text{GeV}$
and take the KK mass range between $600\,\text{GeV}$ and $2000\,\text{GeV}$.
We use the values of {the} W boson mass $m_W$ and {the} top quark mass $m_t$, which are
$m_W = 80.3\,\text{GeV},  m_t = 173\,\text{GeV}$.

{From top to bottom, the green, blue, red curves represent the results of
${PS}$, $S^2/Z_2$, $T^2/Z_4$ {with $m_h = 120\,\GeV$}, providing no threshold correction, {respectively}.
Each black dashed line near the lines for $m_h = 120\,\text{GeV}$ corresponds to that with $m_h = 145\,\text{GeV}$. The left, right are these with {the} high, low cutoff choices, respectively.
It is noted that the $\mathcal{R}_{2g \rightarrow h^{(0)}} = 1$ shows the SM predictions and there are few difference{s} between $m_h = 120\,\text{GeV}$ case and $m_h = 145\,\text{GeV}$ case in {all the} models.}
Contrast to the case of little Higgs~\cite{Han:2003gf,Dib:2003zj,Chen:2006cs}
or gauge-Higgs unification~\cite{Falkowski:2007hz,Maru:2007xn,Maru:2008cu}, 
the contribution from KK fermions 
is constructive and the results of UED cases are {enhanced compared} to the {SM} prediction. 
These results are naturally understood
because the number of the intermediate particles are much greater than {these} of the SM.\footnote{These results are consistent with the results in {Ref.}~\cite{Maru:2009cu}.}
It is expected that
6D UED models predict a significant collider signature {in the} Higgs production at the LHC,
especially in {the} ${PS}$ case.
The origin of the remarkable enhancement in {the} ${PS}$ case is that {numerous} fermions contribute to the production process in each KK level.
{Besides, we can find the fact that when we choose the higher cutoff, the larger number of KK top modes
propagate in the triangle loop and therefore the deviation from the SM gets significant.
This tendency do not depend on the type of the 6D UED models.}
\begin{figure}[h]
\centering
\includegraphics[width=\columnwidth, clip]{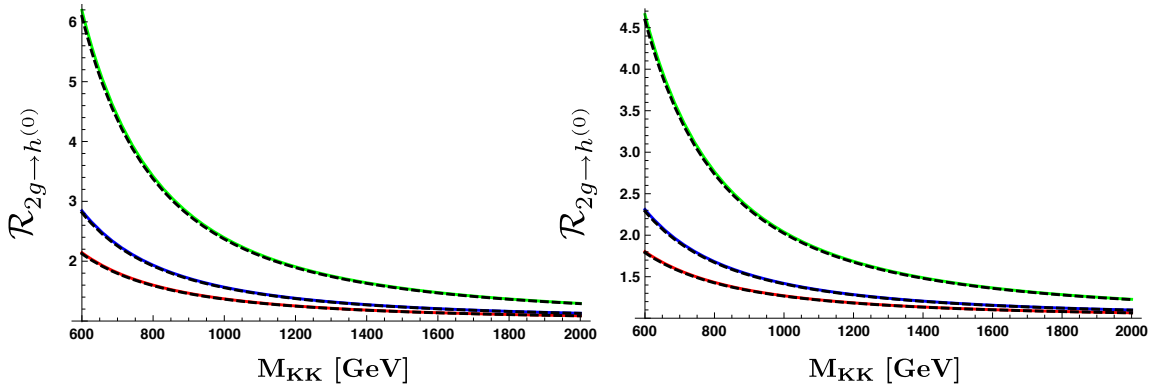}
\caption{
{These plots represent the ratios of the Higgs boson production cross sections
via gluon fusion to the SM prediction $\mathcal{R}_{2g \rightarrow h^{(0)}}$ in 6D UED on ${PS}$(green), $S^2/Z_2$(blue), 
$T^2/Z_4$(red) with $m_h = 120\,\text{GeV}$ providing no threshold correction from top to bottom.
Each black dashed line near the lines for $m_h = 120\,\text{GeV}$ corresponds to that with $m_h = 145\,\text{GeV}$. The left, right are these with {the} high, low cutoff choices, respectively.}
}
\label{gluonfusiontotal}
\end{figure}

The numerical results of {the} ratios of the rate of Higgs decay into two photons
$\mathcal{R}_{h^{(0)} \rightarrow 2\gamma}$ are also given as functions of the first KK
mass scale {($M_{\text{KK}}$)} in a unit of GeV  in Fig.~\ref{twophotondecaytotal}.
From bottom to top, the green, blue, {red curves} represent the results of
${PS}$, $S^2/Z_2$, $T^2/Z_4$ {with $m_h = 120\,\GeV$}, providing no threshold correction, respevtively.
Each black dashed line located above the lines for $m_h = 120\,\text{GeV}$ corresponds to that with $m_h = 145\,\text{GeV}$. The left, right are these with {the} high, low cutoff choices, respectively.}
Differently from the production, the ratios are {suppressed} because the contributions
from quarks and gauge bosons are destructive each other.
The reason of the large reduction in {the} ${PS}$ case is understood as the results of the enormous effects
of KK top quarks, which we discussed before.
In any type of 6D UED model{s}, this ratio takes the lower value than 5D mUED one.
{We can find some differences between $m_h = 120\, \text{GeV}$ and $m_h = 145\, \text{GeV}$ in each case of 6D UED model, which are sizable in particular at the KK mass range between $600\,\text{GeV}$ and $1200\,\text{GeV}$.}
\begin{figure}[h]
\centering
\includegraphics[width=\columnwidth, clip]{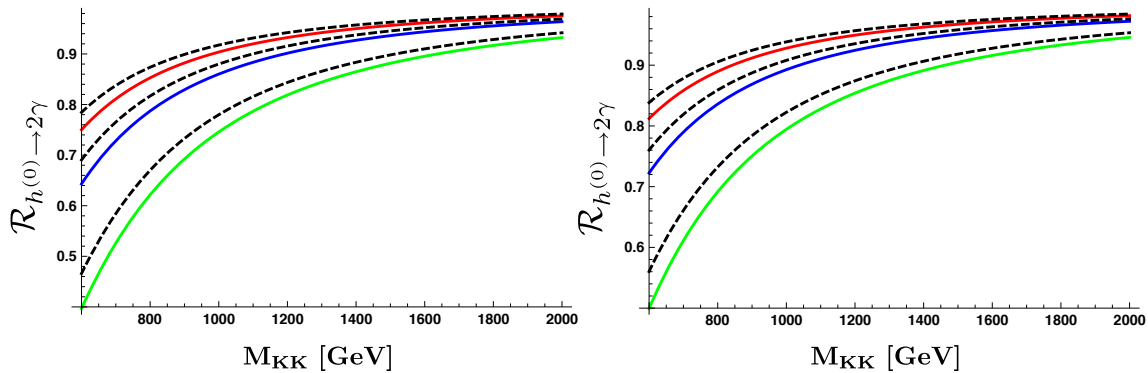}
\caption{
{These plots represent the ratios of the Higgs boson decay width to two photons to the SM prediction $\mathcal{R}_{h^{(0)} \rightarrow 2\gamma}$ in 6D UED on ${PS}$(green), $S^2/Z_2$(blue), 
$T^2/Z_4$(red) with $m_h = 120\,\text{GeV}$ providing no threshold correction from bottom to top.
Each black dashed line located above the lines for $m_h = 120\,\text{GeV}$ corresponds to that with $m_h = 145\,\text{GeV}$. The left, right are these with {the} high, low cutoff choices, respectively.}
}
\label{twophotondecaytotal}
\end{figure}

We now define {a value defined as} 
\al{
\Delta \equiv \mathcal{R}_{2g \rightarrow h^{(0)}}
\times \mathcal{R}_{h^{(0)} \rightarrow 2\gamma},
}
which shows {the ``total ratio" of} the deviation of 
the $h^{(0)} \rightarrow 2\gamma$ signals coming from the $2g \rightarrow h^{(0)}$ Higgs
production.
At the LHC, the Higgs production process through gluon fusion is dominant and
the value of $\Delta$ is considered to be an appropriate approximation of the $h^{(0)} \rightarrow 2\gamma$ signal deviation coming from all the Higgs production processes.
Of course the numerical results of $\Delta$ are given as functions of the first KK
mass scale {($M_{\text{KK}}$)} in a unit of GeV and are shown in Fig.~\ref{totalratiototal}.
It should be noted that in 6D UED models, the collider signal deviations from the SM
{in} the $h^{(0)} \rightarrow 2\gamma$
process take the {greater} values than in the 5D mUED case.
When we take the reference value as ${M_{\text{KK}}} = 800$ GeV {in the $m_h = 120$ GeV and each high cutoff case,
approximately $40\% (T^2/Z_4)$, $60\% (S^2/Z_2)$, $110\% ({PS})$ enhancements}
from the SM expectation value can be seen.
{It should be mentioned that the shapes of $\Delta$ in each model do not have large dependence on the value of the UED cutoff $\Lambda_{\text{UED}}$.
This reason can be considered that the behavior of the ratios of the gluon fusion Higgs production and the Higgs decay to two photons is opposite when we change the value of the cutoff
and a large part of the distinctions due to the value of the cutoff are cancelled out.
This property is accidental but do not depend on the type of the background geometry and thereby we consider that
this is one of the interesting aspects of 6D UED model.} 
The difference between the above results and the SM expectation value is significant
and we hope that this could be tested at the LHC experiments in the near future.

{Finally, we comment on the up-to-date collider experimental results at the LHC.
The ATLAS group  announced their results, which conclude the upper limit of the cross section of the $h^{(0)} \rightarrow 2\gamma$
process in the form of the ratio to the SM result $(\sigma/\sigma_{\text{SM}})$ based on the $1.7\,\text{fb}^{-1}$ data within the $95\%$ confidence level {in the August of 2011}.
According to~\cite{ATLAS-CONF-2011-135}, the value of the upper bound of $(\sigma/\sigma_{\text{SM}})$ is about $3.5$ $(5.0)$
at the point of $m_h = 120\,\text{GeV}$ ($m_h = 145\,\text{GeV}$).
The CMS group also  announced their results, which says that the value of the upper bound of $(\sigma/\sigma_{\text{SM}})$ is about $3.5$ $(4.0)$
at the point of $m_h = 120\,\text{GeV}$ ($m_h = 145\,\text{GeV}$)~\cite{CMS_PAS_HIG-11-022}.
{And at the December of 2011, the new results have been published by both the ALTAS and CMS.
The ATLAS claims that there is an excess of events close to 126 GeV with a 3.6 $\sigma$ confidence~\cite{ATLAS:2012ae}.
On the other hand, the excess also have been observed by the CMS, but the location of the peak is
124 GeV with a 3.1 $\sigma$ confidence~\cite{Chatrchyan:2012tx}.
It is noted that both results are these before taking looking-elsewhere effect.
The allowed region of the SM Higgs becomes highly constrained as $115.5\,\GeV < m_h < 127\,\GeV$
except the unexplored high mass region $m_h > 600\,\GeV$.
}

We do not execute detailed analysis in this paper on this topics but we can conclude from our result in Fig.~\ref{totalratiototal} that {the} $T^2/Z_4$, $S^2/Z_2$, and $PS$ 6D UED with $m_h = 120\,\GeV$ still survive {only} in the KK mass region above ${M_{\text{KK}} = 600, 750, 1150\,\text{GeV}}${, respectively, when we consider the high cutoffs,
judging from the constraint on the value of $\sigma/\sigma_{\text{SM}}$ in the December's ALTAS and CMS results, whose maximum value is roughly 1.6.{\footnote{
{We note that the newer CMS diphoton data set includes vector boson fusion
(VBF) events that occurs at the tree level in the SM. The VBF Higgs production process is not significantly enhanced by the UED loop effects.}
} }

It is obvious that the possibility of $m_h = 145\,\GeV$ is discarded in all the 6D UED models since
the signals are expected to be greater than these in the SM.}

\begin{figure}[h]
\centering
\includegraphics[width=\columnwidth, clip]{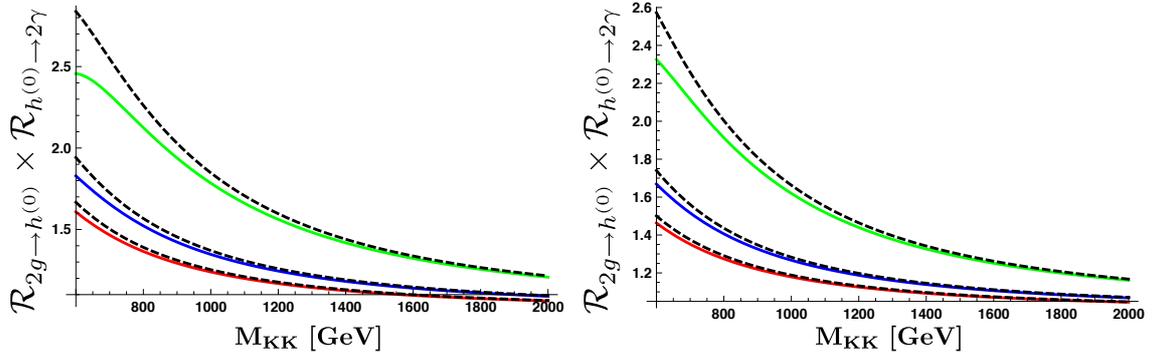}
\caption{
{These plots represent the values of $\Delta$ (total ratio) in 6D UED on ${PS}$(green), $S^2/Z_2$(blue), 
$T^2/Z_4$(red) with $m_h = 120\,\text{GeV}$ providing each high cutoff case and no threshold correction from top to bottom.
Each black dashed line located above the lines for $m_h = 120\,\text{GeV}$ corresponds to that with $m_h = 145\,\text{GeV}$. The left, right are these with {the} high, low cutoff choices, respectively.}
}
\label{totalratiototal}
\end{figure}


\subsection{{Results with threshold corrections}}

{When we switch on the threshold corrections accompanying the processes of $2g \rightarrow h^{(0)}$ and
$h^{(0)} \rightarrow 2 \gamma$, the shapes of the ratios ${\{}\mathcal{R}_{2g \rightarrow h^{(0)}}, \mathcal{R}_{h^{(0)} \rightarrow 2\gamma}{\}}$ and the total ratios $\Delta$ are modified forcefully.
There are two dimensionless new parameters in Eqs.~(\ref{TCcoefficient_gluonfusion},\ref{TCcoefficient_twophoton}) $C_{hgg}, C_{h\gamma\gamma}$, which describe the threshold correction in the process of $2g \rightarrow h^{(0)}$ or $h^{(0)} \rightarrow 2\gamma$, respectively.
In this paper, we only consider some extremal choices of $C_{hgg}, C_{h\gamma\gamma}$ as
\al{
C_{hgg} = 0, \pm 1, \quad C_{h\gamma\gamma} = 0, \pm 1.
}
We mention that the plus/minus sign of $C_{hgg}, C_{h\gamma\gamma}$ determines the direction of interference
effects
to the UED part.
We show the results of our numerical calculations in Figs.~\ref{production_T2Z4}-\ref{combined_PS}.
We write down our convention about the color/shape of curves in Figs.~\ref{combined_T2Z4},\ref{combined_S2Z2},\ref{combined_PS} (total ratios) {in Table~\ref{colorcodes}}.
{In the range of our approximation, the values of $\alpha_{s}^{-1}(m_h)$ and $\alpha_{\text{EM}}^{-1}(m_h)$ only
appears in the terms describing the threshold corrections in Eqs.~(\ref{TCcoefficient_gluonfusion},\ref{TCcoefficient_twophoton}) and we adopt these values at the Z boson mass pole as $\alpha_{s}^{-1}(m_Z) = 8.44, \alpha_{\text{EM}}^{-1}(m_Z) = 127$ with ignoring the renormalization group effects between $m_Z$ and $m_h$ $(= 120\ \text{or}\ 145\,\GeV)$.
We make several comments in order.}
%
\begin{itemize}
\item
In the gluon fusion process in each model, due to the $(-1)$ factor which originates from Fermi statistics
in $F_t^{\text{SM}}$, the interference term between the threshold correction and the UED effect is
destructive (constructive) in case of $C_{hgg} = +1\ (-1)$, respectively.
{It is considered that the degree of a threshold correction is inversely proportional to the value of 
a cutoff.}
When we look at the PS case with its high cutoff choice in Fig.~\ref{production_PS}, we notice that
the threshold correction works a little  {compared to the cases of the $T^2/Z_4$ or $S^2/Z_2$}.
{We can find some differences between the cases of $PS$ and $S^2/Z_2$ with a same cutoff value, which stem from the differences in the corresponding UED contributions.}
{We mention that the threshold correction is still observable in the many cases with $M_{\text{KK}} = 2\,\text{TeV}$.}
By contraries, in all the low cutoff cases in Figs.~\ref{production_T2Z4},\ref{production_S2Z2},\ref{production_PS}, the threshold correction plays {a very} important role.
Here we mention that the cases with $m_h = 145\,\text{GeV}$ are almost identical with these {with} $m_h = 120\,\text{GeV}$. 
\item
In the Higgs decay to two photons in each model, unlike with the previous gluon fusion case,
the interference term between the threshold correction and the UED effect is
constructive (destructive) in case of $C_{h\gamma\gamma} = +1\ (-1)$, respectively.
In this case, the degree of the effect which only comes from the threshold correction is also smaller than the others.
It is an interesting point that in some cases with $C_{g\gamma\gamma} = +1$, the value of the ratio $(\mathcal{R}_{h^{(0)} \rightarrow 2\gamma})$ exceeds one, which we never find in the no-threshold-correction cases {in the range of the parameter region of $M_{\text{KK}}$ which we consider}.
Another remarkable point compared to the gluon fusion, the cases with $m_h = 145\,\text{GeV}$ are not identical with these {with} $m_h = 120\,\text{GeV}$ but this difference is still not so significant since the other effects (cutoff scale, threshold correction and so on) are more effective.
Of course, in all the low cutoff cases in Figs.~\ref{decay_T2Z4},\ref{decay_S2Z2},\ref{decay_PS}, the threshold correction works very well.
{We mention that we can find the $10 \sim 20\,\%$ deviations from the no-threshold-correction cases even with $M_{\text{KK}} = 2\,\text{TeV}$ in all the 6D UED models.}
\item
After combining the above two results, we can estimate the total ratio $\Delta$ in each 6D UED model with the extremal threshold corrections.
In this analysis, we only consider the $m_h=120\,\text{GeV}$ cases.
There are nine curves in each graph and our convention about the color/shape of curves is summarized in Table~\ref{colorcodes}.
Here we would like to only focus on a few important topics.
Firstly, we can find the tendency that every orange line $(C_{hgg} = -1, C_{h \gamma \gamma} = +1)$ is located at
the top of each graph and any cyan one $(C_{hgg} = +1, C_{h \gamma \gamma} = -1)$ is located at
the bottom of each graph.
The reason is that the two threshold corrections function toward maximally enhancing (suppressing) the process
in the former (latter) case.
Secondly, the deviation from the no-threshold-correction case (black dot-dashed curve) is noteworthy, in particular,
in the $M_{\text{KK}}$ range below {$1\,\text{TeV}$}.
All the results tend to converge with the no-threshold-correction curve proportional to the value of $M_{\text{KK}}$
and it is notable {even around $M_{\text{KK}} = 2\,\text{TeV}$ in many choices of $C_{hgg}$ and $C_{h \gamma \gamma}$ because the tens of percents of the deviations still remain}.
Thirdly, in the low cutoff cases, the interference effects dominate the whole process and the predictions about
two photon signals via the gluon fusion Higgs production possibly become extraordinary.
Finally we comment on the constraint from the LHC results briefly.
We also do not execute detailed analysis in this case
but we can conclude that some cases which
predict too great value of the total ratio are already excluded.
{On the other hand, the total ratio can be suppressed in some choices of the parameters describing the threshold corrections, and in this case the possibility with $m_h = 145\,\GeV$ is not totally rejected.}

\end{itemize}
}

{At the end of this section, we give comments for more precise analysis.} We need to take into account the correction from QCD
(parton distribution function and K-factor)~\cite{Rai:2005vy}.
However the KK contributions would receive almost the same QCD corrections
as in the case of the SM
and this deviation from the SM result would not be large.
Actually, to get the ratios of event rates in  6D UED models to that in the SM, the partial decay width in $\Delta$ should be replaced by corresponding branching ratios. We,  however, expect that the effects of heavy KK particles to the leading decay processes at the tree level are small  because of decoupling. Thus $\Delta$ is expected to be enough for crude estimation. 
But there are two other one-loop leading decay processes of
$h^{(0)} \rightarrow 2g$ and $h^{(0)} \rightarrow \gamma Z$,
which may possibly give considerable contribution to the Branching Ratio.
These points are beyond the scope of this paper and are left for future work~{\cite{Nishiwaki:2011gk}}.

\section{Summary
\label{Summary}}
 
In this paper, we have discussed the main Higgs production process through gluon fusion
and the important one-loop leading decay channel to two photons at the LHC in various 6D
UED models. 
{The} Higgs production cross sections in 6D UED models are much enhanced than the prediction
of the SM or the 5D mUED.
In contrast, the decay width in 6D UED models are decreased because of the
destructive contribution between quarks and gauge bosons.
In both cases, the results of ${PS}$ model are significant.
This is because {numerous} fermions contribute to the process in each level of KK modes.
{We also have discussed the threshold corrections in the processes and their effects become significant
{even when we take the higher cutoff and/or a heavy KK scale}. Some parameter region are already excluded by the current LHC experimental {results} {obviously}.}
{By use of the data announced by the ALTAS and CMS experiments in the December of 2011,
we can estimate the lower bounds of the KK scale as $M_{\text{KK}} = 600\,(T^2/Z_4), 750\,(S^2/Z_2), 1150\,(PS)\,\text{GeV}$ when we consider the high cutoffs with no threshold correction}.
{These results are modified by the threshold corrections substantially.}
{The SM with the $145\,\GeV$ Higgs boson is rejected but 6D UED with the Higgs mass parameter
is still survived in some ranges of the parameters describing the threshold corrections.}

Our results are affected by ultraviolet physics because the calculation has
logarithmic cutoff scale dependence.
There seem to be some ambiguities coming from this fact.
We expect that our prediction would be verified by forthcoming LHC {experimental results}.
{Detailed analysis of the final states in the single Higgs production at the LHC is important for discriminating UED from the other models.}

The collider physics and particle cosmology of $S^2$-based 6D UED models are
unexplored and we would like to pursue these topics in future work~{\cite{Nishiwaki:2011gm,Nishiwaki:2011gk}}.

\vspace{5mm}
\noindent{\large \bf Acknowledgments}
\vspace{3.5mm}

\noindent
We are most grateful to C. S. Lim and Kin-ya Oda for valuable comments and discussions.
In particular C. S. Lim red the manuscript carefully and gave us very useful comments.
And we also thank Nobuhito Maru, Naoya Okuda, Makoto Sakamoto,
Joe Sato, Takashi Shimomura, Ryoutaro Watanabe and Masato Yamanaka
for fruitful discussions.
Yasuhiro Okada and Hideo Ito suggest the recent Tevatron experimental results to us.
We express our appreciation to them very much.
{We again appreciate C. S. Lim and Kin-ya Oda for advising me in the revision.}
{Finally we appreciate the referee for giving a lot of useful comments.}



\begin{figure}[H]
\centering
\includegraphics[width=\columnwidth, clip]{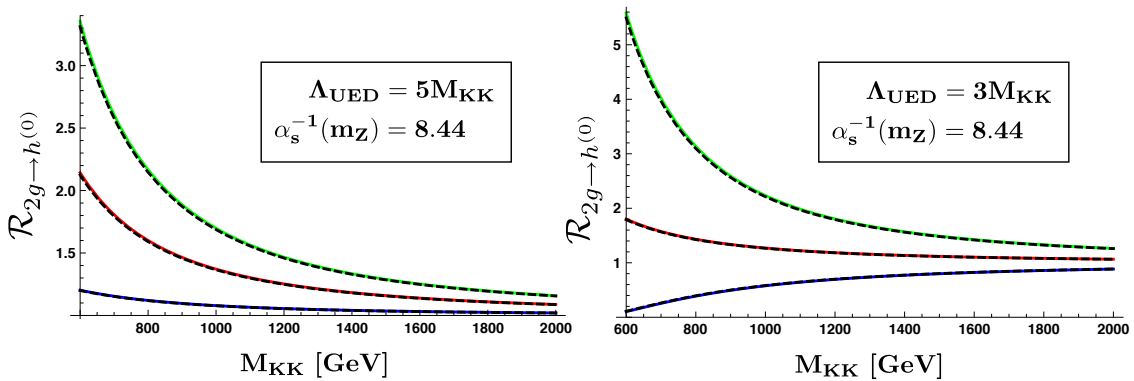}
\caption{
{These plots represent the values of the ratios $\mathcal{R}_{2g \rightarrow h^{(0)}}$ in 6D UED on $T^2/Z_4$ with/without threshold correction.
The red, blue, green curves show these with $C_{hgg} = 0, +1, -1$, respectively.
Each black dashed line near the lines for $m_h = 120\,\text{GeV}$ corresponds to that with $m_h = 145\,\text{GeV}$.
{In the left and right plots, which correspond to the high and low cutoff cases, respectively, we take the value of the QCD coupling strength as that at the Z boson mass scale.}
}
}
\label{production_T2Z4}
\end{figure}
\begin{figure}[H]
\centering
\includegraphics[width=\columnwidth, clip]{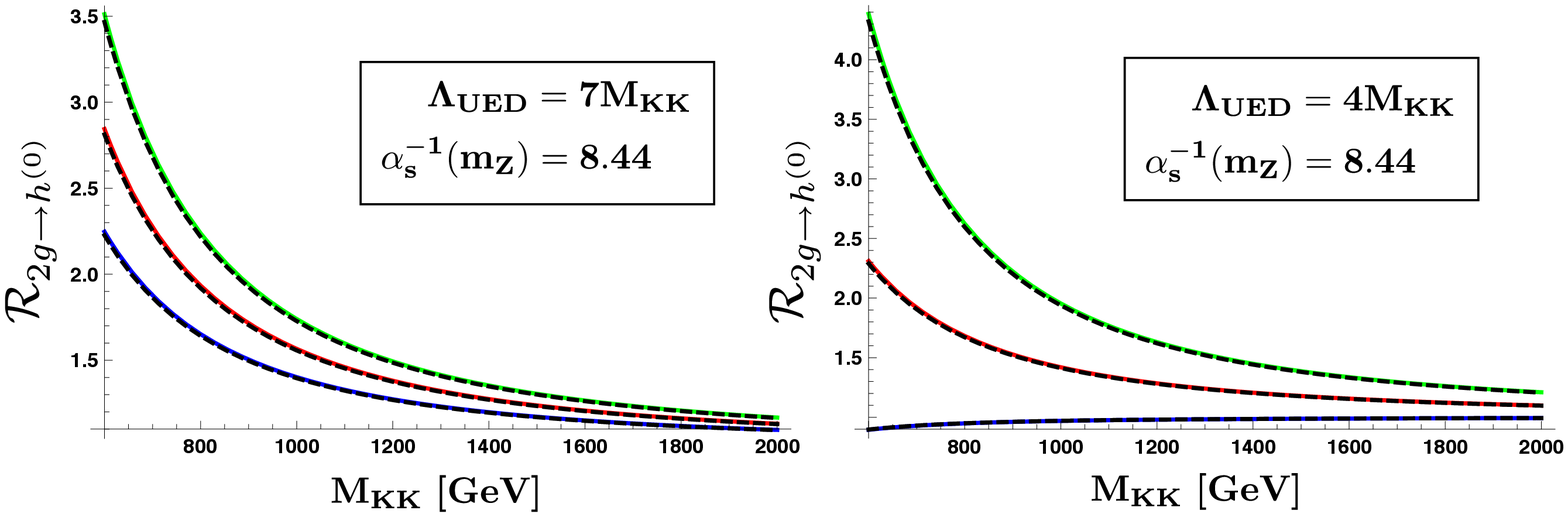}
\caption{
These plots represent the values of the ratios $\mathcal{R}_{2g \rightarrow h^{(0)}}$ in 6D UED on $S^2/Z_2$ with/without threshold correction.
The red, blue, green curves show these with $C_{hgg} = 0, +1, -1$, respectively.
Each black dashed line near the lines for $m_h = 120\,\text{GeV}$ corresponds to that with $m_h = 145\,\text{GeV}$.
{In the left and right plots, which correspond to the high and low cutoff cases, respectively, we take the value of the QCD coupling strength as that at the Z boson mass scale.}
}
\label{production_S2Z2}
\end{figure}
\begin{figure}[H]
\centering
\includegraphics[width=\columnwidth, clip]{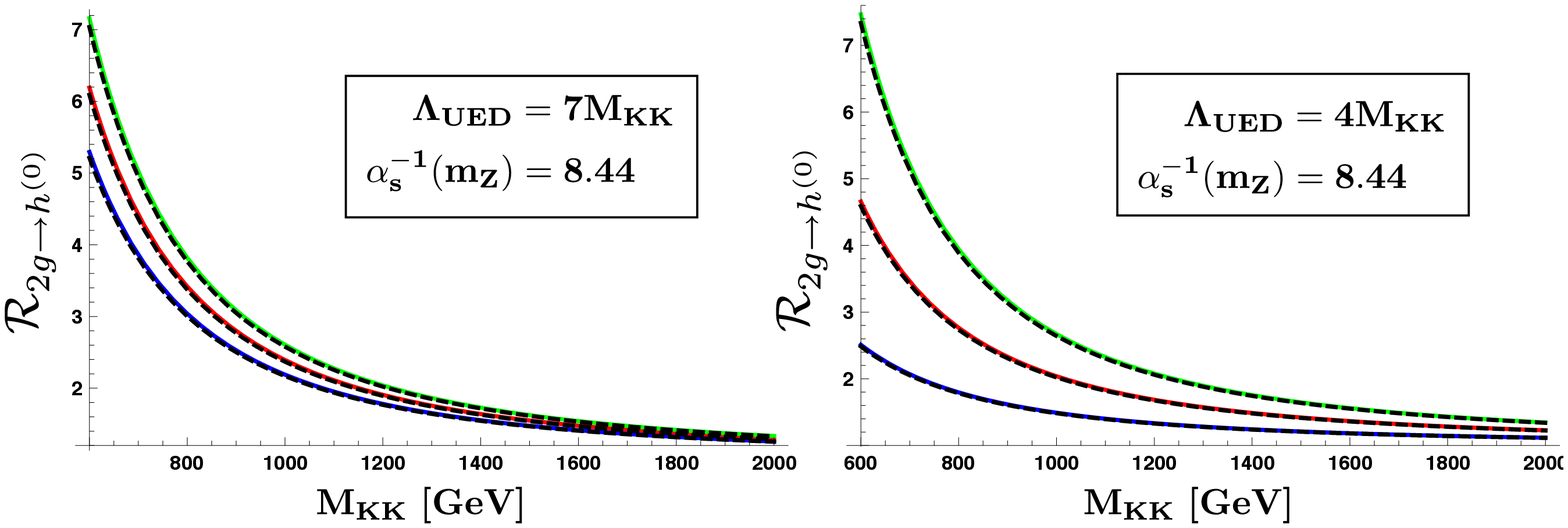}
\caption{
These plots represent the values of the ratios $\mathcal{R}_{2g \rightarrow h^{(0)}}$ in 6D UED on $PS$ with/without threshold correction.
The red, blue, green curves show these with $C_{hgg} = 0, +1, -1$, respectively.
Each black dashed line near the lines for $m_h = 120\,\text{GeV}$ corresponds to that with $m_h = 145\,\text{GeV}$.
{In the left and right plots, which correspond to the high and low cutoff cases, respectively, we take the value of the QCD coupling strength as that at the Z boson mass scale.}
}
\label{production_PS}
\end{figure}
\begin{figure}[H]
\centering
\includegraphics[width=\columnwidth, clip]{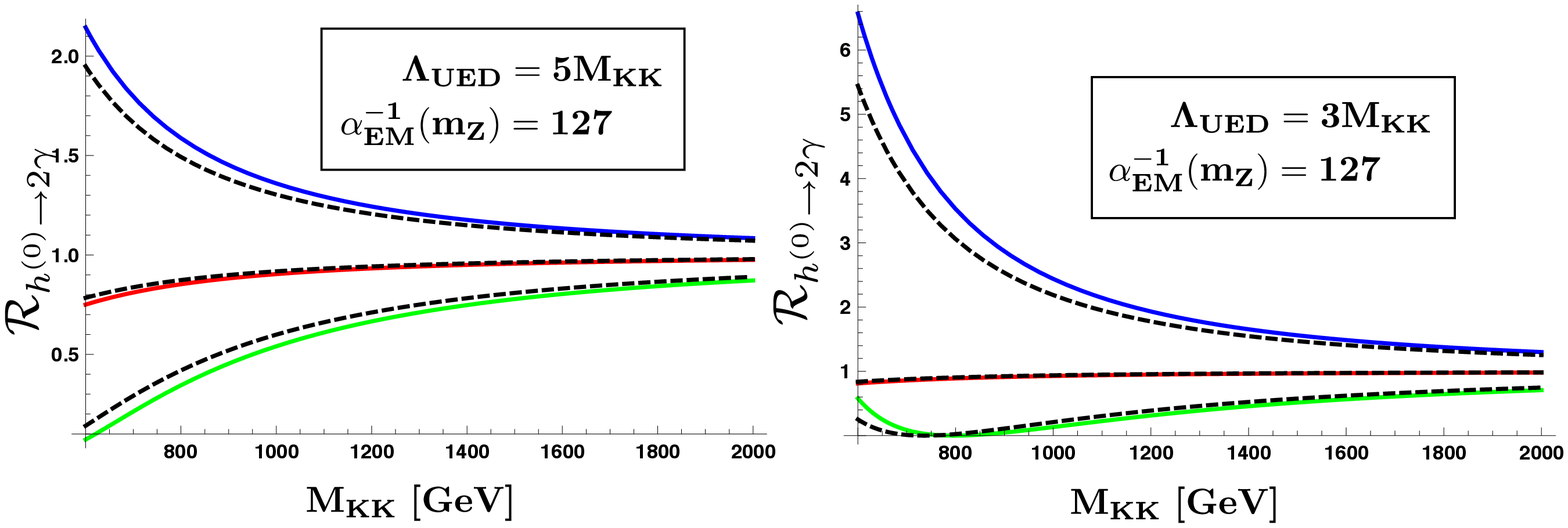}
\caption{
These plots represent the values of the ratios $\mathcal{R}_{h^{(0)} \rightarrow 2\gamma}$ in 6D UED on $T^2/Z_4$ with/without threshold correction.
The red, blue, green curves show these with $C_{h\gamma\gamma} = 0, +1, -1$, respectively.
Each black dashed line near the lines for $m_h = 120\,\text{GeV}$ corresponds to that with $m_h = 145\,\text{GeV}$.
{In the left and right plots, which correspond to the high and low cutoff cases, respectively, we take the value of the electromagnetic coupling strength as that at the Z boson mass scale.}
}
\label{decay_T2Z4}
\end{figure}
\begin{figure}[H]
\centering
\includegraphics[width=\columnwidth, clip]{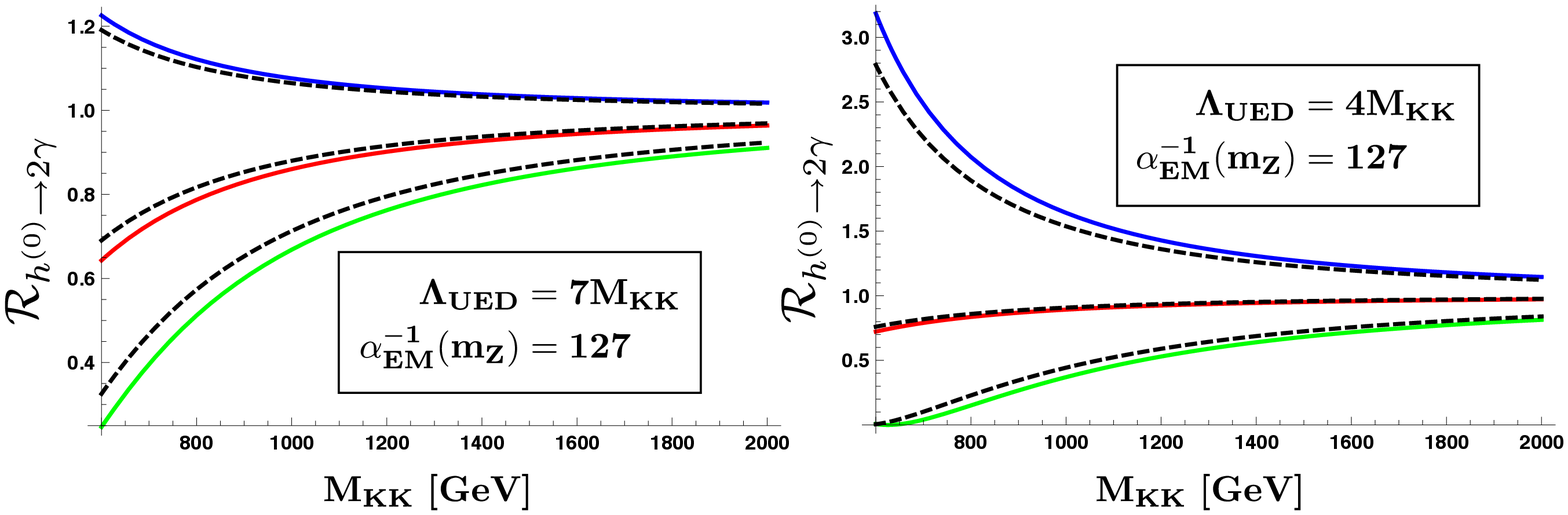}
\caption{
These plots represent the values of the ratios $\mathcal{R}_{h^{(0)} \rightarrow 2\gamma}$ in 6D UED on $S^2/Z_2$ with/without threshold correction.
The red, blue, green curves show these with $C_{h\gamma\gamma} = 0, +1, -1$, respectively.
Each black dashed line near the lines for $m_h = 120\,\text{GeV}$ corresponds to that with $m_h = 145\,\text{GeV}$.
{In the left and right plots, which correspond to the high and low cutoff cases, respectively, we take the value of the electromagnetic coupling strength as that at the Z boson mass scale.}
}
\label{decay_S2Z2}
\end{figure}
\begin{figure}[H]
\centering
\includegraphics[width=\columnwidth, clip]{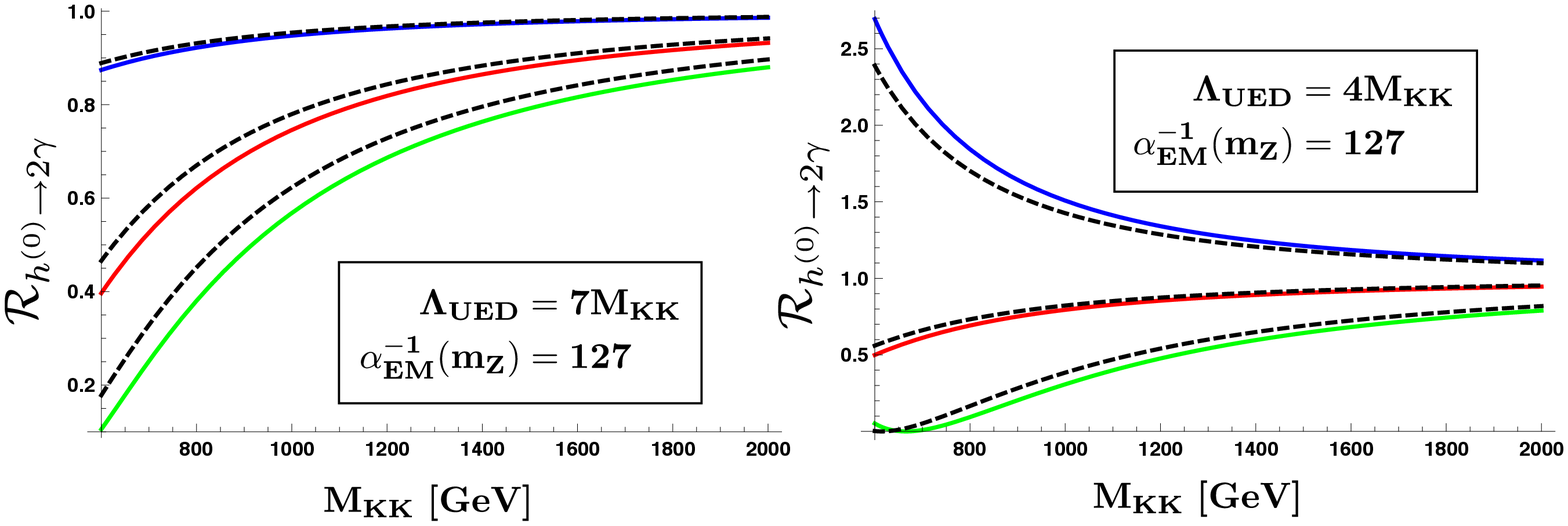}
\caption{
These plots represent the values of the ratios $\mathcal{R}_{h^{(0)} \rightarrow 2\gamma}$ in 6D UED on $PS$ with/without threshold correction.
The red, blue, green curves show these with $C_{h\gamma\gamma} = 0, +1, -1$, respectively.
Each black dashed line near the lines for $m_h = 120\,\text{GeV}$ corresponds to that with $m_h = 145\,\text{GeV}$.
{In the left and right plots, which correspond to the high and low cutoff cases, respectively, we take the value of the electromagnetic coupling strength as that at the Z boson mass scale.}
}
\label{decay_PS}
\end{figure}
\begin{table}[H]
{
\begin{center}
\begin{tabular}{|l|l||c|}
\hline
value of $C_{hgg}$ & value of $C_{h\gamma\gamma}$ & color/shape of curve \\ \hline
$C_{hgg}=0$ & $C_{h\gamma\gamma}=0$ & black, dot-dashed \\ \hline
$C_{hgg}=+1$ & $C_{h\gamma\gamma}=0$ & red \\ \hline
$C_{hgg}=0$ & $C_{h\gamma\gamma}=+1$ & blue \\ \hline
$C_{hgg}=-1$ & $C_{h\gamma\gamma}=0$ & green \\ \hline
$C_{hgg}=0$ & $C_{h\gamma\gamma}=-1$ & magenta \\ \hline
$C_{hgg}=+1$ & $C_{h\gamma\gamma}=+1$ & yellow, dotted \\ \hline
$C_{hgg}=-1$ & $C_{h\gamma\gamma}=+1$ & orange, dotted \\ \hline
$C_{hgg}=+1$ & $C_{h\gamma\gamma}=-1$ & cyan, dotted \\ \hline
$C_{hgg}=-1$ & $C_{h\gamma\gamma}=-1$ & brown, dotted \\ \hline
\end{tabular}
\caption{Our convention about the color/shape of curves in Figs.~\ref{combined_T2Z4},\ref{combined_S2Z2},\ref{combined_PS} (total ratios).}
\label{colorcodes}
\end{center}
}
\end{table}
\begin{figure}[H]
\centering
\includegraphics[width=\columnwidth, clip]{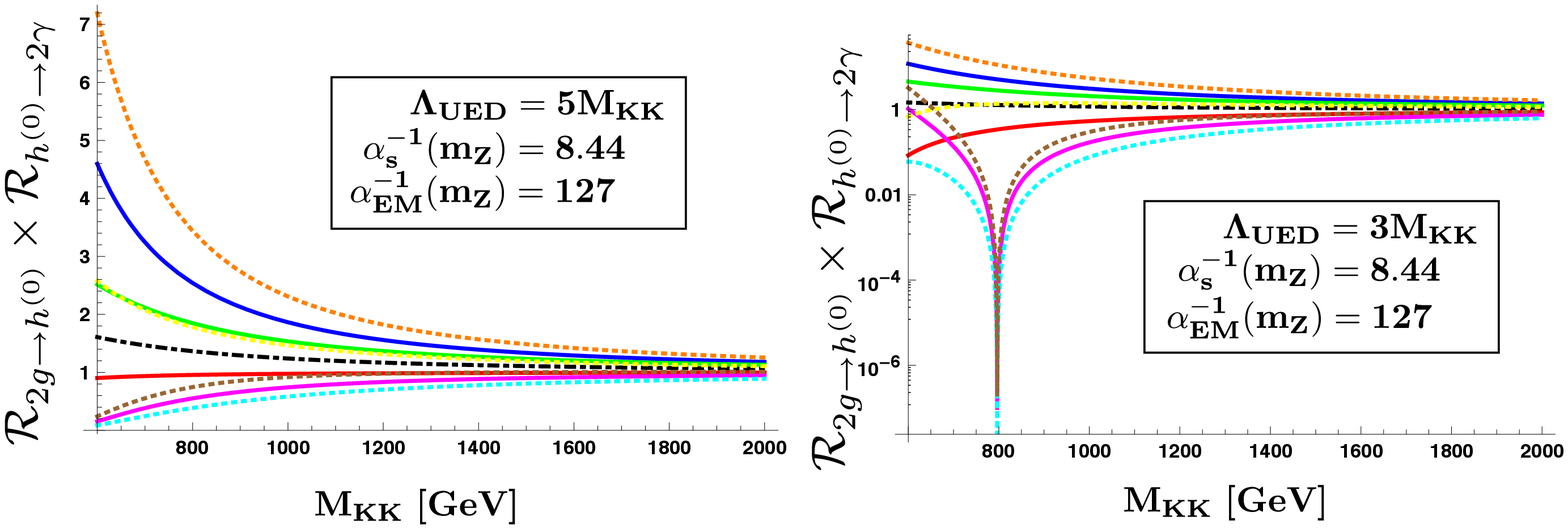}
\caption{
These plots represent the values of the total ratios $\Delta$ in 6D UED on $T^2/Z_4$ with/without threshold corrections with $m_h = 120\,\text{GeV}$.
The color/shape convention is summarized in Table~\ref{colorcodes}.
{In the left and right plots, which correspond to the high and low cutoff cases, respectively, we take the values of the QCD and electromagnetic coupling strengths as that at the Z boson mass scale.}
}
\label{combined_T2Z4}
\end{figure}
\begin{figure}[H]
\centering
\includegraphics[width=\columnwidth, clip]{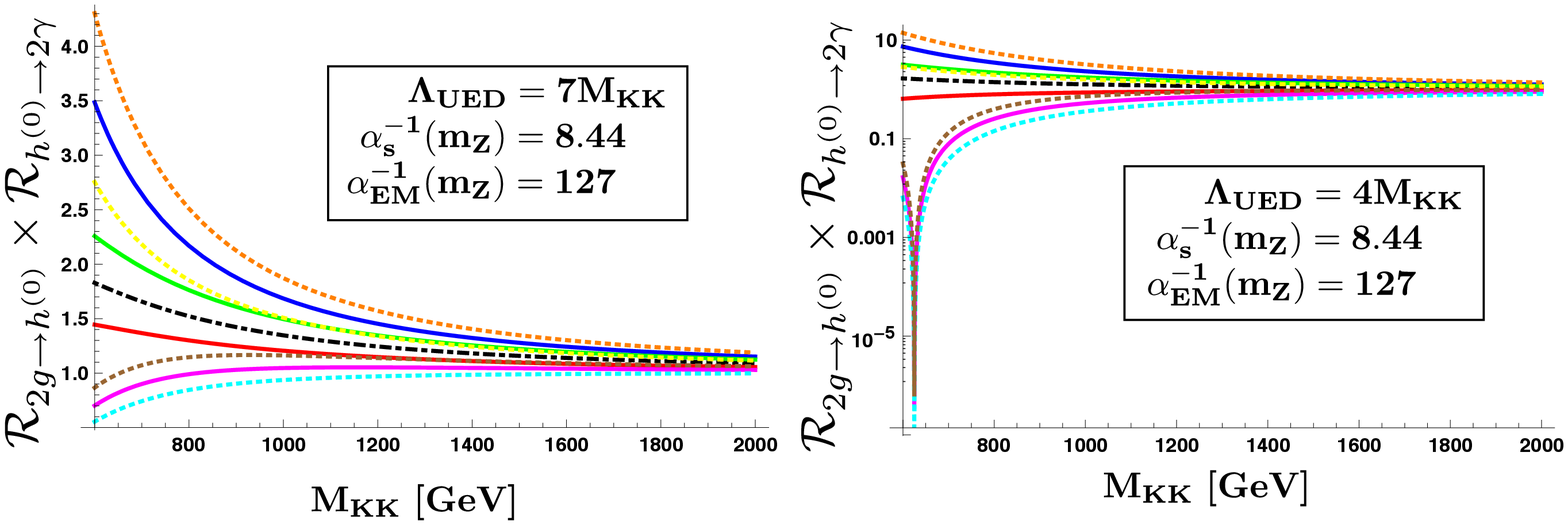}
\caption{
These plots represent the values of the total ratios $\Delta$ in 6D UED on $S^2/Z_2$ with/without threshold corrections with $m_h = 120\,\text{GeV}$.
The color/shape convention is summarized in Table~\ref{colorcodes}.
In the left and right plots, which correspond to the high and low cutoff cases, respectively, we take the values of the QCD and electromagnetic coupling strengths as that at the Z boson mass scale.
}
\label{combined_S2Z2}
\end{figure}
\begin{figure}[H]
\centering
\includegraphics[width=\columnwidth, clip]{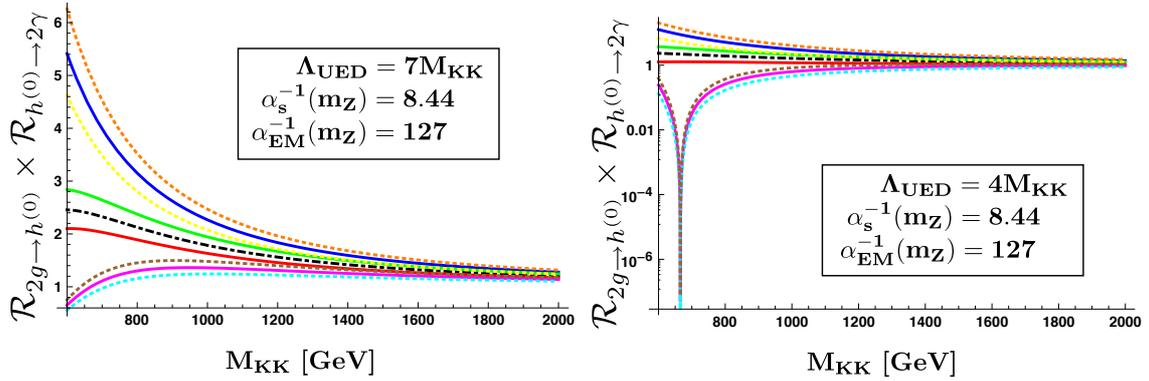}
\caption{
These plots represent the values of the total ratios $\Delta$ in 6D UED on $PS$ with/without threshold corrections with $m_h = 120\,\text{GeV}$.
The color/shape convention is summarized in Table~\ref{colorcodes}.
{In the left and right plots, which correspond to the high and low cutoff cases, respectively, we take the values of the QCD and electromagnetic coupling strengths as that at the Z boson mass scale.}
}
\label{combined_PS}
\end{figure}

\appendix

\section{Feynman Rules containing scalar particle
\label{Feynman Rules containing scalar particle}}

In this appendix, we list the Feynman rules containing scalar particle
in the {}'t Hooft-Feynman gauge.
We omit the rules containing no scalar particle, which are the same with the
corresponding rules of the SM for the zero modes alone.
In the vertices all momenta $(k_1,k_2)$ and directions of propagation are considered as incoming.
{$g^{(2)}$ and $e$ are $SU(2)_L$ the {4D} gauge coupling and the {4D} elementary electric charge, respectively.}
\vspace{10mm}

\fbox{\includegraphics[width=30mm, clip]{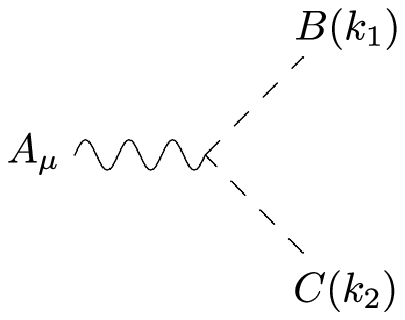}}
\raisebox{12mm}{$\displaystyle = -i (k_1 - k_2)_{\mu} \mathcal{F}$,}
\hspace{10mm}
\raisebox{10mm}{$
\begin{array}{|c|c|c||c|}
\hline
A_{\mu} & B & C & \mathcal{F} \\
\hline
W^{+(m,n)}_{\mu} & G^{-(m,n)} & h^{(0)} & \frac{g^{(2)}}{2} \frac{-im_W}{m_{W,(m,n)}} \\
\hline
W^{+(m,n)}_{\mu} & a^{-(m,n)} & h^{(0)} &  \frac{g^{(2)}}{2} \frac{m_{(m,n)}}{m_{W,(m,n)}} \\
\hline
W^{-(m,n)}_{\mu} & G^{+(m,n)} & h^{(0)} &  \frac{g^{(2)}}{2} \frac{-im_W}{m_{W,(m,n)}} \\
\hline
W^{-(m,n)}_{\mu} & a^{+(m,n)} & h^{(0)} &  \frac{g^{(2)}}{2} \frac{-m_{(m,n)}}{m_{W,(m,n)}} \\
\hline
A^{(0)}_{\mu} & G^{+(m,n)} & G^{-(m,n)} & e\\
\hline
A^{(0)}_{\mu} & a^{+(m,n)} & a^{-(m,n)} & e\\
\hline
A^{(0)}_{\mu} & H^{+(m,n)} & H^{-(m,n)} & e\\
\hline
\end{array}
$}

\vspace{10mm}

\fbox{\includegraphics[width=30mm, clip]{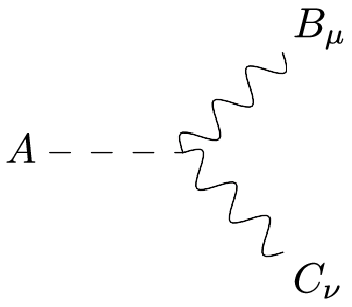}}
\raisebox{12mm}{$\displaystyle = i \eta_{\mu \nu} \mathcal{F}$,}
\hspace{10mm}
\raisebox{15mm}{$
\begin{array}{|c|c|c||c|}
\hline
A & B_{\mu} & C_{\nu} & \mathcal{F} \\
\hline
h^{(0)} & W^{+(m,n)}_{\mu} & W^{-(m,n)}_{\nu} & m_W g^{(2)} \\
\hline
G^{+(m,n)} & W^{-(m,n)}_{\mu} & A^{(0)}_{\nu} & -i e m_{W,(m,n)} \\
\hline
G^{-(m,n)} & W^{+(m,n)}_{\mu} & A^{(0)}_{\nu} & i e m_{W,(m,n)} \\
\hline
\end{array}
$}

\vspace{10mm}

\fbox{\includegraphics[width=30mm, clip]{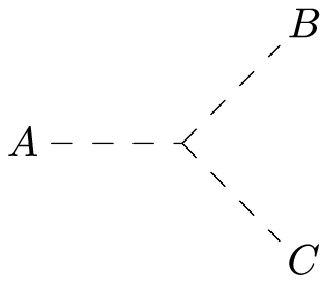}}
\raisebox{12mm}{$\displaystyle = -i m_W g^{(2)} \mathcal{F}$,} \\
\begin{center}
$
\begin{array}{|c|c|c||c|}
\hline
A & B & C & \mathcal{F} \\
\hline
h^{(0)} & G^{+(m,n)} & G^{-(m,n)} & \frac{m_h^2}{2m_{W,(m,n)}^2} \\
\hline
h^{(0)} & G^{+(m,n)} & a^{-(m,n)} & i \frac{m_{(m,n)}}{2m_{W,(m,n)}^2}
\left( \frac{m_h^2}{m_W}  -  \frac{m_{W,(m,n)}^2}{m_W}  \right) \\
\hline
h^{(0)} & a^{+(m,n)} & G^{-(m,n)} & i \frac{m_{(m,n)}}{2m_{W,(m,n)}^2}
\left( - \frac{m_h^2}{m_W}  +  \frac{m_{W,(m,n)}^2}{m_W}  \right) \\
\hline
h^{(0)} & a^{+(m,n)} & a^{-(m,n)} & \frac{1}{2m_{W,(m,n)}^2}
\left( \frac{m_h^2}{m_W^2} m_{(m,n)}^2  + 2 {m_{W,(m,n)}^2} \right) \\
\hline
h^{(0)} & H^{+(m,n)} & H^{-(m,n)} & 1 \\
\hline
\end{array}
$
\end{center}

\vspace{10mm}

\fbox{\includegraphics[width=30mm, clip]{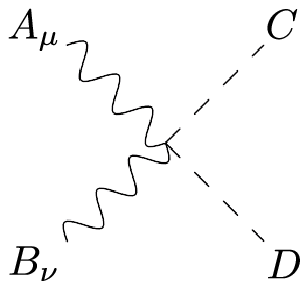}}
\raisebox{12mm}{$\displaystyle = i \eta_{\mu \nu} \mathcal{F}$,}
\hspace{10mm}
\raisebox{15mm}{$
\begin{array}{|c|c|c|c||c|}
\hline
A_{\mu} & B_{\nu} & C & D & \mathcal{F} \\
\hline
W_{\mu}^{+(m,n)} & A_{\nu}^{(0)} & G^{-(m,n)} & h^{(0)} & \frac{eg^{(2)}}{2} 
\left( \frac{i m_W}{m_{W,(m,n)}} \right) \\
\hline
W_{\mu}^{-(m,n)} & A_{\nu}^{(0)} & G^{+(m,n)} & h^{(0)} & \frac{eg^{(2)}}{2} 
\left( \frac{-i m_W}{m_{W,(m,n)}} \right) \\
\hline
W_{\mu}^{+(m,n)} & A_{\nu}^{(0)} & a^{-(m,n)} & h^{(0)} & \frac{eg^{(2)}}{2} 
\left( \frac{-m_{(m,n)}}{m_{W,(m,n)}} \right) \\
\hline
W_{\mu}^{-(m,n)} & A_{\nu}^{(0)} & a^{+(m,n)} & h^{(0)} & \frac{eg^{(2)}}{2} 
\left( \frac{-m_{(m,n)}}{m_{W,(m,n)}} \right) \\
\hline
A_{\mu}^{(0)} & A_{\nu}^{(0)} & G^{+(m,n)} & G^{-(m,n)} & 2 e^2 \\
\hline
A_{\mu}^{(0)} & A_{\nu}^{(0)} & a^{+(m,n)} & a^{-(m,n)} & 2 e^2 \\
\hline
A_{\mu}^{(0)} & A_{\nu}^{(0)} & H^{+(m,n)} & H^{-(m,n)} & 2 e^2 \\
\hline
\end{array}
$}

\section{{Detail on threshold correction}
\label{Detail on threshold correction}}
{In this Appendix, we explain the concrete forms of threshold corrections in the gluon fusion $(2g \rightarrow h^{(0)})$ and the Higgs decay to two photons $(h^{(0)} \rightarrow 2\gamma)$.
The parts of the Lagrangian describing the former ($\mathcal{L}_{hgg}$) and the latter ($\mathcal{L}_{h\gamma\gamma}$) processes
are defined as
\al{
\mathcal{L}_{hgg} &= - \frac{1}{4} \frac{C_{hgg}}{{\Lambda_{\text{UED}}}^2} V_2 F_{MN}^{[\text{QCD}]} F^{[\text{QCD}]MN} H^{\dagger} H,\\
\mathcal{L}_{h\gamma\gamma} &= - \frac{1}{4} \frac{C_{h\gamma\gamma}}{{\Lambda_{\text{UED}}}^2} V_2 F_{MN}^{[\text{QED}]} F^{[\text{QED}]MN} H^{\dagger} H,
}
where $C_{hgg}$ and $C_{h\gamma\gamma}$ are dimensionless coefficients characterizing the processes,
$\Lambda_{\text{UED}}$ is 6D UED cutoff, $V_2$ is the volume of the two extra dimensions, $H$ is the 6D Higgs doublet, and
$F_{MN}^{[\text{QCD}]}$ ($F_{MN}^{[\text{QED}]}$) is the 6D field strength of gluon (photon), respectively.
It is an important thing that the Higgs doublet should be introduced in these effective operators
in a bilinear form of $H^{\dagger} H$ because the electroweak symmetry breaking (EWSB) is realized by the usual Higgs mechanism in 6D UED models.
After EWSB and KK reduction, the Higgs doublet can acquire the VEV as $\langle H \rangle = 
\paren{0,v}^{\text{T}} /\sqrt{2V_2}$, where $v \simeq 246\,\text{GeV}$,
and we would like to focus on the parts, which are
\al{
\mathcal{L}_{hgg} &\supset - \frac{v/\sqrt{2}}{4} \frac{C_{hgg}}{{\Lambda_{\text{UED}}}^2} F_{\mu\nu}^{(0)[\text{QCD}]} F^{(0)[\text{QCD}]\mu\nu}  h^{(0)}, \label{EO_hgg}\\
\mathcal{L}_{h\gamma\gamma} &\supset  - \frac{v/\sqrt{2}}{4} \frac{C_{h\gamma\gamma}}{{\Lambda_{\text{UED}}}^2}  F_{\mu\nu}^{(0)[\text{QED}]} F^{(0)[\text{QED}]\mu\nu} h^{(0)}. \label{EO_g2gamma}
}
The superscript ``$(0)$" means that the fields are zero modes and we take integration toward the two extra spacial directions in Eqs.~(\ref{EO_hgg}) and (\ref{EO_g2gamma}).
The two operators are understood as dimension-six operators in 4D point of view.
Finally, we write down the concrete forms of Feynman rules as follows:
\al{
\fbox{\includegraphics[width=50mm, clip]{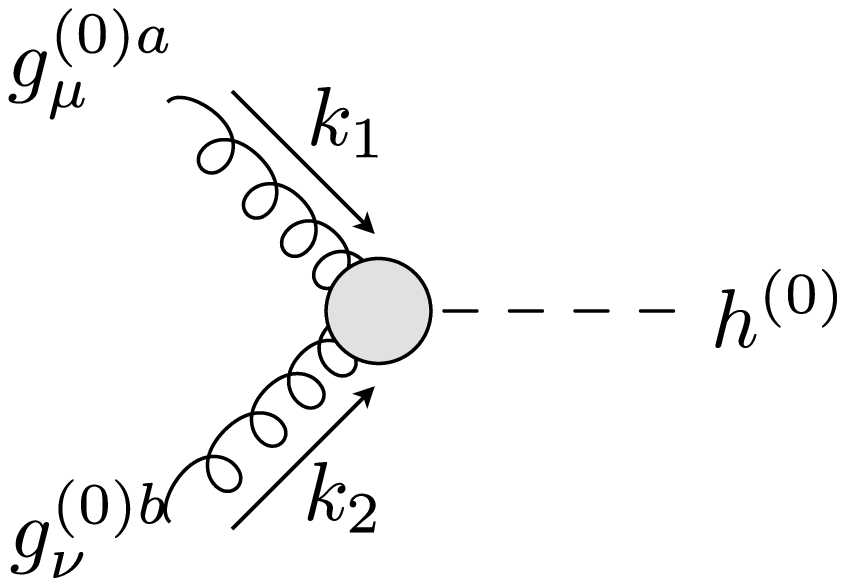}}
&\raisebox{16mm}{$\displaystyle \quad = -i \frac{C_{hgg}v/\sqrt{2}}{{\Lambda_{\text{UED}}}^2} \paren{k_{2\mu} k_{1\nu} - (k_1 \cdot k_2) \eta_{\mu\nu}} \delta^{ab}$,} \\
\fbox{\includegraphics[width=50mm, clip]{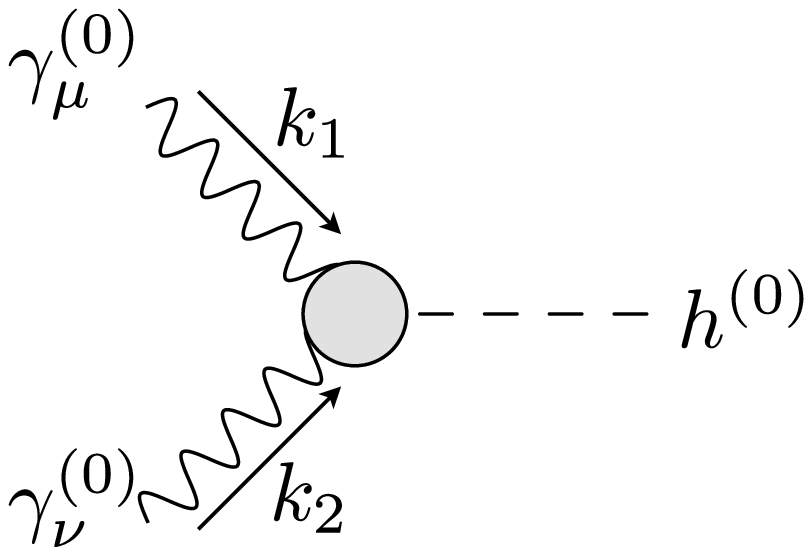}}
&\raisebox{16mm}{$\displaystyle \quad = -i \frac{C_{h\gamma\gamma}v/\sqrt{2}}{{\Lambda_{\text{UED}}}^2} \paren{k_{2\mu} k_{1\nu} - (k_1 \cdot k_2) \eta_{\mu\nu}}$,}
}
where $a,b$ are gluon indices.

}

\bibliographystyle{JHEP}
\bibliography{maintextref}

\providecommand{\href}[2]{#2}\begingroup\raggedright\begin{thebibliography}{10}

\bibitem{ArkaniHamed:1998rs}
N.~Arkani-Hamed, S.~Dimopoulos, and G.~R. Dvali, {\it {The hierarchy problem
  and new dimensions at a millimeter}},  {\em Phys. Lett.} {\bf B429} (1998)
  263--272, [\href{http://xxx.lanl.gov/abs/hep-ph/9803315}{{\tt
  hep-ph/9803315}}].

\bibitem{Randall:1999ee}
L.~Randall and R.~Sundrum, {\it {A large mass hierarchy from a small extra
  dimension}},  {\em Phys. Rev. Lett.} {\bf 83} (1999) 3370--3373,
  [\href{http://xxx.lanl.gov/abs/hep-ph/9905221}{{\tt hep-ph/9905221}}].

\bibitem{Antoniadis:1990ew}
I.~Antoniadis, {\it {A Possible new dimension at a few TeV}},  {\em Phys.
  Lett.} {\bf B246} (1990) 377--384.

\bibitem{Appelquist:2000nn}
T.~Appelquist, H.-C. Cheng, and B.~A. Dobrescu, {\it {Bounds on universal extra
  dimensions}},  {\em Phys. Rev.} {\bf D64} (2001) 035002,
  [\href{http://xxx.lanl.gov/abs/hep-ph/0012100}{{\tt hep-ph/0012100}}].

\bibitem{Agashe:2001ra}
K.~Agashe, N.~G. Deshpande, and G.~H. Wu, {\it {Can extra dimensions accessible
  to the SM explain the recent measurement of anomalous magnetic moment of the
  muon?}},  {\em Phys. Lett.} {\bf B511} (2001) 85--91,
  [\href{http://xxx.lanl.gov/abs/hep-ph/0103235}{{\tt hep-ph/0103235}}].

\bibitem{Agashe:2001xt}
K.~Agashe, N.~G. Deshpande, and G.~H. Wu, {\it {Universal extra dimensions and
  {b $\rightarrow$ s gamma}}},  {\em Phys. Lett.} {\bf B514} (2001) 309--314,
  [\href{http://xxx.lanl.gov/abs/hep-ph/0105084}{{\tt hep-ph/0105084}}].

\bibitem{Appelquist:2001jz}
T.~Appelquist and B.~A. Dobrescu, {\it {Universal extra dimensions and the muon
  magnetic moment}},  {\em Phys. Lett.} {\bf B516} (2001) 85--91,
  [\href{http://xxx.lanl.gov/abs/hep-ph/0106140}{{\tt hep-ph/0106140}}].

\bibitem{Appelquist:2002wb}
T.~Appelquist and H.-U. Yee, {\it {Universal extra dimensions and the Higgs
  boson mass}},  {\em Phys. Rev.} {\bf D67} (2003) 055002,
  [\href{http://xxx.lanl.gov/abs/hep-ph/0211023}{{\tt hep-ph/0211023}}].

\bibitem{Oliver:2002up}
J.~F. Oliver, J.~Papavassiliou, and A.~Santamaria, {\it {Universal extra
  dimensions and {Z $\rightarrow$ b anti-b}}},  {\em Phys. Rev.} {\bf D67}
  (2003) 056002, [\href{http://xxx.lanl.gov/abs/hep-ph/0212391}{{\tt
  hep-ph/0212391}}].

\bibitem{Chakraverty:2002qk}
D.~Chakraverty, K.~Huitu, and A.~Kundu, {\it {Effects of Universal Extra
  Dimensions on $B^0 - {\bar B^0}$ Mixing}},  {\em Phys. Lett.} {\bf B558}
  (2003) 173--181, [\href{http://xxx.lanl.gov/abs/hep-ph/0212047}{{\tt
  hep-ph/0212047}}].

\bibitem{Buras:2002ej}
A.~J. Buras, M.~Spranger, and A.~Weiler, {\it {The Impact of Universal Extra
  Dimensions on the Unitarity Triangle and Rare K and B Decays}},  {\em Nucl.
  Phys.} {\bf B660} (2003) 225--268,
  [\href{http://xxx.lanl.gov/abs/hep-ph/0212143}{{\tt hep-ph/0212143}}].

\bibitem{Colangelo:2006vm}
P.~Colangelo, F.~De~Fazio, R.~Ferrandes, and T.~N. Pham, {\it {Exclusive {$B
  \rightarrow K^{(\ast)} l_+ l_-$, $B \rightarrow K^{(\ast)} \nu \bar{\nu}$ and
  $B \rightarrow K^{\ast} \gamma$ } transitions in a scenario with a single
  universal extra dimension}},  {\em Phys. Rev.} {\bf D73} (2006) 115006,
  [\href{http://xxx.lanl.gov/abs/hep-ph/0604029}{{\tt hep-ph/0604029}}].

\bibitem{Gogoladze:2006br}
I.~Gogoladze and C.~Macesanu, {\it {Precision electroweak constraints on
  Universal Extra Dimensions revisited}},  {\em Phys. Rev.} {\bf D74} (2006)
  093012, [\href{http://xxx.lanl.gov/abs/hep-ph/0605207}{{\tt
  hep-ph/0605207}}].

\bibitem{Cheng:2002ej}
H.-C. Cheng, J.~L. Feng, and K.~T. Matchev, {\it {Kaluza-Klein dark matter}},
  {\em Phys. Rev. Lett.} {\bf 89} (2002) 211301,
  [\href{http://xxx.lanl.gov/abs/hep-ph/0207125}{{\tt hep-ph/0207125}}].

\bibitem{Servant:2002aq}
G.~Servant and T.~M.~P. Tait, {\it {Is the lightest Kaluza-Klein particle a
  viable dark matter candidate?}},  {\em Nucl. Phys.} {\bf B650} (2003)
  391--419, [\href{http://xxx.lanl.gov/abs/hep-ph/0206071}{{\tt
  hep-ph/0206071}}].

\bibitem{Kakizaki:2005en}
M.~Kakizaki, S.~Matsumoto, Y.~Sato, and M.~Senami, {\it {Significant effects of
  second KK particles on LKP dark matter physics}},  {\em Phys. Rev.} {\bf D71}
  (2005) 123522, [\href{http://xxx.lanl.gov/abs/hep-ph/0502059}{{\tt
  hep-ph/0502059}}].

\bibitem{Matsumoto:2005uh}
S.~Matsumoto and M.~Senami, {\it {Efficient coannihilation process through
  strong Higgs self-coupling in LKP dark matter annihilation}},  {\em Phys.
  Lett.} {\bf B633} (2006) 671--674,
  [\href{http://xxx.lanl.gov/abs/hep-ph/0512003}{{\tt hep-ph/0512003}}].

\bibitem{Burnell:2005hm}
F.~Burnell and G.~D. Kribs, {\it {The abundance of Kaluza-Klein dark matter
  with coannihilation}},  {\em Phys. Rev.} {\bf D73} (2006) 015001,
  [\href{http://xxx.lanl.gov/abs/hep-ph/0509118}{{\tt hep-ph/0509118}}].

\bibitem{Kakizaki:2006dz}
M.~Kakizaki, S.~Matsumoto, and M.~Senami, {\it {Relic abundance of dark matter
  in the minimal universal extra dimension model}},  {\em Phys. Rev.} {\bf D74}
  (2006) 023504, [\href{http://xxx.lanl.gov/abs/hep-ph/0605280}{{\tt
  hep-ph/0605280}}].

\bibitem{Kong:2005hn}
K.~Kong and K.~T. Matchev, {\it {Precise calculation of the relic density of
  Kaluza-Klein dark matter in universal extra dimensions}},  {\em JHEP} {\bf
  01} (2006) 038, [\href{http://xxx.lanl.gov/abs/hep-ph/0509119}{{\tt
  hep-ph/0509119}}].

\bibitem{Matsumoto:2007dp}
S.~Matsumoto, J.~Sato, M.~Senami, and M.~Yamanaka, {\it {Relic abundance of
  dark matter in universal extra dimension models with right-handed
  neutrinos}},  {\em Phys. Rev.} {\bf D76} (2007) 043528,
  [\href{http://xxx.lanl.gov/abs/0705.0934}{{\tt arXiv:0705.0934}}].

\bibitem{Kakizaki:2005uy}
M.~Kakizaki, S.~Matsumoto, Y.~Sato, and M.~Senami, {\it {Relic abundance of LKP
  dark matter in UED model including effects of second KK resonances}},  {\em
  Nucl. Phys.} {\bf B735} (2006) 84--95,
  [\href{http://xxx.lanl.gov/abs/hep-ph/0508283}{{\tt hep-ph/0508283}}].

\bibitem{Belanger:2010yx}
G.~Belanger, M.~Kakizaki, and A.~Pukhov, {\it {Dark matter in UED: The Role of
  the second KK level}},  {\em JCAP} {\bf 1102} (2011) 009,
  [\href{http://xxx.lanl.gov/abs/1012.2577}{{\tt arXiv:1012.2577}}].

\bibitem{Hisano:2010yh}
J.~Hisano, K.~Ishiwata, N.~Nagata, and M.~Yamanaka, {\it {Direct Detection of
  Vector Dark Matter}},  {\em Prog.Theor.Phys.} {\bf 126} (2011) 435--456,
  [\href{http://xxx.lanl.gov/abs/1012.5455}{{\tt arXiv:1012.5455}}].

\bibitem{Cheng:2002ab}
H.-C. Cheng, K.~T. Matchev, and M.~Schmaltz, {\it {Bosonic supersymmetry?
  Getting fooled at the CERN LHC}},  {\em Phys. Rev.} {\bf D66} (2002) 056006,
  [\href{http://xxx.lanl.gov/abs/hep-ph/0205314}{{\tt hep-ph/0205314}}].

\bibitem{Datta:2005zs}
A.~Datta, K.~Kong, and K.~T. Matchev, {\it {Discrimination of supersymmetry and
  universal extra dimensions at hadron colliders}},  {\em Phys. Rev.} {\bf D72}
  (2005) 096006, [\href{http://xxx.lanl.gov/abs/hep-ph/0509246}{{\tt
  hep-ph/0509246}}].

\bibitem{Matsumoto:2009tb}
S.~Matsumoto, J.~Sato, M.~Senami, and M.~Yamanaka, {\it {Productions of second
  Kaluza-Klein gauge bosons in the minimal universal extra dimension model at
  LHC}},  {\em Phys. Rev.} {\bf D80} (2009) 056006,
  [\href{http://xxx.lanl.gov/abs/0903.3255}{{\tt arXiv:0903.3255}}].

\bibitem{Dobrescu:2001ae}
B.~A. Dobrescu and E.~Poppitz, {\it {Number of fermion generations derived from
  anomaly cancellation}},  {\em Phys. Rev. Lett.} {\bf 87} (2001) 031801,
  [\href{http://xxx.lanl.gov/abs/hep-ph/0102010}{{\tt hep-ph/0102010}}].

\bibitem{Appelquist:2001mj}
T.~Appelquist, B.~A. Dobrescu, E.~Ponton, and H.-U. Yee, {\it {Proton stability
  in six dimensions}},  {\em Phys. Rev. Lett.} {\bf 87} (2001) 181802,
  [\href{http://xxx.lanl.gov/abs/hep-ph/0107056}{{\tt hep-ph/0107056}}].

\bibitem{ArkaniHamed:2000hv}
N.~Arkani-Hamed, H.-C. Cheng, B.~A. Dobrescu, and L.~J. Hall, {\it
  {Self-breaking of the standard model gauge symmetry}},  {\em Phys. Rev.} {\bf
  D62} (2000) 096006, [\href{http://xxx.lanl.gov/abs/hep-ph/0006238}{{\tt
  hep-ph/0006238}}].

\bibitem{Hashimoto:2003ve}
M.~Hashimoto, M.~Tanabashi, and K.~Yamawaki, {\it {Topped MAC with extra
  dimensions?}},  {\em Phys. Rev.} {\bf D69} (2004) 076004,
  [\href{http://xxx.lanl.gov/abs/hep-ph/0311165}{{\tt hep-ph/0311165}}].

\bibitem{Hashimoto:2004xz}
M.~Hashimoto and D.~K. Hong, {\it {Topcolor breaking through boundary
  conditions}},  {\em Phys. Rev.} {\bf D71} (2005) 056004,
  [\href{http://xxx.lanl.gov/abs/hep-ph/0409223}{{\tt hep-ph/0409223}}].

\bibitem{Cheng:2002iz}
H.-C. Cheng, K.~T. Matchev, and M.~Schmaltz, {\it {Radiative corrections to
  Kaluza-Klein masses}},  {\em Phys. Rev.} {\bf D66} (2002) 036005,
  [\href{http://xxx.lanl.gov/abs/hep-ph/0204342}{{\tt hep-ph/0204342}}].

\bibitem{Ponton:2005kx}
E.~Ponton and L.~Wang, {\it {Radiative effects on the chiral square}},  {\em
  JHEP} {\bf 11} (2006) 018,
  [\href{http://xxx.lanl.gov/abs/hep-ph/0512304}{{\tt hep-ph/0512304}}].

\bibitem{Burdman:2006gy}
G.~Burdman, B.~A. Dobrescu, and E.~Ponton, {\it {Resonances from two universal
  extra dimensions}},  {\em Phys. Rev.} {\bf D74} (2006) 075008,
  [\href{http://xxx.lanl.gov/abs/hep-ph/0601186}{{\tt hep-ph/0601186}}].

\bibitem{Dobrescu:2007xf}
B.~A. Dobrescu, K.~Kong, and R.~Mahbubani, {\it {Leptons and photons at the
  LHC: Cascades through spinless adjoints}},  {\em JHEP} {\bf 07} (2007) 006,
  [\href{http://xxx.lanl.gov/abs/hep-ph/0703231}{{\tt hep-ph/0703231}}].

\bibitem{Dobrescu:2007ec}
B.~A. Dobrescu, D.~Hooper, K.~Kong, and R.~Mahbubani, {\it {Spinless photon
  dark matter from two universal extra dimensions}},  {\em JCAP} {\bf 0710}
  (2007) 012, [\href{http://xxx.lanl.gov/abs/0706.3409}{{\tt
  arXiv:0706.3409}}].

\bibitem{Freitas:2007rh}
A.~Freitas and K.~Kong, {\it {Two universal extra dimensions and spinless
  photons at the ILC}},  {\em JHEP} {\bf 02} (2008) 068,
  [\href{http://xxx.lanl.gov/abs/0711.4124}{{\tt arXiv:0711.4124}}].

\bibitem{Freitas:2008vh}
A.~Freitas and U.~Haisch, {\it {Anti-B $\rightarrow$ X(s) gamma in two
  universal extra dimensions}},  {\em Phys. Rev.} {\bf D77} (2008) 093008,
  [\href{http://xxx.lanl.gov/abs/0801.4346}{{\tt arXiv:0801.4346}}].

\bibitem{Ghosh:2008dp}
K.~Ghosh and A.~Datta, {\it {Probing two Universal Extra Dimensions at
  International Linear Collider}},  {\em Phys. Lett.} {\bf B665} (2008)
  369--373, [\href{http://xxx.lanl.gov/abs/0802.2162}{{\tt arXiv:0802.2162}}].

\bibitem{Bertone:2009cb}
G.~Bertone, C.~B. Jackson, G.~Shaughnessy, T.~M.~P. Tait, and A.~Vallinotto,
  {\it {The WIMP Forest: Indirect Detection of a Chiral Square}},  {\em Phys.
  Rev.} {\bf D80} (2009) 023512, [\href{http://xxx.lanl.gov/abs/0904.1442}{{\tt
  arXiv:0904.1442}}].

\bibitem{Blennow:2009ag}
M.~Blennow, H.~Melbeus, and T.~Ohlsson, {\it {Neutrinos from Kaluza-Klein dark
  matter in the Sun}},  {\em JCAP} {\bf 1001} (2010) 018,
  [\href{http://xxx.lanl.gov/abs/0910.1588}{{\tt arXiv:0910.1588}}].

\bibitem{Dobrescu:2004zi}
B.~A. Dobrescu and E.~Ponton, {\it {Chiral compactification on a square}},
  {\em JHEP} {\bf 03} (2004) 071,
  [\href{http://xxx.lanl.gov/abs/hep-th/0401032}{{\tt hep-th/0401032}}].

\bibitem{Burdman:2005sr}
G.~Burdman, B.~A. Dobrescu, and E.~Ponton, {\it {Six-dimensional gauge theory
  on the chiral square}},  {\em JHEP} {\bf 02} (2006) 033,
  [\href{http://xxx.lanl.gov/abs/hep-ph/0506334}{{\tt hep-ph/0506334}}].

\bibitem{Cacciapaglia:2009pa}
G.~Cacciapaglia, A.~Deandrea, and J.~Llodra-Perez, {\it {A Dark Matter
  candidate from Lorentz Invariance in 6 Dimensions}},  {\em JHEP} {\bf 03}
  (2010) 083, [\href{http://xxx.lanl.gov/abs/0907.4993}{{\tt
  arXiv:0907.4993}}].

\bibitem{Maru:2009wu}
N.~Maru, T.~Nomura, J.~Sato, and M.~Yamanaka, {\it {The Universal Extra
  Dimensional Model with $S^2/Z_2$ extra-space}},  {\em Nucl. Phys.} {\bf B830}
  (2010) 414--433, [\href{http://xxx.lanl.gov/abs/0904.1909}{{\tt
  arXiv:0904.1909}}].

\bibitem{Dohi:2010vc}
H.~Dohi and K.-y. Oda, {\it {Universal Extra Dimensions on Real Projective
  Plane}},  {\em Phys. Lett.} {\bf B692} (2010) 114--120,
  [\href{http://xxx.lanl.gov/abs/1004.3722}{{\tt arXiv:1004.3722}}].

\bibitem{Flacke:2008ne}
T.~Flacke, A.~Menon, and D.~J. Phalen, {\it {Non-minimal universal extra
  dimensions}},  {\em Phys. Rev.} {\bf D79} (2009) 056009,
  [\href{http://xxx.lanl.gov/abs/0811.1598}{{\tt arXiv:0811.1598}}].

\bibitem{Park:2009cs}
S.~C. Park and J.~Shu, {\it {Split-UED and Dark Matter}},  {\em Phys. Rev.}
  {\bf D79} (2009) 091702, [\href{http://xxx.lanl.gov/abs/0901.0720}{{\tt
  arXiv:0901.0720}}].

\bibitem{Csaki:2010az}
C.~Csaki, J.~Heinonen, J.~Hubisz, S.~C. Park, and J.~Shu, {\it {5D UED: Flat
  and Flavorless}},  {\em JHEP} {\bf 1101} (2011) 089,
  [\href{http://xxx.lanl.gov/abs/1007.0025}{{\tt arXiv:1007.0025}}].

\bibitem{Haba:2009uu}
N.~Haba, K.-y. Oda, and R.~Takahashi, {\it {Top Yukawa Deviation in Extra
  Dimension}},  {\em Nucl. Phys.} {\bf B821} (2009) 74--128,
  [\href{http://xxx.lanl.gov/abs/0904.3813}{{\tt arXiv:0904.3813}}].

\bibitem{Haba:2009pb}
N.~Haba, K.-y. Oda, and R.~Takahashi, {\it {Dirichlet Higgs in extra-dimension,
  consistent with electroweak data}},  {\em Acta Phys.Polon.} {\bf B42} (2011)
  33--44, [\href{http://xxx.lanl.gov/abs/0910.3356}{{\tt arXiv:0910.3356}}].

\bibitem{Haba:2009wa}
N.~Haba, K.-y. Oda, and R.~Takahashi, {\it {Diagonal Kaluza-Klein expansion
  under brane localized potential}},  {\em Acta Phys. Polon.} {\bf B41} (2010)
  1291--1316, [\href{http://xxx.lanl.gov/abs/0910.4528}{{\tt
  arXiv:0910.4528}}].

\bibitem{Haba:2010xz}
N.~Haba, K.-y. Oda, and R.~Takahashi, {\it {Phenomenological Aspects of
  Dirichlet Higgs Model from Extra-Dimension}},  {\em JHEP} {\bf 07} (2010)
  079, [\href{http://xxx.lanl.gov/abs/1005.2306}{{\tt arXiv:1005.2306}}].

\bibitem{Nishiwaki:2010te}
K.~Nishiwaki and K.-y. Oda, {\it {Unitarity in Dirichlet Higgs Model}},  {\em
  Eur.Phys.J.} {\bf C71} (2011) 1786,
  [\href{http://xxx.lanl.gov/abs/1011.0405}{{\tt arXiv:1011.0405}}].

\bibitem{ATLAS-CONF-2011-135}
{\bf The ATLAS collaboration} Collaboration, {\it {Update of the Combination of
  Higgs Boson Searches in 1.0 to 2.3 $\text{fb}^{-1}$ of $pp$ Collisions Data
  Taken at $\sqrt{s} = 7$ TeV with the ATLAS Experiment at the LHC}}, . ATLAS
  NOTE, ATLAS-CONF-2011-135.

\bibitem{CMS_PAS_HIG-11-022}
{\bf The CMS collaboration} Collaboration, {\it {Search for standard model
  Higgs boson in pp collisions at $\sqrt{s} = 7$ TeV and integrated luminosity
  up to 1.7 $\text{fb}^{-1}$}}, . CMS PAS HIG-11-022.

\bibitem{ATLAS:2012ae}
{\bf ATLAS Collaboration} Collaboration, G.~Aad {\em et.~al.}, {\it {Combined
  search for the Standard Model Higgs boson using up to 4.9 fb$^{-1}$ of $pp$
  collision data at $\sqrt{s}=7$ TeV with the ATLAS detector at the LHC}},
  {\em Phys.Lett.} {\bf B710} (2012) 49--66,
  [\href{http://xxx.lanl.gov/abs/1202.1408}{{\tt arXiv:1202.1408}}].

\bibitem{Chatrchyan:2012tx}
{\bf CMS Collaboration} Collaboration, S.~Chatrchyan {\em et.~al.}, {\it
  {Combined results of searches for the standard model Higgs boson in $pp$
  collisions at $\sqrt{s}=7$ TeV}},  {\em Phys.Lett.} {\bf B710} (2012) 26--48,
  [\href{http://xxx.lanl.gov/abs/1202.1488}{{\tt arXiv:1202.1488}}].

\bibitem{Georgi:2000ks}
H.~Georgi, A.~K. Grant, and G.~Hailu, {\it {Brane couplings from bulk loops}},
  {\em Phys. Lett.} {\bf B506} (2001) 207--214,
  [\href{http://xxx.lanl.gov/abs/hep-ph/0012379}{{\tt hep-ph/0012379}}].

\bibitem{Lim:2009pj}
C.~S. Lim, N.~Maru, and K.~Nishiwaki, {\it {CP Violation due to
  Compactification}},  {\em Phys.Rev.} {\bf D81} (2010) 076006,
  [\href{http://xxx.lanl.gov/abs/0910.2314}{{\tt arXiv:0910.2314}}].

\bibitem{Scrucca:2003ut}
C.~A. Scrucca, M.~Serone, L.~Silvestrini, and A.~Wulzer, {\it {Gauge-Higgs
  Unification in Orbifold Models}},  {\em JHEP} {\bf 02} (2004) 049,
  [\href{http://xxx.lanl.gov/abs/hep-th/0312267}{{\tt hep-th/0312267}}].

\bibitem{Georgi:1977gs}
H.~M. Georgi, S.~L. Glashow, M.~E. Machacek, and D.~V. Nanopoulos, {\it {Higgs
  Bosons from Two Gluon Annihilation in Proton Proton Collisions}},  {\em Phys.
  Rev. Lett.} {\bf 40} (1978) 692.

\bibitem{Rizzo:1979mf}
T.~G. Rizzo, {\it {GLUON FINAL STATES IN HIGGS BOSON DECAY}},  {\em Phys. Rev.}
  {\bf D22} (1980) 178.

\bibitem{Petriello:2002uu}
F.~J. Petriello, {\it {Kaluza-Klein effects on Higgs physics in universal extra
  dimensions}},  {\em JHEP} {\bf 05} (2002) 003,
  [\href{http://xxx.lanl.gov/abs/hep-ph/0204067}{{\tt hep-ph/0204067}}].

\bibitem{Maru:2009cu}
N.~Maru, T.~Nomura, J.~Sato, and M.~Yamanaka, {\it {Higgs Production via Gluon
  Fusion in a Six Dimensional Universal Extra Dimension Model on $S^2/Z_2$}},
  {\em Eur. Phys. J.} {\bf C66} (2010) 283--287,
  [\href{http://xxx.lanl.gov/abs/0905.4554}{{\tt arXiv:0905.4554}}].

\bibitem{Passarino:1978jh}
G.~Passarino and M.~J.~G. Veltman, {\it {One Loop Corrections for e+ e-
  Annihilation Into mu+ mu- in the Weinberg Model}},  {\em Nucl. Phys.} {\bf
  B160} (1979) 151.

\bibitem{Denner:1991kt}
A.~Denner, {\it {Techniques for calculation of electroweak radiative
  corrections at the one loop level and results for W physics at LEP-200}},
  {\em Fortschr. Phys.} {\bf 41} (1993) 307--420,
  [\href{http://xxx.lanl.gov/abs/0709.1075}{{\tt arXiv:0709.1075}}].

\bibitem{Ellis:1975ap}
J.~R. Ellis, M.~K. Gaillard, and D.~V. Nanopoulos, {\it {A Phenomenological
  Profile of the Higgs Boson}},  {\em Nucl. Phys.} {\bf B106} (1976) 292.

\bibitem{RandjbarDaemi:1982hi}
S.~Randjbar-Daemi, A.~Salam, and J.~A. Strathdee, {\it {Spontaneous
  Compactification in Six-Dimensional Einstein- Maxwell Theory}},  {\em Nucl.
  Phys.} {\bf B214} (1983) 491--512.

\bibitem{Newman:1966ub}
E.~T. Newman and R.~Penrose, {\it {Note on the Bondi-Metzner-Sachs group}},
  {\em J. Math. Phys.} {\bf 7} (1966) 863--870.

\bibitem{Smith:1999cr}
G.~L. Smith {\em et.~al.}, {\it {Short range tests of the equivalence
  principle}},  {\em Phys. Rev.} {\bf D61} (2000) 022001.

\bibitem{Scrucca:2003ra}
C.~A. Scrucca, M.~Serone, and L.~Silvestrini, {\it {Electroweak symmetry
  breaking and fermion masses from extra dimensions}},  {\em Nucl. Phys.} {\bf
  B669} (2003) 128--158, [\href{http://xxx.lanl.gov/abs/hep-ph/0304220}{{\tt
  hep-ph/0304220}}].

\bibitem{Dienes:1998vh}
K.~R. Dienes, E.~Dudas, and T.~Gherghetta, {\it {Extra spacetime dimensions and
  unification}},  {\em Phys. Lett.} {\bf B436} (1998) 55--65,
  [\href{http://xxx.lanl.gov/abs/hep-ph/9803466}{{\tt hep-ph/9803466}}].

\bibitem{Dienes:1998vg}
K.~R. Dienes, E.~Dudas, and T.~Gherghetta, {\it {Grand unification at
  intermediate mass scales through extra dimensions}},  {\em Nucl. Phys.} {\bf
  B537} (1999) 47--108, [\href{http://xxx.lanl.gov/abs/hep-ph/9806292}{{\tt
  hep-ph/9806292}}].

\bibitem{Nishiwaki:2011gm}
K.~Nishiwaki, K.-y. Oda, N.~Okuda, and R.~Watanabe, {\it {Heavy Higgs at
  Tevatron and LHC in Universal Extra Dimension Models}},  {\em Phys.Rev.} {\bf
  D85} (2012) 035026, [\href{http://xxx.lanl.gov/abs/1108.1765}{{\tt
  arXiv:1108.1765}}].

\bibitem{Han:2003gf}
T.~Han, H.~E. Logan, B.~McElrath, and L.-T. Wang, {\it {Loop induced decays of
  the little Higgs: H $\rightarrow$ g g, gamma gamma}},  {\em Phys. Lett.} {\bf
  B563} (2003) 191--202, [\href{http://xxx.lanl.gov/abs/hep-ph/0302188}{{\tt
  hep-ph/0302188}}].

\bibitem{Dib:2003zj}
C.~Dib, R.~Rosenfeld, and A.~Zerwekh, {\it {Higgs production and decay in the
  little Higgs model}},  \href{http://xxx.lanl.gov/abs/hep-ph/0302068}{{\tt
  hep-ph/0302068}}.

\bibitem{Chen:2006cs}
C.-R. Chen, K.~Tobe, and C.~P. Yuan, {\it {Higgs boson production and decay in
  little Higgs models with T-parity}},  {\em Phys. Lett.} {\bf B640} (2006)
  263--271, [\href{http://xxx.lanl.gov/abs/hep-ph/0602211}{{\tt
  hep-ph/0602211}}].

\bibitem{Falkowski:2007hz}
A.~Falkowski, {\it {Pseudo-Goldstone Higgs Production via Gluon Fusion}},  {\em
  Phys. Rev.} {\bf D77} (2008) 055018,
  [\href{http://xxx.lanl.gov/abs/0711.0828}{{\tt arXiv:0711.0828}}].

\bibitem{Maru:2007xn}
N.~Maru and N.~Okada, {\it {Gauge-Higgs Unification at LHC}},  {\em Phys. Rev.}
  {\bf D77} (2008) 055010, [\href{http://xxx.lanl.gov/abs/0711.2589}{{\tt
  arXiv:0711.2589}}].

\bibitem{Maru:2008cu}
N.~Maru, {\it {Finite Gluon Fusion Amplitude in the Gauge-Higgs Unification}},
  {\em Mod. Phys. Lett.} {\bf A23} (2008) 2737--2750,
  [\href{http://xxx.lanl.gov/abs/0803.0380}{{\tt arXiv:0803.0380}}].

\bibitem{Rai:2005vy}
S.~K. Rai, {\it {UED effects on Higgs signals at LHC}},  {\em Int. J. Mod.
  Phys.} {\bf A23} (2008) 823--834,
  [\href{http://xxx.lanl.gov/abs/hep-ph/0510339}{{\tt hep-ph/0510339}}].

\bibitem{Nishiwaki:2011gk}
K.~Nishiwaki, K.-y. Oda, N.~Okuda, and R.~Watanabe, {\it {A Bound on Universal
  Extra Dimension Models from up to $\mathrm{2fb}^{-1}$ of LHC Data at 7TeV}},
  {\em Phys.Lett.} {\bf B707} (2012) 506--511,
  [\href{http://xxx.lanl.gov/abs/1108.1764}{{\tt arXiv:1108.1764}}].

\end{thebibliography}\endgroup


\end{document}